\documentclass[a4paper,11pt]{article}
\usepackage{amsmath,amssymb}
\usepackage{threeparttable}
\usepackage{multirow}
\usepackage{makecell}
\usepackage{array}

\usepackage[colorlinks=true
,urlcolor=blue
,anchorcolor=blue
,citecolor=blue
,filecolor=blue
,linkcolor=blue
,menucolor=blue
,linktocpage=true
,pdfproducer=medialab
,pdfa=true
]{hyperref}

\usepackage[
  backend=biber,
  style=authoryear,
  maxbibnames=99,
  doi=true,
  url=true,
  eprint=true
]{biblatex}
\usepackage{bm}
\usepackage{graphicx}
\usepackage{color}
\usepackage{booktabs}

\title{Negative Convexity in Pass-Through MBS:\\Mechanisms and Conditions}
\author{
    Kaname Imagawa
    \thanks{
        Nomura Asset Management Co., Ltd. Innovation Lab Dept., Full Address: 2-2-1 Toyosu, Koto-ku, Tokyo 135-0061, Japan, Email: k-imagawa@nomura-am.co.jp /  Hitotsubashi University Business School, Full Address: National Center of Sciences 2-1-2 Hitotsubashi, Chiyoda-ku Tokyo 101-8439, Japan, Email: jtvjeap@gmail.com
    }
    \thanks{
        Opinions expressed in this paper are those of the author, and do not necessarily reflect the views of Nomura Asset Management.
    }
    \thanks{
        Acknowledgments: 
        I am grateful to Professor Toshinao Yoshiba of Tokyo Metropolitan University for his valuable comments at the 63rd JAFEE Conference (Summer 2025).
        I would also like to express my sincere gratitude to Professor Hidetoshi Nakagawa of Hitotsubashi University Business School for his invaluable guidance throughout this research.
        I also thank the faculty members of Hitotsubashi University Business School for their valuable feedback during the doctoral summer camp.
        Furthermore, I thank Takashi Ohmoto, Yoshiteru Yamagishi, and the members of the Fixed Income Quantitative Research Team in the Innovation Lab Dept. at Nomura Asset Management for their many helpful comments and discussions.
    }
    \thanks{
        Funding: 
        This study was supported by JSPS KAKENHI Grant Number JP23K04285.
    }
}
\date{\today}

\begin{document}
\sloppy
\maketitle

\begin{abstract}
    This paper investigates the mechanisms and conditions underlying negative convexity in pass-through mortgage-backed securities (MBSs) backed by fixed-rate mortgages.
    We first derive an exact decomposition of MBS prices into three components: the present value of interest-rate-independent cash flows, the present value of principal repaid early through interest-rate-dependent prepayments, and an interest rate option component capturing the reduction in future cash flows due to prepayments.
    Under a first-order approximation, the latter two components resemble portfolios of floorlets and Asian-style interest rate options, respectively.
    We then apply a cumulant expansion to derive closed-form valuation approximations under the one-factor Hull--White model.
    Numerical results show that including the second cumulant improves accuracy relative to a first-cumulant approximation.
    Building on the price decomposition, we decompose duration and convexity into component-level contributions and express the condition for negative convexity in terms of their second-order interest rate sensitivities.
    Finally, phase diagrams across interest rate levels and refinancing sensitivities identify the regions where negative convexity occurs and the corresponding sign boundaries.
    The negative-convexity region expands as the long-run mean of refinancing sensitivity increases.
    Negative convexity can arise even when no refinancing incentive exists initially, while its magnitude may vary nonmonotonically with the interest rate level.
    These findings explain negative MBS convexity through the sensitivities of its price components and quantitatively characterize the conditions governing its occurrence.
    {\flushleft{{\bf Keywords:} Pass-Through MBS, Prepayment, Interest Rate Options, Duration, Negative Convexity, Cumulant Expansion}}
\end{abstract}

\newpage


\section{Introduction}

The future cash flows of pass-through mortgage-backed securities (MBSs) depend on the future path of interest rates through borrowers' prepayment behavior, making their prices and interest-rate sensitivities more complex than those of straight bonds.
In particular, declining interest rates tend to increase prepayments through mortgage refinancing and other mechanisms, thereby limiting the increase in MBS prices.
As a result, MBSs exhibit distinctive interest-rate sensitivities, such as duration shortening and negative convexity.
However, the presence of a prepayment option does not necessarily imply that MBS convexity is always negative.
Therefore, to understand the interest-rate sensitivities of MBSs, it is important to clarify the mechanisms through which negative convexity arises and the conditions under which it occurs.

In this paper, we investigate this issue from two perspectives: a component-wise decomposition of MBS prices and their interest-rate sensitivities, and phase diagrams of MBS convexity.
Specifically, we decompose the MBS price into three components: a component corresponding to interest-rate-independent cash flows, a component corresponding to the present value of principal repaid early through interest-rate-dependent prepayments, and an interest-rate-linked option component reflecting the reduction in future cash flows due to increased prepayments.
Furthermore, we derive an approximate closed-form expression for the MBS price and analytically evaluate the first- and second-order interest-rate sensitivities of each price component, thereby decomposing MBS duration and convexity into the contributions of these components.
This decomposition allows us to quantitatively examine which price components contribute to the emergence of negative convexity through their interest-rate sensitivities.
We also use phase diagrams of convexity with respect to the interest-rate level and refinancing sensitivity to numerically identify the parameter regions in which negative convexity occurs and the boundaries separating positive and negative convexity.
Through these analyses, we provide a unified examination of the mechanisms and conditions underlying the emergence of negative convexity.

\subsection{Pass-Through MBS}

Mortgage-backed securities (MBSs) are securitized products backed by principal and interest payments generated by pools of residential mortgage loans.
The cash flows of MBSs have a more complex structure than those of government and corporate bonds because they are affected by borrowers' early repayments, known as prepayments.

In this paper, we focus on MBSs backed by fixed-rate residential mortgage loans.
Under fixed-rate mortgages, borrowers generally make constant monthly principal and interest payments, which serve as the source of cash flows distributed to MBS investors.
Therefore, if all borrowers continue to make their contractual payments until their loans mature, the principal and interest cash flows received by MBS investors are predetermined.

On the other hand, under typical mortgage contracts, borrowers are allowed to repay all or part of the outstanding principal before the scheduled maturity date.
Such early repayments are referred to as prepayments.
When prepayments occur, principal is repaid ahead of schedule in addition to the scheduled principal and interest payments, thereby altering the amounts of subsequent principal and interest payments.
Consequently, the future cash flows of MBSs depend on borrowers' prepayment behavior, making their valuation more complex than that of straight bonds.

Pass-through MBSs are securities that distribute principal and interest payments collected from the underlying mortgage pool to investors in proportion to their ownership shares, after deducting servicing fees and other expenses.
In addition to scheduled principal and interest payments, principal collected through borrowers' prepayments is generally passed through directly to investors.
In this paper, we investigate the valuation of such pass-through MBSs.

\subsection{Previous Studies on MBS Optionality and Interest-Rate Sensitivities}

Under fixed-rate mortgage contracts, borrowers can repay the outstanding principal before the contractual maturity date, allowing the prepayment right to be interpreted as an option held by the borrower.
Therefore, investors in MBSs backed by such mortgages can be viewed as effectively holding a short position in the prepayment option.
The optionality associated with prepayments and its effects on MBS prices and interest-rate sensitivities have been extensively studied in the literature.

\textcite{Hendershott_Van_Order:1987} examine the valuation of fixed-rate mortgages within an option-pricing framework by relating the value of a prepayable mortgage to that of an otherwise equivalent nonprepayable mortgage and the values of contingent claims, including the prepayment option granted to the borrower.
As interest rates decline, borrowers have stronger incentives to refinance, increasing the value of the prepayment option and thereby limiting the increase in the value of prepayable mortgages relative to nonprepayable mortgages.
They also emphasize the importance of modeling actual prepayment behavior as a function of economic variables, including interest rates, rather than relying solely on a simple optimal exercise boundary.

\textcite{Jacob_et_al:1987} present a general framework for valuing securities whose future cash flows depend on state variables such as interest rates.
They demonstrate that securities whose cash flows change in response to future interest-rate movements require a valuation approach different from that used for fixed-rate bonds with predetermined cash flows.
Since the future cash flows of MBSs also depend on interest rates through prepayments, their framework is relevant to the present study in that it accounts for changes in prepayments and cash flows induced by interest-rate movements, rather than considering only changes in discount rates when evaluating MBS interest-rate sensitivities.

\textcite{Brazil:1988} is an example of a study that incorporates prepayment optionality into the valuation of MBS prices and their interest-rate sensitivities.
The study points out that changes in interest rates affect MBS prices not only through changes in discount rates but also through changes in future cash flows induced by changes in prepayment rates.
Consequently, conventional modified duration, which assumes fixed cash flows, cannot adequately capture the interest-rate sensitivities of MBSs, making it necessary to account for changes in prepayment rates and future cash flows in response to interest-rate movements.
The study evaluates MBS prices using the option-adjusted spread (OAS) and assesses MBS interest-rate risk using a measure of duration that incorporates changes in prepayment rates and future cash flows induced by interest-rate movements.

\textcite{Ronn_et_al:1995} provide a more direct examination of the relationship between MBS prices and the value of the embedded prepayment option.
Focusing on fixed-rate MBSs issued by the Government National Mortgage Association (Ginnie Mae MBSs), they construct synthetic Ginnie Mae MBSs under the assumption of no prepayments by replicating the scheduled cash flows using portfolios of U.S. Treasury securities.
They then estimate the market value of the embedded prepayment option as the difference between the value of the synthetic Ginnie Mae MBS without prepayments and the observed market price of the Ginnie Mae MBS.
In other words, they extract the value of the prepayment option based on no-arbitrage principles by comparing the value of a hypothetical security without prepayments with the actual MBS price.
This approach is closely related to the price decomposition developed in the present study, as both examine MBS prices in terms of the value of a hypothetical security unaffected by prepayments and the value of the prepayment option.
However, whereas their study extracts the option value from the difference between market prices and the values of replicating portfolios, the present study explicitly models the prepayment process and decomposes the theoretical MBS price itself into multiple components with distinct economic interpretations.

\textcite{Anderson_et_al:1993} provide a more analytical examination of the relationship between prepayments and MBS duration.
They derive an expression for MBS duration that accounts for the interest-rate sensitivity of prepayment rates and explicitly demonstrate how this sensitivity modifies conventional duration.
In particular, by introducing the elasticity of the fraction of mortgages remaining outstanding after prepayments with respect to interest rates, they analytically characterize the effects of changes in prepayment rates and future cash flows on MBS duration.
Furthermore, by defining duration with respect to the prevailing market mortgage rate rather than the cash-flow yield specific to each MBS, they provide a measure of interest-rate sensitivity that is comparable across MBSs with different contractual mortgage rates.
Their numerical examples demonstrate that when the market mortgage rate falls below the contractual mortgage rate, duration measures that ignore the interest-rate sensitivity of prepayments substantially overestimate MBS duration.
They further show that, in such regions, duration shortens as interest rates decline, resulting in negative convexity.

More recent studies have suggested that the interest-rate sensitivities of MBSs arising from prepayment optionality can affect not only the pricing characteristics of individual securities but also the broader bond market.
\textcite{Hanson:2014} examines the effects of MBS convexity on the bond market, focusing on how prepayments of fixed-rate mortgages cause MBS duration to vary with the level of interest rates.
When interest rates decline, stronger refinancing incentives lead to increased prepayments and shorter MBS duration, whereas rising interest rates reduce prepayments and extend duration.
The study demonstrates that these changes in MBS duration affect bond risk premia by altering the amount of interest-rate risk borne by investors in the bond market.
Therefore, MBS duration and convexity arising from prepayments are important not only for the valuation and risk management of individual MBSs but also for understanding price formation in the broader bond market.

Similarly, \textcite{Malkhozov_et_al:2016} investigate the relationship between interest-rate risk in the MBS market and the yield curve from both theoretical and empirical perspectives.
They focus on how changes in MBS duration and convexity arising from interest-rate-dependent prepayment behavior alter market participants' exposure to interest-rate risk.
In particular, they show that interest-rate risk associated with MBS duration is related to expected excess returns on bonds and that MBS convexity is associated with interest-rate volatility.
These findings suggest that the interest-rate sensitivities of MBSs arising from prepayments are closely related not only to the pricing characteristics of MBSs themselves but also to movements in the yield curve and the broader interest-rate market.

As discussed above, previous studies have extensively examined the interpretation of borrowers' prepayment rights as options and the significant effects of prepayment optionality on MBS prices, duration, and convexity.
In particular, the economic mechanism through which declining interest rates strengthen refinancing incentives and increase prepayments, thereby limiting the increase in MBS prices and shortening duration, has been widely recognized in the literature.
Furthermore, as demonstrated by \textcite{Hanson:2014} and \textcite{Malkhozov_et_al:2016}, these interest-rate sensitivities of MBSs are important not only for the valuation and risk management of individual securities but also for understanding risk premia and interest-rate movements in the broader bond market.

However, at least among the previous studies reviewed in this section, none has decomposed the theoretical MBS price within a single pricing expression into a component corresponding to interest-rate-independent cash flows, a component corresponding to principal repaid early through interest-rate-dependent prepayments, and a component reflecting the reduction in future cash flows due to prepayments, while also comparing the contributions of these components to the first- and second-order interest-rate sensitivities in a manner consistent with the price decomposition.
Furthermore, although it is widely recognized that negative MBS convexity arises from prepayment optionality, how the sign of convexity varies with different combinations of interest-rate levels and refinancing sensitivity remains an area for further investigation.
Focusing on these issues, this paper decomposes the MBS price into distinct components, analyzes their contributions to duration and convexity, and examines the boundaries between positive and negative convexity as well as the dependence of convexity on the interest-rate level.

\subsection{Previous Studies on Analytical MBS Valuation and Prepayment Models}

We next review previous studies on MBS valuation methods.
Historical prepayment data indicate that MBS prepayment rates respond nonlinearly to interest rates, and nonlinear functions of interest rates have been widely used to model prepayment rates.
However, this nonlinearity makes it difficult to derive closed-form expressions for MBS prices, and previous studies have proposed various methods to enable analytical or semi-analytical valuation while preserving the nonlinear dependence of prepayment rates on interest rates.

One of the studies most closely related to the present paper is \textcite{Kolbe_Zagst:2009}.
They develop an approximate closed-form valuation model for the efficient pricing of fixed-rate MBSs based on a reduced-form approach.
They decompose the prepayment rate into a refinancing-related component and a baseline component, and model the former using $\max$ and $\min$ functions of the difference between the weighted-average coupon rate and the 10-year interest rate, thereby capturing the S-shaped nonlinear relationship between prepayment rates and refinancing incentives.
They also model the baseline prepayment rate as a stochastic process driven by economic factors and derive approximate closed-form valuation formulas for pass-through MBSs, as well as interest-only (IO) and principal-only (PO) securities, by approximating exponential functions under a CIR-type interest-rate model.
The present study shares their approach of deriving closed-form expressions for MBS prices through suitable approximations while preserving the nonlinear dependence of prepayment rates on interest rates.

\textcite{Ozeki_et_al:2009} is also relevant to the approximation method employed in the present study.
They adopt an intensity-based approach in which prepayment times are modeled using a stochastic hazard rate and formulate the instantaneous prepayment rate using a proportional hazards model, thereby capturing the nonlinear increase in prepayment rates as interest rates decline.
They then reduce the valuation of MBSs to the evaluation of expectations of mortgage survival probabilities under forward measures associated with individual cash flow dates, where the survival probabilities are expressed as exponential functions of cumulative hazards.
By applying a cumulant expansion to the cumulative hazard, they obtain approximate valuation formulas for residential mortgage-backed securities (RMBSs), interest-only (IO) securities, and principal-only (PO) securities.

We next review other studies on the theoretical frameworks and functional forms used to model prepayment rates in MBS valuation.
\textcite{Chernov_et_al:2017} develop an arbitrage-free valuation model for Fannie Mae agency MBSs in which the prepayment function under the risk-neutral measure is estimated to be consistent with the cross-section of TBA prices across different coupons.
They decompose the instantaneous prepayment rate into a component driven by factors other than interest rates and a component that responds to refinancing incentives, and formulate the latter using a $\max$ function so that the prepayment rate responds to interest rates only when the refinancing incentive is positive.
This specification directly captures the nonlinear response of prepayment rates to interest rates through the functional form.

\textcite{Diep_et_al:2021} explain the cross-section of interest-rate-hedged returns on agency pass-through MBSs in terms of their exposure to prepayment risk and the associated risk premium.
They model the prepayment rate using an interest-rate-dependent $\max$ function and decompose its variation into a prepayment level factor and a factor capturing sensitivity to refinancing incentives.
Furthermore, they estimate shocks to these factors using the difference between predicted and realized prepayment rates and show that discount MBSs have positive sensitivity to prepayment shocks, whereas premium MBSs have negative sensitivity.

\textcite{Boyarchenko_et_al:2019} analyze the smile-shaped pattern observed across coupons in the option-adjusted spreads (OAS) of U.S. agency MBSs, as well as the factors driving their common time-series variation.
They exploit the fact that interest-only (IO) and principal-only (PO) strips associated with the same mortgage pool have opposite sensitivities to prepayments and estimate risk-neutral prepayment rates that equate the OAS of the two strips.
They specify a reduced-form prepayment model in which the aggregate prepayment rate is a weighted average of the prepayment rates of two groups of borrowers with different sensitivities to refinancing incentives.
The prepayment rate for each borrower group consists of an interest-rate-independent component that depends on loan age and a logistic function of the refinancing incentive.

Other approaches that enable analytical valuation while capturing the nonlinear dependence of prepayment rates on interest rates include \textcite{Rom_Poulsen:2007} and \textcite{Qian_et_al:2012}.
\textcite{Rom_Poulsen:2007} models the instantaneous prepayment rate as a quadratic function of the state variables driving interest rates and enables semi-analytical valuation of MBSs under a multifactor Gaussian interest-rate model by reducing the evaluation of expectations of exponential-quadratic functions to the solution of a system of ordinary differential equations.
In addition, \textcite{Qian_et_al:2012} assume that the instantaneous prepayment rate is inversely proportional to the short rate and derive an explicit pricing formula for pass-through MBSs by modeling both the short rate and non-interest-rate prepayment factors as CIR processes.

\subsection{Research Framework}

Building on the previous studies discussed above, this paper aims to clarify the mechanisms through which negative convexity, a characteristic feature of pass-through MBSs, arises and the conditions under which it occurs.
Specifically, we investigate the sources of negative convexity by decomposing MBS prices and their interest-rate sensitivities into economically interpretable components, and analyze the conditions under which negative convexity occurs using phase diagrams of convexity with respect to the interest-rate level and refinancing sensitivity.

To this end, we adopt a prepayment model that explicitly captures the nonlinear dependence of prepayment rates on interest rates.
Specifically, following \textcite{Chernov_et_al:2017} and \textcite{Boyarchenko_et_al:2019}, we formulate the prepayment process under the risk-neutral measure.
Furthermore, drawing on the prepayment functions employed by \textcite{Chernov_et_al:2017} and \textcite{Diep_et_al:2021}, we adopt a prepayment function involving the positive part, represented by a $\max$ function, to directly capture the nonlinear response to refinancing incentives.
In addition, following \textcite{Chernov_et_al:2017}, we model the sensitivity of prepayment rates to refinancing incentives as a stochastic process.

Under this framework, we first decompose the MBS price into three price components.
The first component corresponds to the value of an MBS whose cash flows are independent of interest rates.
The second component corresponds to the present value of interest-rate-dependent prepayments of principal, accounting for the reduction in outstanding principal, and makes a nonnegative contribution to the MBS price when the prepayment intensity is nonnegative.
In contrast, the third component represents the present value of baseline cash flows lost due to interest-rate-dependent prepayments.
This component takes the form of a short position in interest-rate-linked options whose number of units is independent of interest rates, and makes a nonpositive contribution to the MBS price when the prepayment intensity is nonnegative.
Furthermore, under a first-order approximation, the second component can be interpreted as a portfolio of interest-rate options resembling interest-rate floors, while the third component can be interpreted as a portfolio of Asian-style interest-rate options.
This price decomposition allows us to explicitly characterize the interest-rate optionality embedded in MBSs through prepayments in terms of multiple price contributions with distinct economic characteristics.

Furthermore, to quantitatively analyze the interest-rate sensitivities of each component based on this price decomposition, we apply a cumulant expansion to the expectations arising in MBS valuation and derive a closed-form approximation.
This allows us to analytically evaluate the first- and second-order interest-rate sensitivities of each price component and decompose the overall MBS duration and convexity into the contributions of these components.
Thus, the closed-form approximation serves not only as a means of efficiently valuing MBSs but also as a tool for understanding the sources of negative convexity through the interest-rate sensitivities of the individual price components.

In addition, to clarify the conditions under which negative convexity occurs, we construct phase diagrams of MBS convexity with respect to the interest-rate level and the initial value of refinancing sensitivity.
Furthermore, by varying the long-run mean of refinancing sensitivity, we analyze how the mean-reversion level of refinancing sensitivity affects the boundaries between positive and negative convexity.
This allows us to identify the parameter regions in which negative convexity occurs and examine its dependence on the interest-rate level.

Based on the foregoing discussion, this paper makes two main contributions.
First, we investigate the mechanisms through which negative convexity arises in MBSs and the conditions under which it occurs from two perspectives: a component-wise decomposition of MBS prices and their interest-rate sensitivities, and phase diagrams of convexity.
As discussed in the previous studies, the interpretation of borrowers' prepayment rights as options and the effects of prepayment optionality on duration shortening and negative convexity have been extensively examined in the literature.
Furthermore, studies such as \textcite{Ronn_et_al:1995} interpret the difference between the value of a hypothetical security without prepayments and the actual MBS price as the value of the prepayment option.
In contrast, the present study explicitly models the prepayment process and provides an exact decomposition of the theoretical MBS price into three components: a component corresponding to interest-rate-independent cash flows, a component corresponding to the present value of principal repaid early through interest-rate-dependent prepayments, and an interest-rate-linked option component reflecting the reduction in future cash flows due to increased prepayments.
Furthermore, we evaluate the first- and second-order interest-rate sensitivities of each price component and decompose the overall MBS duration and convexity into the contributions of these components.
This allows us to distinguish between principal repaid early through interest-rate-dependent prepayments and the reduction in future cash flows that would otherwise have been paid in the absence of such prepayments, and to quantify how the interest-rate sensitivities of the individual price components, accounting for the reduction in outstanding principal, contribute to the overall MBS duration and convexity.
In particular, we explicitly show that the occurrence of negative convexity is determined by the relative magnitudes of the second-order interest-rate sensitivities of the individual price components.

Furthermore, to clarify the conditions under which negative convexity occurs, we examine phase diagrams of MBS convexity with respect to the interest-rate level and the initial value of refinancing sensitivity.
We also analyze how the mean-reversion level of refinancing sensitivity affects the boundaries between positive and negative convexity by varying its long-run mean.
This allows us to numerically identify the parameter regions in which negative convexity occurs and examine the nonmonotonic dependence of convexity on the interest-rate level.
Thus, this paper identifies the mechanisms underlying negative convexity through the decomposition of MBS prices and their interest-rate sensitivities, and the conditions under which it occurs through phase diagrams of convexity.

Second, to enable the analysis of the contributions of individual price components to interest-rate sensitivities based on the above price decomposition, we derive closed-form approximations for pass-through MBS prices and their interest-rate sensitivities under a prepayment model that incorporates both nonlinear responses to refinancing incentives and stochastic variation in refinancing sensitivity itself.
As discussed above, \textcite{Chernov_et_al:2017} and \textcite{Boyarchenko_et_al:2019} formulate prepayment processes under the risk-neutral measure, emphasizing consistency with observed market prices.
Furthermore, \textcite{Chernov_et_al:2017} and \textcite{Diep_et_al:2021} employ prepayment functions involving a $\max$ function to capture the nonlinear response of prepayment rates to refinancing incentives.
In addition, \textcite{Chernov_et_al:2017} model the sensitivity of prepayment rates to refinancing incentives itself as a stochastic process.
Incorporating these features, we adopt a prepayment model that directly captures the nonlinear response to interest rates under the risk-neutral measure while modeling refinancing sensitivity as a stochastic process.
In particular, our prepayment model can be viewed as an extension of the model proposed by \textcite{Kolbe_Zagst:2009}, which assumes constant refinancing sensitivity in the interest-rate-dependent component, to the case where refinancing sensitivity varies stochastically.
We show that, by applying a cumulant expansion as in \textcite{Ozeki_et_al:2009}, a closed-form approximation for MBS prices can be derived even under this prepayment model.

The significance of deriving a closed-form approximation lies not only in providing a means of efficiently valuing MBSs but also in enabling an analytical assessment of the contributions to interest-rate sensitivities based on the price decomposition described above.
Although MBS prices and their interest-rate sensitivities can also be evaluated using Monte Carlo simulation, analytically evaluating the interest-rate sensitivities of individual price components under a nonlinear prepayment model is generally difficult.
In contrast, by combining a cumulant expansion with the one-factor Hull--White model, we derive analytical approximations for the first- and second-order interest-rate sensitivities of each price component, allowing us to compare their respective contributions.
We also assess the accuracy of the approximations through comparisons with Monte Carlo simulation results.
Thus, the closed-form approximation developed in this paper serves not only as a method for valuing MBSs but also as a tool for quantitatively understanding the mechanisms underlying negative convexity through the interest-rate sensitivities of the individual price components.

The remainder of this paper is organized as follows.
Section~\ref{sec:mbs_valuation_model} formulates the cash flows and prepayment process of MBSs and presents the basic framework for MBS valuation.
Section~\ref{sec:mbs_price_decomposition} decomposes the MBS price into three price components and provides an interpretation of the components associated with interest-rate-dependent prepayments in terms of interest-rate options based on a first-order approximation.
Section~\ref{sec:mbs_approximation} applies a cumulant expansion to the expectations arising in MBS valuation and derives a closed-form approximation under the one-factor Hull--White model.
Section~\ref{sec:mbs_numerical_analysis} examines the accuracy of the approximation through comparisons with Monte Carlo simulation results, decomposes MBS duration and convexity into the contributions of individual price components, and investigates the mechanisms underlying negative convexity.
Section~\ref{sec:convexity_phase_diagram} uses phase diagrams of convexity with respect to the interest-rate level and refinancing sensitivity to numerically identify the parameter regions in which negative convexity occurs and the boundaries between positive and negative convexity.
Furthermore, we examine the nonmonotonic behavior of convexity as the interest-rate level declines using a simplified model.
Section~\ref{sec:conclusion_discussion} presents the conclusions of this paper and discusses directions for future research.

\section{Valuation Model}
\label{sec:mbs_valuation_model}

In this section, we formulate the basic framework for valuing pass-through MBSs backed by fixed-rate mortgages.
In Section~\ref{subsec:mbs_price}, we derive an MBS valuation formula that accounts for cash flows and prepayments, and reformulate the expectations arising in the valuation using a change of measure to forward measures.
In Section~\ref{subsec:prepayment_process}, we decompose the prepayment rate into interest-rate-independent and interest-rate-dependent components, formulate a prepayment process that responds nonlinearly to refinancing incentives, and derive an expression for the MBS price based on this process.

Let $(\Omega, \mathcal{F}, \mathbb{Q})$ be a probability space, where $\mathbb{Q}$ denotes the risk-neutral probability measure.
We also assume that $\{\mathcal{F}_t\}$ is a complete filtration generated by a collection of Brownian motions, whose details are specified below.

\subsection{MBS Price}
\label{subsec:mbs_price}

In the following, we present only the results concerning MBS cash flows; for details of the derivations and mathematical formulation, see, for example, \textcite{Chernov_et_al:2017} and \textcite{Imagawa_Nakagawa:2025}.
Consider an MBS with a fixed mortgage rate $m$ applied to the underlying mortgage pool, maturity $T$, and a combined principal and interest payment rate per unit time $c$ (hereafter referred to as the payment rate).
For simplicity, we disregard servicing and guarantee fees.
These three parameters satisfy the relation $c=m/(1-e^{-mT})$, so only two of them can be specified independently.
We assume that MBS investors receive cash flows continuously over time.
In the absence of prepayments, the outstanding principal balance at time $t$, denoted by $\tilde{P}(t)$, is given by $(c/m)(1-e^{-m(T-t)})$.

When prepayments occur in the underlying mortgage pool, we assume that a fraction $\lambda(t)\mathrm{d}t$ of the outstanding principal at time $t$ is prepaid over the infinitesimal time interval $(t,t+\mathrm{d}t]$.
Here, $\lambda(t)$ is a stochastic process representing the instantaneous prepayment rate.
Therefore, the fraction of principal remaining outstanding at time $t$ due solely to prepayments is given by $e^{-\int_0^t\lambda(u)\mathrm{d}u}$.

Let $\left\{r(t)\right\}_{0\leq t\leq T+10}$ denote the short-rate process.
The specification of a particular interest-rate model is required only when evaluating the expectations arising in MBS valuation, as discussed later.
Therefore, at this stage, we assume only that the short-rate process follows a one-factor model driven by a one-dimensional Brownian motion $\left\{W_r^{\mathbb{Q}}(t)\right\}_{0\leq t\leq T+10}$ under $\mathbb{Q}$.
We define the short-rate process up to time $T+10$ because the prepayment rate introduced later depends on the 10-year zero-coupon yield $R_{10}(t)$ at time $t$.

The MBS price $V(t,T)$ at time $t$ ($0\leq t\leq T$) is given by
\begin{align}
    V(t,T)
    =&
    e^{-\int_0^t\lambda(u)\mathrm{d}u}
    \int_t^T
    \mathbb{E}^{\mathbb{Q}}
    \left[
        \left.
        e^{-\int_t^s(r(u)+\lambda(u))\mathrm{d}u}
        \left(
            c+\tilde{P}(s)\lambda(s)
        \right)
        \right|
        \mathcal{F}_t
    \right]
    \mathrm{d}s.
    \label{eq:MBS_value}
\end{align}
Since $e^{-\int_0^t\lambda(u)\mathrm{d}u}$, which represents the fraction of outstanding principal remaining at time $t$, is $\mathcal{F}_t$-measurable, it can be taken outside the conditional expectation given $\mathcal{F}_t$.

To evaluate the expectation on the right-hand side of \eqref{eq:MBS_value}, we define the $s$-forward probability measure $\mathbb{Q}_s$ for $s\geq t$ as follows:
\begin{align}
    L(t)
    :=
    \left.
        \frac{\mathrm{d}\mathbb{Q}_s}{\mathrm{d}\mathbb{Q}}
    \right|_{\mathcal{F}_t}
    =&
    \frac{D(t,s)/D(0,s)}{M(t)/M(0)},
    \label{eq:def_RND}
    \\
    \text{where} \quad
    M(t)
    :=&
    e^{\int_0^t r(u)\mathrm{d}u},
    \nonumber
    \\
    D(t,s)
    :=&
    \mathbb{E}^{\mathbb{Q}}
    \left[
        \left.
        e^{-\int_t^s r(u)\mathrm{d}u}
        \right|
        \mathcal{F}_t
    \right].
    \nonumber
\end{align}
Here, $M(t)$ denotes the value of the bank account (money-market account) at time $t$, and $D(t,s)$ denotes the price at time $t$ of a zero-coupon bond maturing at time $s$.

In the following, we fix the cash flow date $s$.
For a fixed $s$, we define the following auxiliary function for $t\leq v\leq s$:
\begin{align}
    Z^{(s)}(t,v)
    :=
    \mathbb{E}^{\mathbb{Q}_s}
    \left[
        \left.
        e^{-\int_t^v\lambda(u)\mathrm{d}u}
        \right|
        \mathcal{F}_t
    \right].
    \label{eq:MBS_value_result_05}
\end{align}
We also denote the diagonal value corresponding to the actual cash flow date $s$ by
\begin{align*}
    Z(t,s)
    :=
    Z^{(s)}(t,s).
\end{align*}

Furthermore, we define the effective prepayment rate at time $s$ as follows:
\begin{align}
    \bar{\lambda}^{\mathbb{Q}_s}(t,s)
    :=&
    -
    \left.
        \frac{\partial}{\partial v}
        \ln Z^{(s)}(t,v)
    \right|_{v=s}
    \nonumber
    \\
    =&
    \frac{
        \mathbb{E}^{\mathbb{Q}_s}
        \left[
            \left.
            e^{-\int_t^s\lambda(u)\mathrm{d}u}
            \lambda(s)
            \right|
            \mathcal{F}_t
        \right]
    }{
        \mathbb{E}^{\mathbb{Q}_s}
        \left[
            \left.
            e^{-\int_t^s\lambda(u)\mathrm{d}u}
            \right|
            \mathcal{F}_t
        \right]
    }.
    \label{eq:MBS_value_result_04}
\end{align}
Here, the partial derivative with respect to $v$ in \eqref{eq:MBS_value_result_04} is taken while holding $s$ and the corresponding $s$-forward probability measure $\mathbb{Q}_s$ fixed.
Therefore, this partial derivative does not account for changes in the probability measure itself as $s$ varies.

Using the $s$-forward probability measure $\mathbb{Q}_s$, the MBS price can be expressed as follows (see Appendix~\ref{app:derivation_mbs_value} for a proof):
\begin{align}
    V(t,T)
    =&
    e^{-\int_0^t\lambda(u)\mathrm{d}u}
    \int_t^T
    D(t,s)
    Z(t,s)
    \left(
        c
        +
        \tilde{P}(s)
        \bar{\lambda}^{\mathbb{Q}_s}(t,s)
    \right)
    \mathrm{d}s
    \nonumber
    \\
    =&
    \int_t^T
    V_{\mathrm{d}}(t,s)
    \mathrm{d}s.
    \label{eq:MBS_value_result_01}
\end{align}
For any cash flow date $v\in[t,T]$, we use the $v$-forward probability measure $\mathbb{Q}_v$, associated with maturity $v$, to write
\begin{align*}
    V_{\mathrm{d}}(t,v)
    :=&
    \bar{c}(t,v)
    D(t,v)
    Z(t,v),
    \\
    \bar{c}(t,v)
    :=&
    e^{-\int_0^t\lambda(u)\mathrm{d}u}
    \left(
        c
        +
        \tilde{P}(v)
        \bar{\lambda}^{\mathbb{Q}_v}(t,v)
    \right),
\end{align*}
where
\begin{align*}
    Z(t,v)
    :=
    Z^{(v)}(t,v),
    \qquad
    \bar{\lambda}^{\mathbb{Q}_v}(t,v)
    :=
    -
    \left.
        \frac{\partial}{\partial u}
        \ln Z^{(v)}(t,u)
    \right|_{u=v}.
\end{align*}
The derivative with respect to $u$ is taken while holding $\mathbb{Q}_v$ fixed.

From Equation~\eqref{eq:MBS_value_result_04}, we see that $\bar{\lambda}^{\mathbb{Q}_s}(t,s)$ is a conditional expectation of $\lambda(s)$ weighted by $e^{-\int_t^s\lambda(u)\mathrm{d}u}$.
Therefore, $\bar{\lambda}^{\mathbb{Q}_s}(t,s)$ can be interpreted as an effective prepayment rate that accounts for the reduction in outstanding principal due to prepayments up to time $s$.

Furthermore,
\begin{align*}
    Z(t,s)\bar{c}(t,s)
    =
    e^{-\int_0^t\lambda(u)\mathrm{d}u}
    \mathbb{E}^{\mathbb{Q}_s}
    \left[
        \left.
        e^{-\int_t^s\lambda(u)\mathrm{d}u}
        \left(
            c+\tilde{P}(s)\lambda(s)
        \right)
        \right|
        \mathcal{F}_t
    \right]
\end{align*}
holds.
Therefore, $Z(t,s)\bar{c}(t,s)$ corresponds to the conditional expectation of the MBS cash flow at time $s$, evaluated under $\mathbb{Q}_s$ given the information available at time $t$.
In this sense, $\bar{c}(t,s)$ can be interpreted as representing the magnitude of the cash flow at time $s$, excluding the effect of the reduction in outstanding principal due to prepayments, which is separately captured by $Z(t,s)$.

Furthermore, $Z(t,s)$ can be interpreted as an effective outstanding principal factor that reflects the reduction in outstanding principal due to prepayments between times $t$ and $s$.
Therefore, $V_{\mathrm{d}}(t,s)$ represents the contribution of the MBS cash flow at time $s$ to its present value at time $t$.
Thus, Equation~\eqref{eq:MBS_value_result_01} shows that the MBS price $V(t,T)$ can be expressed by integrating these basic components $\left\{V_{\mathrm{d}}(t,s)\right\}_{t\leq s\leq T}$ from time $t$ to maturity $T$.
This representation is analogous to expressing a conventional coupon-bearing government bond as a portfolio of zero-coupon bonds maturing on its coupon payment dates and principal repayment date.

$e^{-\int_0^t\lambda(u)\mathrm{d}u}$ is $\mathcal{F}_t$-measurable, and $\tilde{P}(t)$ is deterministic.
Furthermore, $D(t,s)$ is the discount factor either observed in the market at time $t$ or determined using an appropriate interest-rate model.
Therefore, Equation~\eqref{eq:MBS_value_result_01} shows that the main computational challenge in MBS valuation lies in evaluating $Z(t,s)$.

\subsection{Prepayment Process}
\label{subsec:prepayment_process}

In this subsection, we outline the prepayment process employed in this paper and express $Z(t,s)$ in terms of this process.
Since the prepayment model used in this paper is similar to those adopted by \textcite{Chernov_et_al:2017} and \textcite{Imagawa_Nakagawa:2025}, we omit the background and details of the model and present only the formulation necessary for our analysis.

We consider a model in which the prepayment process $\lambda(t)$ is decomposed into two components as follows:
\begin{align*}
    \lambda(t)
    =&
    \lambda_{\mathrm{c}}(t)
    +
    \lambda_{\mathrm{r}}(t),
    \\
    \text{where} \quad
    \lambda_{\mathrm{r}}(t)
    :=&
    y(t)
    \left(
        m-a-bR_{10}(t)
    \right)^+,
    \\
    R_{10}(t)
    :=&
    -
    \frac{1}{10}
    \ln D(t,t+10).
\end{align*}
Here, $a$ and $b$ are constants, and $R_{10}(t)$ denotes the 10-year zero-coupon yield at time $t$.
Furthermore, we model $\lambda_{\mathrm{c}}(t)$ and $y(t)$ as stochastic processes satisfying the following stochastic differential equations:
\begin{align}
    \mathrm{d}\lambda_{\mathrm{c}}(t)
    =&
    \kappa_{\lambda}
    \left(
        \bar{\lambda}
        -
        \lambda_{\mathrm{c}}(t)
    \right)
    \mathrm{d}t
    +
    \sigma_{\lambda}
    \sqrt{\lambda_{\mathrm{c}}(t)}
    \mathrm{d}W_{\lambda}^{\mathbb{Q}}(t),
    \label{eq:SDE_lmd_c}
    \\
    \mathrm{d}y(t)
    =&
    \kappa_y
    \left(
        \bar{y}
        -
        y(t)
    \right)
    \mathrm{d}t
    +
    \sigma_y
    \mathrm{d}W_y^{\mathbb{Q}}(t).
    \label{eq:SDE_y}
\end{align}
Here, we allow $W_y^{\mathbb{Q}}$ and $W_r^{\mathbb{Q}}$ to be correlated, satisfying $\mathrm{d}W_y^{\mathbb{Q}}(t)\mathrm{d}W_r^{\mathbb{Q}}(t)=\rho\mathrm{d}t$.
In contrast, we assume that $W_{\lambda}^{\mathbb{Q}}$ is independent of both $W_r^{\mathbb{Q}}$ and $W_y^{\mathbb{Q}}$.
In view of the above, we may also take the filtration $\left\{ \mathcal{F}_t \right\}$ to be the natural filtration generated by the three standard Brownian motions $W_r^{\mathbb{Q}}$, $W_{\lambda}^{\mathbb{Q}}$, and $W_y^{\mathbb{Q}}$.

Note that $y(t)$, governed by Equation~\eqref{eq:SDE_y}, is a Gaussian mean-reverting process and can therefore theoretically take negative values.
Consequently, $y(t)\geq0$ is required for $\lambda_{\mathrm{r}}(t)$ to be interpreted as a nonnegative prepayment intensity in the strict sense, and our specification should therefore be regarded as a Gaussian approximation in this respect.
However, in the numerical analysis presented later, we choose parameter values such that the probability of $y(t)$ taking negative values at the monthly valuation dates is negligibly small.
We verify this property in the numerical analysis.

For a fixed cash flow date $s$, we define the following functions for $t\leq v\leq s$:
\begin{align}
    Z_{\mathrm{r}}^{(s)}(t,v)
    :=&
    \mathbb{E}^{\mathbb{Q}_s}
    \left[
        \left.
        e^{-\int_t^v\lambda_{\mathrm{r}}(u)\mathrm{d}u}
        \right|
        \mathcal{F}_t
    \right]
    \nonumber
    \\
    =&
    \mathbb{E}^{\mathbb{Q}_s}
    \left[
        \left.
        e^{
            -\int_t^v
            y(u)
            \left(
                m-a-bR_{10}(u)
            \right)^+
            \mathrm{d}u
        }
        \right|
        \mathcal{F}_t
    \right],
    \label{eq:Z_closed_form_03}
    \\
    Z_{\mathrm{c}}^{(s)}(t,v)
    :=&
    \mathbb{E}^{\mathbb{Q}_s}
    \left[
        \left.
        e^{-\int_t^v\lambda_{\mathrm{c}}(u)\mathrm{d}u}
        \right|
        \mathcal{F}_t
    \right].
    \label{eq:Z_closed_form_02}
\end{align}
The Radon--Nikodym derivative associated with the change of measure from $\mathbb{Q}$ to $\mathbb{Q}_s$ depends only on $W_r^{\mathbb{Q}}$, which drives the interest-rate process.
Furthermore, $W_{\lambda}^{\mathbb{Q}}$ is independent of both $W_r^{\mathbb{Q}}$ and $W_y^{\mathbb{Q}}$.
Therefore, the process $\left\{\lambda_{\mathrm{c}}(t)\right\}$ remains independent of the pair of processes $\left\{r(t)\right\}$ and $\left\{y(t)\right\}$ under $\mathbb{Q}_s$, and the following relation holds:
\begin{align*}
    Z^{(s)}(t,v)
    =
    Z_{\mathrm{c}}^{(s)}(t,v)
    Z_{\mathrm{r}}^{(s)}(t,v)
\end{align*}
Moreover, since the change of measure to $\mathbb{Q}_s$ does not change the dynamics of $\lambda_{\mathrm{c}}$, $Z_{\mathrm{c}}^{(s)}(t,v)$ does not actually depend on $s$.
Hereafter, we denote it simply by $Z_{\mathrm{c}}(t,v)$.
For the interest-rate-dependent component, we denote the diagonal value corresponding to the actual cash flow date by
\begin{align*}
    Z_{\mathrm{r}}(t,s)
    :=
    Z_{\mathrm{r}}^{(s)}(t,s)
\end{align*}
Consequently, we obtain
\begin{align*}
    Z(t,s)
    =
    Z_{\mathrm{c}}(t,s)
    Z_{\mathrm{r}}(t,s)
\end{align*}

The expectation in Equation~\eqref{eq:Z_closed_form_03} captures the dependence of MBS cash flows on interest rates, a characteristic feature of MBSs.
How to efficiently evaluate this expectation is a central issue in deriving a closed-form approximation for MBS prices.
Since it is generally difficult to evaluate this expectation analytically, we approximate it using a cumulant expansion in a later section.
In contrast, the expectation in Equation~\eqref{eq:Z_closed_form_02} admits an analytical expression.
See Section~\ref{subsec:analytic_Zc} for its derivation.

Summarizing the above results, the MBS price can be expressed as follows:
\begin{align}
    V(t,T)
    =&
    \int_t^T
    V_{\mathrm{d}}(t,s)
    \mathrm{d}s,
    \label{eq:Z_r_mc_01}
    \\
    \text{where} \quad
    V_{\mathrm{d}}(t,v)
    =&
    \bar{c}(t,v)
    D(t,v)
    Z_{\mathrm{c}}(t,v)
    Z_{\mathrm{r}}(t,v),
    \label{eq:Z_r_mc_02}
    \\
    \bar{c}(t,v)
    =&
    \bar{c}_0(t,v)
    +
    \bar{c}_{\mathrm{r}}(t,v),
    \label{eq:Z_r_mc_03}
    \\
    \bar{c}_0(t,v)
    :=&
    e^{-\int_0^t\lambda(u)\mathrm{d}u}
    \left(
        c
        +
        \tilde{P}(v)
        \bar{\lambda}_{\mathrm{c}}(t,v)
    \right),
    \\
    \bar{c}_{\mathrm{r}}(t,v)
    :=&
    e^{-\int_0^t\lambda(u)\mathrm{d}u}
    \tilde{P}(v)
    \bar{\lambda}_{\mathrm{r}}^{\mathbb{Q}_v}(t,v),
    \\
    \bar{\lambda}_{\mathrm{c}}(t,v)
    :=&
    -
    \frac{\partial}{\partial v}
    \ln Z_{\mathrm{c}}(t,v)
    \nonumber
    \\
    =&
    A_{\mathrm{c}}^{\prime}(v-t)
    \lambda_{\mathrm{c}}(t)
    +
    B_{\mathrm{c}}^{\prime}(v-t),
    \label{eq:Z_r_mc_04}
    \\
    \bar{\lambda}_{\mathrm{r}}^{\mathbb{Q}_v}(t,v)
    :=&
    -
    \left.
        \frac{\partial}{\partial u}
        \ln Z_{\mathrm{r}}^{(v)}(t,u)
    \right|_{u=v}
    \nonumber
    \\
    =&
    \frac{
        \mathbb{E}^{\mathbb{Q}_v}
        \left[
            \left.
            e^{-\int_t^v\lambda_{\mathrm{r}}(u)\mathrm{d}u}
            \lambda_{\mathrm{r}}(v)
            \right|
            \mathcal{F}_t
        \right]
    }{
        \mathbb{E}^{\mathbb{Q}_v}
        \left[
            \left.
            e^{-\int_t^v\lambda_{\mathrm{r}}(u)\mathrm{d}u}
            \right|
            \mathcal{F}_t
        \right]
    }.
    \label{eq:Z_r_mc_05}
\end{align}
Here, $Z_{\mathrm{r}}(t,v):=Z_{\mathrm{r}}^{(v)}(t,v)$.
The key expectation in the above expressions that has not been evaluated analytically is $Z_{\mathrm{r}}^{(s)}(t,v)$.
In particular, evaluating the diagonal value $Z_{\mathrm{r}}(t,s)=Z_{\mathrm{r}}^{(s)}(t,s)$, which appears directly in the MBS price, is especially important.

\section{MBS Price Decomposition and Interest Rate Options}
\label{sec:mbs_price_decomposition}

In the previous section, we formulated the MBS price and identified $Z_{\mathrm{r}}(t,v)$ as the key expectation capturing the interest-rate dependence of MBS cash flows.
In Section~\ref{subsec:mbs_price_decomposition}, we decompose the MBS price into three price components: a component corresponding to cash flows that do not depend on interest rates, a component corresponding to the present value of principal repaid early through interest-rate-dependent prepayments, and an interest rate option component reflecting the reduction in future cash flows due to prepayments.
In Section~\ref{subsec:first_order_interest_rate_options}, we use a first-order approximation to identify the payoffs corresponding to the latter two price components and interpret them as a portfolio of interest rate options resembling interest rate floorlets and a portfolio of Asian-style interest rate options, respectively.
Furthermore, we clarify the economic interpretation of each price component in terms of MBS cash flows.

\subsection{MBS Price Decomposition and Interest Rate Option Portfolios}
\label{subsec:mbs_price_decomposition}

In this subsection, we show that the MBS price $V(t,T)$ can be decomposed into three components: a component corresponding to cash flows that do not depend on interest rates, a component corresponding to the present value of principal repaid through interest-rate-dependent prepayments, and a component corresponding to a portfolio of interest rate options.
To this end, we first focus on $V_{\mathrm{d}}(t,v)=\bar{c}(t,v)D(t,v)Z(t,v)$, a basic component of the MBS price, and examine its decomposition.

First, for simplicity, we consider the case where $\lambda_{\mathrm{c}}(t)\equiv0$.
In this case, $D(t,v)Z(t,v)$ can be expressed as follows:
\begin{align}
    D(t,v)Z(t,v)
    &=
    D(t,v)-\tilde{Z}_{\mathrm{op}}(t,v),
    \label{eq:decomp_DZ}
    \\
    \text{where} \quad
    \tilde{Z}_{\mathrm{op}}(t,v)
    &:=
    \mathbb{E}^{\mathbb{Q}}
    \left[
        \left.
        e^{-\int_t^v r(u)\mathrm{d}u}
        \left(
            1-e^{-\int_t^v\lambda_{\mathrm{r}}(u)\mathrm{d}u}
        \right)
        \right|
        \mathcal{F}_t
    \right]
    \label{eq:Z_op_def_1}
    \\
    &=
    \mathbb{E}^{\mathbb{Q}}
    \left[
        \left.
        e^{-\int_t^v r(u)\mathrm{d}u}
        \left(
            1-e^{-\int_t^v y(u)(m-a-bR_{10}(u))^{+}\mathrm{d}u}
        \right)
        \right|
        \mathcal{F}_t
    \right].
    \label{eq:Z_op_def_2}
\end{align}
Equations~\eqref{eq:Z_op_def_1} and \eqref{eq:Z_op_def_2} imply that $\tilde{Z}_{\mathrm{op}}(t,v)$ can be interpreted as the time-$t$ value of an option that pays $1-\exp\left(-\int_t^v\lambda_{\mathrm{r}}(u)\mathrm{d}u\right)$ at time $v$.
On the other hand, the left-hand side of Equation~\eqref{eq:decomp_DZ}, $D(t,v)Z(t,v)$, represents the value of a zero-coupon bond paying $1$ at time $v$, adjusted to reflect the reduction in outstanding principal due to interest-rate-dependent prepayments between times $t$ and $v$.
Therefore, Equation~\eqref{eq:decomp_DZ} shows that this value can be represented as the value of a position that is long one unit of a zero-coupon bond maturing at time $v$ and short one unit of the option associated with these prepayments.
Note that this interpretation as an option strictly requires $\lambda_{\mathrm{r}}(u)\geq0$.
As discussed in the previous section, this nonnegativity is not theoretically guaranteed under the Gaussian refinancing-sensitivity model employed in this paper, and this interpretation therefore applies to parameter regions in which the effect of negative values of $y(t)$ is sufficiently small, as in the numerical setting considered in this paper.

Using Equation~\eqref{eq:decomp_DZ}, $V_{\mathrm{d}}(t,v)$ can be decomposed as follows:
\begin{align*}
    V_{\mathrm{d}}(t,v)
    =
    e^{-\int_0^t\lambda(u)\mathrm{d}u}cD(t,v)
    +
    \bar{c}_{\mathrm{r}}(t,v)D(t,v)Z_{\mathrm{r}}(t,v)
    -
    \bar{c}_0(t,v)\tilde{Z}_{\mathrm{op}}(t,v).
\end{align*}
The second term on the right-hand side of the above expression can be written as follows:
\begin{align}
    \bar{c}_{\mathrm{r}}(t,v)D(t,v)Z_{\mathrm{r}}(t,v)
    =
    e^{-\int_0^t\lambda(u)\mathrm{d}u}
    \tilde{P}(v)
    \mathbb{E}^{\mathbb{Q}}
    \left[
        \left.
        e^{-\int_t^v r(u)\mathrm{d}u}
        e^{-\int_t^v\lambda_{\mathrm{r}}(u)\mathrm{d}u}
        \lambda_{\mathrm{r}}(v)
        \right|
        \mathcal{F}_t
    \right].
    \label{eq:c_r_D}
\end{align}
The expectation on the right-hand side of the above equation can be interpreted as the time-$t$ value of an interest rate option that pays $e^{-\int_t^v\lambda_{\mathrm{r}}(u)\mathrm{d}u}\lambda_{\mathrm{r}}(v)$ at time $v$.
Therefore, the quantity represented by Equation~\eqref{eq:c_r_D} corresponds to the present value of the principal cash flow repaid through interest-rate-dependent prepayments at time $v$, taking into account the reduction in outstanding principal due to such prepayments up to time $v$.
Here, the exponential factor reflects cumulative prepayments, making the payoff dependent on the path of interest rates.
Furthermore, this quantity represents the present value of the cash flow per unit time, and the present value corresponding to the payment over an infinitesimal interval $(v,v+\mathrm{d}v]$ is obtained by multiplying this quantity by $\mathrm{d}v$.

In this case, the MBS price can be decomposed as follows:
\begin{align}
    V(t,T)
    =&
    \tilde{V}_0(t,T)
    +
    \tilde{V}_{\mathrm{r}}(t,T)
    -
    \tilde{V}_{\mathrm{op}}(t,T),
    \label{eq:lmd_c_zero_V_decomp}
    \\
    \text{where} \quad
    \tilde{V}_0(t,T)
    :=&
    \int_t^T
    e^{-\int_0^t\lambda(u)\mathrm{d}u}
    cD(t,s)
    \mathrm{d}s,
    \nonumber
    \\
    \tilde{V}_{\mathrm{r}}(t,T)
    :=&
    \int_t^T
    \bar{c}_{\mathrm{r}}(t,s)D(t,s)Z_{\mathrm{r}}(t,s)
    \mathrm{d}s,
    \label{eq:lmd_c_zero_V_r}
    \\
    \tilde{V}_{\mathrm{op}}(t,T)
    :=&
    \int_t^T
    \bar{c}_0(t,s)\tilde{Z}_{\mathrm{op}}(t,s)
    \mathrm{d}s.
    \label{eq:lmd_c_zero_V_op}
\end{align}
Here, the first term on the right-hand side of Equation~\eqref{eq:lmd_c_zero_V_decomp}, $\tilde{V}_0(t,T)$, represents the present value of an MBS for which no future prepayments occur and only fixed principal and interest payments are made.
The second term, $\tilde{V}_{\mathrm{r}}(t,T)$, can be interpreted as the integral, from time $t$ to maturity $T$, of the present value $\bar{c}_{\mathrm{r}}(t,s)D(t,s)Z_{\mathrm{r}}(t,s)\mathrm{d}s$ of principal cash flows arising from interest-rate-dependent prepayments over the infinitesimal interval $(s,s+\mathrm{d}s]$.
Furthermore, the third term, $\tilde{V}_{\mathrm{op}}(t,T)$, represents the present value of fixed principal and interest cash flows lost due to interest-rate-dependent prepayments.
This can be expressed as the value of a portfolio holding $\bar{c}_0(t,s)\mathrm{d}s$ units of the option $\tilde{Z}_{\mathrm{op}}(t,s)$ for each maturity $s$, where the number of units does not depend on interest rates.

Next, we consider the case where $\lambda_{\mathrm{c}}(t)$ follows the stochastic differential equation given in Equation~\eqref{eq:SDE_lmd_c}.
As in the previous case, the following relation holds:
\begin{align}
    D(t,v)Z(t,v)
    &=
    D(t,v)Z_{\mathrm{c}}(t,v)
    -
    Z_{\mathrm{op}}(t,v),
    \label{eq:lmd_c_nonzero_DZ_decomp}
    \\
    \text{where} \quad
    Z_{\mathrm{op}}(t,v)
    &:=
    \mathbb{E}^{\mathbb{Q}}
    \left[
        \left.
        e^{-\int_t^v(r(u)+\lambda_{\mathrm{c}}(u))\mathrm{d}u}
        \left(
            1-e^{-\int_t^v\lambda_{\mathrm{r}}(u)\mathrm{d}u}
        \right)
        \right|
        \mathcal{F}_t
    \right]
    \nonumber
    \\
    &=
    \mathbb{E}^{\mathbb{Q}}
    \left[
        \left.
        e^{-\int_t^v(r(u)+\lambda_{\mathrm{c}}(u))\mathrm{d}u}
        \left(
            1-e^{-\int_t^v y(u)(m-a-bR_{10}(u))^{+}\mathrm{d}u}
        \right)
        \right|
        \mathcal{F}_t
    \right]
    \nonumber
\end{align}
Here, $Z_{\mathrm{c}}(t,v)$ can be interpreted as an effective discount factor that reflects prepayments arising from $\lambda_{\mathrm{c}}(t)$.
Using Equation~\eqref{eq:lmd_c_nonzero_DZ_decomp}, the basic component of the MBS price, $V_{\mathrm{d}}(t,v)$, can be decomposed exactly as follows:
\begin{align}
    V_{\mathrm{d}}(t,v)
    =&
    \bar{c}_0(t,v)D(t,v)Z_{\mathrm{c}}(t,v)
    +
    \bar{c}_{\mathrm{r}}(t,v)D(t,v)Z_{\mathrm{c}}(t,v)Z_{\mathrm{r}}(t,v)
    \nonumber
    \\
    &-
    \bar{c}_0(t,v)Z_{\mathrm{op}}(t,v).
    \label{eq:lmd_c_nonzero_Vd_decomp}
\end{align}
Equation~\eqref{eq:lmd_c_nonzero_Vd_decomp} has the same structure as the decomposition of $V_{\mathrm{d}}(t,v)$ obtained for the case where $\lambda_{\mathrm{c}}\equiv0$.
The first term on the right-hand side corresponds to cash flows that do not depend on interest rates, while the second term corresponds to the present value of principal repaid through interest-rate-dependent prepayments.
The third term corresponds to the present value of baseline cash flows lost due to interest-rate-dependent prepayments and is represented by a short position in $\bar{c}_0(t,v)$ units of the option $Z_{\mathrm{op}}(t,v)$, where the number of units does not depend on interest rates.
The main difference from the case where $\lambda_{\mathrm{c}}\equiv0$ is that the first and second terms are multiplied by the effective discount factor $Z_{\mathrm{c}}(t,v)$, which reflects interest-rate-independent prepayments.

Therefore, integrating Equation~\eqref{eq:lmd_c_nonzero_Vd_decomp} from time $t$ to maturity $T$, the MBS price $V(t,T)$ can be decomposed exactly as follows:
\begin{align}
    V(t,T)
    =&
    V_0(t,T)
    +
    V_{\mathrm{r}}(t,T)
    -
    V_{\mathrm{op}}(t,T),
    \label{eq:lmd_c_nonzero_V_decomp}
    \\
    \text{where} \quad
    V_0(t,T)
    :=&
    \int_t^T
    \bar{c}_0(t,s)D(t,s)Z_{\mathrm{c}}(t,s)
    \mathrm{d}s,
    \nonumber
    \\
    V_{\mathrm{r}}(t,T)
    :=&
    \int_t^T
    \bar{c}_{\mathrm{r}}(t,s)D(t,s)Z_{\mathrm{c}}(t,s)Z_{\mathrm{r}}(t,s)
    \mathrm{d}s,
    \nonumber
    \\
    V_{\mathrm{op}}(t,T)
    :=&
    \int_t^T
    \bar{c}_0(t,s)Z_{\mathrm{op}}(t,s)
    \mathrm{d}s.
    \nonumber
\end{align}
The first term on the right-hand side of Equation~\eqref{eq:lmd_c_nonzero_V_decomp}, $V_0(t,T)$, represents the value of an MBS whose cash flows do not depend on interest rates.
Here, we hold fixed the prepayment history up to time $t$ and the outstanding principal at time $t$.
Under this condition, although the baseline cash flows themselves do not depend on the future path of interest rates, their present value depends on interest rates through the discount factor $D(t,s)$.
The second term, $V_{\mathrm{r}}(t,T)$, represents the present value of principal payments arising from interest-rate-dependent prepayments, taking into account the reduction in outstanding principal due to both prepayment components through $Z_{\mathrm{c}}(t,s)Z_{\mathrm{r}}(t,s)$.
Indeed, using Equation~\eqref{eq:Z_r_mc_05} and a change of measure, we obtain the following relation:
\begin{align}
    \bar{c}_{\mathrm{r}}(t,s)D(t,s)Z_{\mathrm{c}}(t,s)Z_{\mathrm{r}}(t,s)
    =&
    e^{-\int_0^t\lambda(u)\mathrm{d}u}\tilde{P}(s)
    \mathbb{E}^{\mathbb{Q}}\left[
        \left.
        e^{-\int_t^s(r(u)+\lambda_{\mathrm{c}}(u)+\lambda_{\mathrm{r}}(u))\mathrm{d}u}
        \lambda_{\mathrm{r}}(s)
        \right|\mathcal{F}_t
    \right].
    \label{eq:c_r_D_general}
\end{align}
Furthermore, the third term, $V_{\mathrm{op}}(t,T)$, represents the present value of baseline cash flows lost due to interest-rate-dependent prepayments.
Here, the baseline cash flows refer to the fixed principal and interest payments and the principal payments arising from the interest-rate-independent prepayment component $\lambda_{\mathrm{c}}$, assuming that the prepayment history up to time $t$ is held fixed and only the interest-rate-dependent component $\lambda_{\mathrm{r}}$ is set to zero after time $t$.
By the independence of the process $\{\lambda_{\mathrm{c}}(u)\}$ from the pair of processes $\{r(u)\}$ and $\{y(u)\}$, the value of these lost cash flows is expressed as $\bar{c}_0(t,s)D(t,s)Z_{\mathrm{c}}(t,s)\{1-Z_{\mathrm{r}}(t,s)\}$.
Therefore, $V_{\mathrm{op}}(t,T)$ represents the value of a portfolio holding $\bar{c}_0(t,s)\mathrm{d}s$ units of the option $Z_{\mathrm{op}}(t,s)$ for each maturity $s$, where the number of units does not depend on the future path of interest rates.
In this sense, the second and third terms separately capture the newly arising principal payments from prepayments and the baseline cash flows lost due to prepayments, respectively.
Note that $V_{\mathrm{op}}(t,T)$ represents the present value of cash flows lost due to interest-rate-dependent prepayments and is distinct from the effect that also accounts for the present value $V_{\mathrm{r}}(t,T)$ of principal repaid early through prepayments.
The net effect of interest-rate-dependent prepayments on the MBS price is given by $V_{\mathrm{r}}(t,T)-V_{\mathrm{op}}(t,T)$.

The above results show that Equation~\eqref{eq:lmd_c_nonzero_V_decomp} provides an exact decomposition of the MBS price into three components: $V_0(t,T)$, corresponding to cash flows that do not depend on interest rates; $V_{\mathrm{r}}(t,T)$, corresponding to the present value of principal repaid through interest-rate-dependent prepayments; and $V_{\mathrm{op}}(t,T)$, representing the value of an interest rate option portfolio that reflects the reduction in outstanding principal due to prepayments.
Under the conditions $c\geq0$, $\tilde{P}(s)\geq0$, and $\lambda_{\mathrm{c}}(u),\lambda_{\mathrm{r}}(u)\geq0$, all three price components are nonnegative.
Therefore, $V_0(t,T)$ and $V_{\mathrm{r}}(t,T)$ make nonnegative contributions to the MBS price, whereas $-V_{\mathrm{op}}(t,T)$ makes a nonpositive contribution.
However, the nonnegativity of these price components does not guarantee the signs of their first- and second-order interest-rate sensitivities.
Furthermore, under the Gaussian model for $y$ employed in this paper, the nonnegativity of $\lambda_{\mathrm{r}}$ is not guaranteed, and the above interpretation of the signs of the price contributions holds under this nonnegativity condition.
The price decomposition itself remains algebraically valid regardless of this condition, provided that the required expectations are finite.

\subsection{Interest Rate Options Under a First-Order Approximation}
\label{subsec:first_order_interest_rate_options}

In this subsection, we apply a first-order approximation to the price components introduced in the previous subsection and interpret the effects of interest-rate-dependent prepayments on MBS prices in terms of interest rate options.
Specifically, we identify the payoffs associated with $V_{\mathrm{r}}(t,T)$, corresponding to the present value of principal repaid early through interest-rate-dependent prepayments, and $V_{\mathrm{op}}(t,T)$, reflecting the reduction in future cash flows due to prepayments.
We then show that the former can be interpreted as a portfolio of interest rate options resembling interest rate floorlets, while the latter can be interpreted as a portfolio of Asian-style interest rate options, and clarify the economic meaning of each in terms of MBS cash flows.

First, we define
\begin{align*}
    \bar{\lambda}_{\mathrm{r},1}^{\mathbb{Q}_s}(u)
    :=
    \mathbb{E}^{\mathbb{Q}_s}
    \left[
        \left.
        \lambda_{\mathrm{r}}(u)
        \right|
        \mathcal{F}_t
    \right].
\end{align*}
We also regard
\begin{align*}
    w
    :=
    -
    \int_t^s
    \lambda_{\mathrm{r}}(u)
    \mathrm{d}u
\end{align*}
as a small quantity and expand Equation~\eqref{eq:Z_closed_form_03} to first order using $e^w\simeq1+w$.
This first-order expansion is related to the approximation method of \textcite{Kolbe_Zagst:2009} in that both approximate an exponential term involving the cumulative prepayment intensity, although the approximations are applied differently.
We further discuss the relationship between these approaches in Section~\ref{subsec:cumulant_expansion}.
We then obtain
\begin{align}
    Z_{\mathrm{r}}(t,s)
    &=
    \mathbb{E}^{\mathbb{Q}_s}
    \left[
        \left.
        e^{-\int_t^s\lambda_{\mathrm{r}}(u)\mathrm{d}u}
        \right|
        \mathcal{F}_t
    \right]
    \nonumber
    \\
    &\simeq
    1-
    \int_t^s
    \bar{\lambda}_{\mathrm{r},1}^{\mathbb{Q}_s}(u)
    \mathrm{d}u.
    \label{eq:taylor_expansion}
\end{align}

Next, we consider the price component corresponding to the present value of principal repaid through interest-rate-dependent prepayments.
We apply a first-order approximation with respect to the magnitude of future interest-rate-dependent prepayments to the payoff $e^{-\int_t^v\lambda_{\mathrm{r}}(u)\mathrm{d}u}\lambda_{\mathrm{r}}(v)$ in Equation~\eqref{eq:c_r_D}.
Specifically, holding the outstanding principal at time $t$ fixed and replacing future values of $\lambda_{\mathrm{r}}$ with $\varepsilon\lambda_{\mathrm{r}}$, we obtain
\begin{align*}
    e^{-\varepsilon\int_t^v\lambda_{\mathrm{r}}(u)\mathrm{d}u}
    \varepsilon\lambda_{\mathrm{r}}(v)
    =\varepsilon\lambda_{\mathrm{r}}(v)+O(\varepsilon^2).
\end{align*}
Thus, to first order, we replace the exponential factor representing the reduction in outstanding principal by $1$.
On the other hand, $1-e^{-\varepsilon\int_t^v\lambda_{\mathrm{r}}(u)\mathrm{d}u}$ becomes $\varepsilon\int_t^v\lambda_{\mathrm{r}}(u)\mathrm{d}u+O(\varepsilon^2)$.
Assuming the integrability conditions required for this expansion and suppressing $\varepsilon$ in what follows, we obtain
\begin{align}
    \bar{c}_{\mathrm{r}}(t,v)D(t,v)Z_{\mathrm{r}}(t,v)
    \simeq &
    e^{-\int_0^t\lambda(u)\mathrm{d}u}
    \tilde{P}(v)
    \mathbb{E}^{\mathbb{Q}}
    \left[
        \left.
        e^{-\int_t^v r(u)\mathrm{d}u}
        \lambda_{\mathrm{r}}(v)
        \right|
        \mathcal{F}_t
    \right]
    \nonumber
    \\
    =&
    e^{-\int_0^t\lambda(u)\mathrm{d}u}
    \tilde{P}(v)
    \mathbb{E}^{\mathbb{Q}}
    \left[
        \left.
        e^{-\int_t^v r(u)\mathrm{d}u}
        y(v)
        \bigl(m-a-bR_{10}(v)\bigr)^{+}
        \right|
        \mathcal{F}_t
    \right].
    \label{eq:c_r_D_first_order}
\end{align}
For $b>0$, the payoff appearing in the expectation in Equation~\eqref{eq:c_r_D_first_order} takes the form of a put-type payoff that is positive when $R_{10}(v)$ falls below $(m-a)/b$, multiplied by the refinancing sensitivity $y(v)$.
Therefore, this expectation can be interpreted as the present value of a contingent claim resembling an interest rate floorlet with $R_{10}(v)$ as its reference rate.
However, this contingent claim is not identical to a standard interest rate floorlet, because the reference rate is the $10$-year zero-coupon yield and $y(v)$ is a stochastic coefficient.
The exact payoff additionally contains a path-dependent factor representing the outstanding principal, and the floorlet-type representation above is obtained under the first-order approximation.
Consequently, $\tilde{V}_{\mathrm{r}}(t,T)$, defined by Equation~\eqref{eq:lmd_c_zero_V_r}, can be interpreted as a portfolio of such interest rate options with different maturities $v$.

From the perspective of MBS cash flows, $\bar{c}_{\mathrm{r}}(t,v)D(t,v)Z_{\mathrm{r}}(t,v)$ corresponds to the present value of principal repaid early through interest-rate-dependent prepayments at time $v$.
Therefore, $\tilde{V}_{\mathrm{r}}(t,T)$ can be interpreted as a price component obtained by aggregating, across maturities, the present values of principal repaid through interest-rate-dependent prepayments at each point in time.
Under the first-order approximation, an increase in the refinancing incentive, holding other conditions fixed, increases the put-type payoff when $y(v)\geq0$.
On the other hand, the exact price component incorporates not only the increase in principal repayments at a given time due to a higher prepayment rate but also the reduction in outstanding principal resulting from earlier prepayments.
Therefore, an increase in the prepayment rate does not necessarily lead to an increase in principal payments at each point in time.
Furthermore, assessing the monotonicity of the overall price component with respect to interest rates requires taking into account changes in the discount factor as well.

Next, we consider the price component reflecting the reduction in future cash flows due to prepayments.
Applying the same first-order approximation as in Equation~\eqref{eq:taylor_expansion} to Equation~\eqref{eq:Z_op_def_1}, we approximate $\tilde{Z}_{\mathrm{op}}(t,v)$ as follows:
\begin{align}
    \tilde{Z}_{\mathrm{op}}(t,v)
    \simeq &
    \mathbb{E}^{\mathbb{Q}}
    \left[
        \left.
        e^{-\int_t^v r(u)\mathrm{d}u}
        \int_t^v
        \lambda_{\mathrm{r}}(u)
        \mathrm{d}u
        \right|
        \mathcal{F}_t
    \right]
    \label{eq:Z_op_KZ_app_01}
    \\
    =&
    \mathbb{E}^{\mathbb{Q}}
    \left[
        \left.
        e^{-\int_t^v r(u)\mathrm{d}u}
        \int_t^v
        y(u)
        \bigl(m-a-bR_{10}(u)\bigr)^{+}
        \mathrm{d}u
        \right|
        \mathcal{F}_t
    \right].
    \label{eq:Z_op_KZ_app_02}
\end{align}
The payoff appearing in Equation~\eqref{eq:Z_op_KZ_app_02} takes the form of an integral, from time $t$ to $v$, of put-type payoffs on $R_{10}(u)$ at each point in time $u$.
Therefore, $\tilde{Z}_{\mathrm{op}}(t,v)$ can be interpreted as the time-$t$ value of an Asian-style interest rate option whose payoff depends on the path of interest rates.
Here, ``Asian-style'' refers to the dependence on the integral of put-type payoffs at each point in time, which differs from a conventional average-price Asian put option, where the payoff is the positive part of the difference between the strike price and the average value of the underlying asset.
Consequently, $\tilde{V}_{\mathrm{op}}(t,T)$, defined by Equation~\eqref{eq:lmd_c_zero_V_op}, can be interpreted as a portfolio of such Asian-style interest rate options with different maturities $v$.

From the perspective of MBS cash flows, $\tilde{V}_{\mathrm{op}}(t,T)$ corresponds to the effect whereby an increase in prepayments following a decline in interest rates reduces the outstanding principal and, consequently, the principal and interest cash flows received thereafter.
Holding other conditions fixed, an increase in the cumulative value of nonnegative $\lambda_{\mathrm{r}}$ increases the payoff $1-e^{-\int_t^v\lambda_{\mathrm{r}}(u)\mathrm{d}u}$, so that $-\tilde{V}_{\mathrm{op}}(t,T)$, which enters Equation~\eqref{eq:lmd_c_zero_V_decomp} as a short position, contributes to limiting increases in the MBS price.
Therefore, $\tilde{V}_{\mathrm{op}}(t,T)$ can be interpreted as a price component representing the effect of interest-rate-dependent prepayments that prevents MBS investors from receiving the principal and interest cash flows they would otherwise have received in the absence of prepayments.

A similar interpretation holds in the general case where the interest-rate-independent prepayment process $\lambda_{\mathrm{c}}(t)$ is present.
Indeed, applying the same first-order approximation as in Equation~\eqref{eq:taylor_expansion}, we obtain
\begin{align*}
    Z_{\mathrm{op}}(t,v)
    \simeq&
    \mathbb{E}^{\mathbb{Q}}
    \left[
        \left.
        e^{-\int_t^v(r(u)+\lambda_{\mathrm{c}}(u))\mathrm{d}u}
        \int_t^v
        \lambda_{\mathrm{r}}(u)
        \mathrm{d}u
        \right|
        \mathcal{F}_t
    \right]
    \\
    =&
    \mathbb{E}^{\mathbb{Q}}
    \left[
        \left.
        e^{-\int_t^v(r(u)+\lambda_{\mathrm{c}}(u))\mathrm{d}u}
        \int_t^v
        y(u)
        \bigl(m-a-bR_{10}(u)\bigr)^{+}
        \mathrm{d}u
        \right|
        \mathcal{F}_t
    \right].
\end{align*}
In this case, $Z_{\mathrm{op}}(t,v)$ can be interpreted as the present value of an Asian-style interest rate option with the same payoff as the options represented by Equations~\eqref{eq:Z_op_KZ_app_01} and \eqref{eq:Z_op_KZ_app_02}, but with a discount factor that also reflects the effect of prepayments arising from $\lambda_{\mathrm{c}}(t)$.
Similarly, applying the first-order approximation to Equation~\eqref{eq:c_r_D_general}, we obtain
\begin{align*}
    \bar{c}_{\mathrm{r}}(t,v)D(t,v)Z_{\mathrm{c}}(t,v)Z_{\mathrm{r}}(t,v)
    \simeq{}&
    e^{-\int_0^t\lambda(u)\mathrm{d}u}\tilde{P}(v)Z_{\mathrm{c}}(t,v)
    \mathbb{E}^{\mathbb{Q}}\left[
        \left.
        e^{-\int_t^v r(u)\mathrm{d}u}
        y(v)\bigl(m-a-bR_{10}(v)\bigr)^+
        \right|\mathcal{F}_t
    \right].
\end{align*}
Here, we use the independence of the process $\{\lambda_{\mathrm{c}}(u)\}$ from the pair of processes $\{r(u)\}$ and $\{y(u)\}$.
Therefore, the price component in the general case can also be interpreted in terms of a contingent claim resembling an interest rate floorlet, with the additional factor $Z_{\mathrm{c}}(t,v)$ reflecting the reduction in outstanding principal due to interest-rate-independent prepayments.
Furthermore, the number of units held in the portfolio $V_{\mathrm{op}}(t,T)$ is $\bar{c}_0(t,v)\mathrm{d}v$, which remains independent of interest rates under this first-order approximation.

The above results show that, under the first-order approximation, the effects of interest-rate-dependent prepayments on MBS prices can be understood in terms of two price components associated with distinct types of interest rate options.
Specifically, $V_{\mathrm{r}}(t,T)$ can be interpreted as a portfolio of interest rate options resembling interest rate floorlets, corresponding to the present value of principal repaid early as interest rates decline.
On the other hand, $V_{\mathrm{op}}(t,T)$ can be interpreted as a portfolio of Asian-style interest rate options, corresponding to the reduction in future principal and interest cash flows resulting from the decrease in outstanding principal due to prepayments.
Thus, interest-rate-dependent prepayments can be understood in terms of the nonnegative price contribution from early principal repayments and the nonpositive price contribution from lost cash flows that do not depend on interest rates.
However, the signs of these price contributions must be distinguished from the signs of their interest-rate sensitivities.
Our price decomposition distinguishes between these two price effects by representing them as $V_{\mathrm{r}}(t,T)$ and $-V_{\mathrm{op}}(t,T)$, respectively.

\section{Closed-Form Approximation of MBS Prices}
\label{sec:mbs_approximation}

In this section, we derive a closed-form approximation for MBS prices to enable efficient valuation.
In Section~\ref{subsec:cumulant_expansion}, we apply a cumulant expansion to $Z_{\mathrm{r}}(t,v)$ and derive approximate expressions using the first and second cumulants, while also clarifying their relationship to the first-order approximation introduced in the previous section.
In Section~\ref{subsec:hull_white_forward_measure}, we adopt the one-factor Hull--White model as a specific interest rate model and derive analytical expressions for zero-coupon bond prices and the dynamics of the stochastic processes under the forward measure.
In Section~\ref{subsec:mbs_price_approximation}, we derive closed-form expressions for $\bar{\lambda}_{\mathrm{r},1}^{\mathbb{Q}_s}(u)$ and $\bar{\lambda}_{\mathrm{r},2}^{\mathbb{Q}_s}(u,\eta)$, which appear in the cumulant expansion, under the Hull--White model.
We then use these analytical expressions to derive a closed-form approximation for $Z_{\mathrm{r}}(t,v)$ and MBS prices.

\subsection{Cumulant Expansion}
\label{subsec:cumulant_expansion}

In the previous section, we introduced a first-order approximation to clarify the properties of MBS prices in terms of interest rate options.
However, efficient valuation of MBS prices requires an approximation of $Z_{\mathrm{r}}(t,s)=Z_{\mathrm{r}}^{(s)}(t,s)$ itself.
In this subsection, following \textcite{Ozeki_et_al:2009}, we apply a cumulant expansion to $Z_{\mathrm{r}}^{(s)}(t,v)$ to derive a closed-form approximation.
Furthermore, we clarify the relationship between the first-order approximation introduced in the previous section and the approximation obtained by truncating the cumulant expansion after the first cumulant, and discuss how our approach differs from the approximation method of \textcite{Kolbe_Zagst:2009}.

For a fixed cash flow date $s$, we define $\kappa_n^{(s)}(t,v)$ as the $n$th conditional cumulant of $\int_t^v\lambda_{\mathrm{r}}(u)\mathrm{d}u$ under $\mathbb{Q}_s$, conditional on $\mathcal{F}_t$, for $t\leq v\leq s$, as follows.
\begin{align*}
    \kappa_n^{(s)}(t,v)
    :=
    \left.
        \frac{\partial^n}{\partial\theta^n}
        \ln
        \mathbb{E}^{\mathbb{Q}_s}
        \left[
            \left.
            e^{
                \theta
                \int_t^v
                \lambda_{\mathrm{r}}(u)
                \mathrm{d}u
            }
            \right|
            \mathcal{F}_t
        \right]
    \right|_{\theta=0}.
\end{align*}
In this definition, $s$ and the corresponding forward measure $\mathbb{Q}_s$ are held fixed as $v$ varies.
Applying the cumulant expansion to $Z_{\mathrm{r}}^{(s)}(t,v)$ then yields
\begin{align*}
    Z_{\mathrm{r}}^{(s)}(t,v)
    =
    \exp
    \left(
        \sum_{n=1}^{\infty}
        \frac{(-1)^n}{n!}
        \kappa_n^{(s)}(t,v)
    \right).
\end{align*}
Therefore, at the actual cash flow date $v=s$, we obtain
\begin{align*}
    Z_{\mathrm{r}}(t,s)
    =
    Z_{\mathrm{r}}^{(s)}(t,s)
    =
    \exp
    \left(
        \sum_{n=1}^{\infty}
        \frac{(-1)^n}{n!}
        \kappa_n^{(s)}(t,s)
    \right).
\end{align*}
Henceforth, for notational simplicity, we write $\kappa_n(t,s) := \kappa_n^{(s)}(t,s)$ only for the diagonal values.
Thus, truncating the cumulant expansion after the second cumulant gives the following approximation for $Z_{\mathrm{r}}(t,s)$.
\begin{align}
    Z_{\mathrm{r}}(t,s)
    \simeq
    e^{
        -\kappa_1(t,s)
        +
        \kappa_2(t,s)/2
    }.
    \label{eq:cumulant_app_for_Zr}
\end{align}

In the following, we examine the expressions for the first and second cumulants.
First, the first cumulant is given by
\begin{align*}
    \kappa_1^{(s)}(t,v)
    =&
    \int_t^v
    \bar{\lambda}_{\mathrm{r},1}^{\mathbb{Q}_s}(u)
    \mathrm{d}u,
    \\
    \text{where} \quad
    \bar{\lambda}_{\mathrm{r},1}^{\mathbb{Q}_s}(u)
    :=&
    \mathbb{E}^{\mathbb{Q}_s}
    \left[
        \left.
        \lambda_{\mathrm{r}}(u)
        \right|
        \mathcal{F}_t
    \right].
    \nonumber
\end{align*}
In particular, setting $v=s$ yields $\kappa_1(t,s) = \int_t^s \bar{\lambda}_{\mathrm{r},1}^{\mathbb{Q}_s}(u) \mathrm{d}u$.
Next, the second cumulant is given by
\begin{align*}
    \kappa_2^{(s)}(t,v)
    =&
    2
    \int_t^v
    \int_u^v
    \left[
        \bar{\lambda}_{\mathrm{r},2}^{\mathbb{Q}_s}(u,\eta)
        -
        \bar{\lambda}_{\mathrm{r},1}^{\mathbb{Q}_s}(u)
        \bar{\lambda}_{\mathrm{r},1}^{\mathbb{Q}_s}(\eta)
    \right]
    \mathrm{d}\eta
    \mathrm{d}u,
    \\
    \text{where} \quad
    \bar{\lambda}_{\mathrm{r},2}^{\mathbb{Q}_s}(u,\eta)
    :=&
    \mathbb{E}^{\mathbb{Q}_s}
    \left[
        \left.
        \lambda_{\mathrm{r}}(u)
        \lambda_{\mathrm{r}}(\eta)
        \right|
        \mathcal{F}_t
    \right].
    \nonumber
\end{align*}
In particular, setting $v=s$ yields
\begin{align*}
    \kappa_2(t,s)
    =
    2
    \int_t^s
    \int_u^s
    \left[
        \bar{\lambda}_{\mathrm{r},2}^{\mathbb{Q}_s}(u,\eta)
        -
        \bar{\lambda}_{\mathrm{r},1}^{\mathbb{Q}_s}(u)
        \bar{\lambda}_{\mathrm{r},1}^{\mathbb{Q}_s}(\eta)
    \right]
    \mathrm{d}\eta
    \mathrm{d}u.
\end{align*}
Furthermore, differentiating with respect to $v$ while holding $s$ and the corresponding forward measure $\mathbb{Q}_s$ fixed yields
\begin{align*}
    \frac{\partial}{\partial v}
    \kappa_1^{(s)}(t,v)
    =&
    \bar{\lambda}_{\mathrm{r},1}^{\mathbb{Q}_s}(v),
    \\
    \frac{\partial}{\partial v}
    \kappa_2^{(s)}(t,v)
    =&
    2
    \int_t^v
    \left[
        \bar{\lambda}_{\mathrm{r},2}^{\mathbb{Q}_s}(u,v)
        -
        \bar{\lambda}_{\mathrm{r},1}^{\mathbb{Q}_s}(u)
        \bar{\lambda}_{\mathrm{r},1}^{\mathbb{Q}_s}(v)
    \right]
    \mathrm{d}u.
\end{align*}
The quantities $\bar{\lambda}_{\mathrm{r},1}^{\mathbb{Q}_s}(u)$ and $\bar{\lambda}_{\mathrm{r},2}^{\mathbb{Q}_s}(u,\eta)$ can be evaluated by specifying an interest rate model.
In particular, closed-form expressions are available under the one-factor Hull--White model.
We derive these expressions in a subsequent subsection.

The cumulant-based approximation in \eqref{eq:cumulant_app_for_Zr} is related to the approximation method employed by \textcite{Kolbe_Zagst:2009}.
In the previous section, we treated $w:=-\int_t^s\lambda_{\mathrm{r}}(u)\mathrm{d}u$ as a small quantity and obtained \eqref{eq:taylor_expansion} by expanding \eqref{eq:Z_closed_form_03} to first order in $w$.
On the other hand, if we truncate the cumulant expansion in \eqref{eq:cumulant_app_for_Zr} after the first cumulant and further assume that $|\kappa_1(t,s)|\ll1$, the resulting approximation for $Z_{\mathrm{r}}(t,s)$ coincides with the approximation on the right-hand side of \eqref{eq:taylor_expansion}.

The approximation presented in \eqref{eq:taylor_expansion} is related to the approach of \textcite{Kolbe_Zagst:2009}, but differs in how the approximation is applied.
This difference allows us to interpret MBS prices as including short positions in interest rate options, as shown in \eqref{eq:Z_op_def_1} and \eqref{eq:Z_op_def_2}.
To clarify the difference between their approach and ours, we present a simplified representation of the main approximation employed by \textcite{Kolbe_Zagst:2009}, using the notation of this paper.
Let
\begin{align*}
    w_{\mathrm{K}}
    &:=
    -\int_t^s r(u)\mathrm{d}u,
    \\
    z_{\mathrm{K}}
    &:=
    -\int_t^s \lambda_{\mathrm{r}}(u)\mathrm{d}u
\end{align*}
The main approximation employed in their study can be expressed as
$e^{w_{\mathrm{K}}+z_{\mathrm{K}}}\simeq e^{w_{\mathrm{K}}}+z_{\mathrm{K}}$,
which leads to the following approximation for $D(t,s)Z_{\mathrm{r}}(t,s)$.
\begin{align*}
    D(t,s)Z_{\mathrm{r}}(t,s)
    =&
    \mathbb{E}^{\mathbb{Q}}
    \left[
        \left.
        e^{-\int_t^s(r(u)+\lambda_{\mathrm{r}}(u))\mathrm{d}u}
        \right|
        \mathcal{F}_t
    \right]
    \\
    \simeq&
    \mathbb{E}^{\mathbb{Q}}
    \left[
        \left.
        e^{-\int_t^s r(u)\mathrm{d}u}
        \right|
        \mathcal{F}_t
    \right]
    -
    \mathbb{E}^{\mathbb{Q}}
    \left[
        \left.
        \int_t^s\lambda_{\mathrm{r}}(u)\mathrm{d}u
        \right|
        \mathcal{F}_t
    \right].
\end{align*}
The second term on the right-hand side is the risk-neutral expectation of the payoff $\int_t^s\lambda_{\mathrm{r}}(u)\mathrm{d}u$ itself, rather than its discounted present value, and therefore cannot be interpreted as an option price.
In contrast, we use the approximation $e^w\simeq1+w$ for $|w|\ll1$ to approximate $Z_{\mathrm{r}}(t,s)$ as in \eqref{eq:taylor_expansion}.
As a result, the first-order term in our approximation can be expressed, as shown in \eqref{eq:Z_op_KZ_app_01}, as the risk-neutral conditional expectation of the product of the stochastic discount factor $\exp(-\int_t^s r(u)\mathrm{d}u)$ and the payoff $\int_t^s\lambda_{\mathrm{r}}(u)\mathrm{d}u$.
Therefore, this term can be interpreted as the present value at time $t$ of an interest rate option.
This interpretation is also consistent with the intuition that MBSs embed interest rate optionality through prepayments.

\subsection{Hull--White Model under the \texorpdfstring{$s$}{s}-Forward Measure}
\label{subsec:hull_white_forward_measure}

As discussed in the previous subsection, approximating $Z_{\mathrm{r}}(t,v)$ and the option values using the first two cumulants requires analytical expressions for $\bar{\lambda}_{\mathrm{r},1}^{\mathbb{Q}_s}(u)$ and $\bar{\lambda}_{\mathrm{r},2}^{\mathbb{Q}_s}(u,\eta)$.
In this subsection, we present the fundamental expressions needed to derive these analytical expressions under the Hull--White model.
Since these expectations are evaluated under the $s$-forward measure $\mathbb{Q}_s$, we also derive the dynamics of the stochastic processes under this measure.

In this subsection, we adopt the following Hull--White model~\parencite{Hull_White:1990} as our interest rate model.
\begin{align}
    \mathrm{d}r(t)
    =
    \left(
        \vartheta(t)-a_r r(t)
    \right)
    \mathrm{d}t
    +
    \sigma_r
    \mathrm{d}W_r^{\mathbb{Q}}(t).
    \label{eq:HW}
\end{align}
Here, $a_r$ and $\sigma_r$ are positive constants.
The function $\vartheta(t)$ is determined so that the theoretical discount curve derived from the short-rate process in \eqref{eq:HW} matches the market-observed discount curve $D(0,t)$ at time $0$.
The zero-coupon bond price is given by the following expressions (see Appendix~\ref{app:hw_bond_price_derivation} for the proof).
\begin{align}
    D(t,T)
    =&
    \mathbb{E}^{\mathbb{Q}}
    \left[
        \left.
        e^{-\int_t^T r(u)\mathrm{d}u}
        \right|
        \mathcal{F}_t
    \right]
    =
    e^{A_r(t,T)-B_r(T-t)r(t)},
    \label{eq:D_sol_HW_01}
    \\
    \text{where} \quad
    A_r(t,T)
    :=&
    \ln\frac{D(0,T)}{D(0,t)}
    -
    B_r(T-t)
    \frac{\partial}{\partial t}
    \ln D(0,t)
    \nonumber
    \\
    &-
    \frac{\sigma_r^2}{4a_r}
    \left(
        1-e^{-2a_rt}
    \right)
    B_r^2(T-t),
    \label{eq:D_sol_HW_02}
    \\
    B_r(\tau)
    :=&
    \frac{1-e^{-a_r\tau}}{a_r},
    \qquad
    \tau\geq0.
    \label{eq:D_sol_HW_03}
\end{align}

Many of the expectations introduced in the previous subsection are defined under the $s$-forward measure $\mathbb{Q}_s$, rather than the risk-neutral measure $\mathbb{Q}$.
Therefore, to evaluate these expectations explicitly, we need to derive the dynamics of the stochastic processes under $\mathbb{Q}_s$.
To this end, we present the relationship between the standard Brownian motions under the risk-neutral measure $\mathbb{Q}$ and those under the $s$-forward measure $\mathbb{Q}_s$.
Specifically, applying Equations~\eqref{eq:relation_W_Q_Qs_01}--\eqref{eq:relation_W_Q_Qs_03} to the stochastic differential equations for $r(t)$, $y(t)$, and $\lambda_{\mathrm{c}}(t)$ under $\mathbb{Q}$ yields the following stochastic differential equations under $\mathbb{Q}_s$ (see Appendix~\ref{app:stochastic_processes_forward_measure} for the proof).
\begin{align}
    \mathrm{d}r(t)
    =&
    \left(
        \vartheta(t)
        -
        \sigma_r^2 B_r(s-t)
        -
        a_r r(t)
    \right)
    \mathrm{d}t
    +
    \sigma_r
    \mathrm{d}W_r^{\mathbb{Q}_s}(t),
    \label{eq:r_t_under_Q_s}
    \\
    \mathrm{d}y(t)
    =&
    \kappa_y
    \left(
        \bar{y}
        -
        \frac{\rho\sigma_r\sigma_y}{\kappa_y}
        B_r(s-t)
        -
        y(t)
    \right)
    \mathrm{d}t
    +
    \sigma_y
    \mathrm{d}W_y^{\mathbb{Q}_s}(t),
    \label{eq:y_t_under_Q_s}
    \\
    \mathrm{d}\lambda_{\mathrm{c}}(t)
    =&
    \kappa_{\lambda}
    \left(
        \bar{\lambda}
        -
        \lambda_{\mathrm{c}}(t)
    \right)
    \mathrm{d}t
    +
    \sigma_{\lambda}
    \sqrt{\lambda_{\mathrm{c}}(t)}
    \mathrm{d}W_{\lambda}^{\mathbb{Q}_s}(t).
    \nonumber
\end{align}

\subsection{Derivation of Approximate MBS Valuation Formulas}
\label{subsec:mbs_price_approximation}

In this subsection, we derive closed-form expressions for $\bar{\lambda}_{\mathrm{r},1}^{\mathbb{Q}_s}(u)$ and $\bar{\lambda}_{\mathrm{r},2}^{\mathbb{Q}_s}(u,\eta)$ under the Hull--White model, and use these expressions to derive approximate valuation formulas for $Z_{\mathrm{r}}(t,v)$ and MBS prices.

From Equations~\eqref{eq:D_sol_HW_01}--\eqref{eq:D_sol_HW_03}, the 10-year zero-coupon yield $R_{10}(t)$ under the Hull--White model can be expressed as follows:
\begin{align*}
    R_{10}(t)
    =&
    -
    \frac{1}{10}
    \ln D(t,t+10)
    \\
    =&
    \frac{1}{10}
    \left(
        -A_{r,10}(t)
        +
        B_{r,10}r(t)
    \right),
    \\
    \text{where} \quad
    A_{r,10}(t)
    :=&
    A_r(t,t+10)
    \\
    =&
    \ln\frac{D(0,t+10)}{D(0,t)}
    -
    B_{r,10}
    \frac{\partial}{\partial t}
    \ln D(0,t)
    -
    \frac{\sigma_r^2}{4a_r}
    \left(
        1-e^{-2a_rt}
    \right)
    B_{r,10}^2,
    \\
    B_{r,10}
    :=&
    B_r(10)
    =
    \frac{1-e^{-10a_r}}{a_r}.
\end{align*}
Using this expression, $\lambda_{\mathrm{r}}(t)$ can be written as follows:
\begin{align}
    \lambda_{\mathrm{r}}(t)
    =&
    \frac{bB_{r,10}}{10}
    y(t)
    \left(
        r_{\mathrm{flr}}(t)-r(t)
    \right)^+,
    \label{eq:lmd_r_written_on_rt}
    \\
    \text{where} \quad
    r_{\mathrm{flr}}(t)
    :=&
    \frac{
        10(m-a)
        +
        bA_{r,10}(t)
    }{
        bB_{r,10}
    }.
    \label{eq:lmd_r_written_on_rt_02}
\end{align}
Equation~\eqref{eq:lmd_r_written_on_rt} shows that, when $y(t)>0$, the prepayment rate $\lambda_{\mathrm{r}}(t)$ becomes positive if the short rate $r(t)$ falls below the strike $r_{\mathrm{flr}}(t)$.
Furthermore, the strike defined in Equation~\eqref{eq:lmd_r_written_on_rt_02} is a deterministic function of time, indicating that, when the prepayment rate is expressed as a function of the short rate $r(t)$, the strike generally exhibits a term structure.

Therefore, $\bar{\lambda}_{\mathrm{r},1}^{\mathbb{Q}_s}(u)$ can be expressed as follows:
\begin{align}
    \bar{\lambda}_{\mathrm{r},1}^{\mathbb{Q}_s}(u)
    =&
    \frac{bB_{r,10}}{10}
    \mathbb{E}^{\mathbb{Q}_s}
    \left[
        \left.
        y(u)
        \left(
            r_{\mathrm{flr}}(u)-r(u)
        \right)^+
        \right|
        \mathcal{F}_t
    \right]
    \nonumber
    \\
    =&
    \frac{bB_{r,10}}{10}
    r_{\mathrm{flr}}(u)
    \mathbb{E}^{\mathbb{Q}_s}
    \left[
        \left.
        y(u)
        \mathbf{1}_{\left\{
            r(u)\leq r_{\mathrm{flr}}(u)
        \right\}}
        \right|
        \mathcal{F}_t
    \right]
    \nonumber
    \\
    &-
    \frac{bB_{r,10}}{10}
    \mathbb{E}^{\mathbb{Q}_s}
    \left[
        \left.
        y(u)r(u)
        \mathbf{1}_{\left\{
            r(u)\leq r_{\mathrm{flr}}(u)
        \right\}}
        \right|
        \mathcal{F}_t
    \right].
    \label{eq:lmd_br_Qs_analytic}
\end{align}
To evaluate the two expectations on the right-hand side of Equation~\eqref{eq:lmd_br_Qs_analytic}, we define the conditional means and standard deviations of $r(u)$ and $y(u)$ under $\mathbb{Q}_s$, for $u\in[t,s]$, as follows:
\begin{align}
    \mu_r^{\mathbb{Q}_s}(t,u)
    :=&
    \mathbb{E}^{\mathbb{Q}_s}
    \left[
        \left.
        r(u)
        \right|
        \mathcal{F}_t
    \right],
    \label{eq:def_r_mean_Qs}
    \\
    \sigma_r^{\mathbb{Q}_s}(t,u)
    :=&
    \sqrt{
        \operatorname{Var}^{\mathbb{Q}_s}
        \left[
            \left.
            r(u)
            \right|
            \mathcal{F}_t
        \right]
    },
    \label{eq:def_r_sd_Qs}
    \\
    \mu_y^{\mathbb{Q}_s}(t,u)
    :=&
    \mathbb{E}^{\mathbb{Q}_s}
    \left[
        \left.
        y(u)
        \right|
        \mathcal{F}_t
    \right],
    \label{eq:def_y_mean_Qs}
    \\
    \sigma_y^{\mathbb{Q}_s}(t,u)
    :=&
    \sqrt{
        \operatorname{Var}^{\mathbb{Q}_s}
        \left[
            \left.
            y(u)
            \right|
            \mathcal{F}_t
        \right]
    }.
    \label{eq:def_y_sd_Qs}
\end{align}
Analytical expressions are available for $\mu_r^{\mathbb{Q}_s}(t,u)$, $\sigma_r^{\mathbb{Q}_s}(t,u)$, $\mu_y^{\mathbb{Q}_s}(t,u)$, and $\sigma_y^{\mathbb{Q}_s}(t,u)$ (see Section~\ref{app:conditional_means_variances} for details).

Using $\mu_r^{\mathbb{Q}_s}(t,u)$, $\sigma_r^{\mathbb{Q}_s}(t,u)$, $\mu_y^{\mathbb{Q}_s}(t,u)$, and $\sigma_y^{\mathbb{Q}_s}(t,u)$, we define $X(u)$ and $Y(u)$ as follows:
\begin{align*}
    X(u)
    :=&
    \frac{
        r(u)-\mu_r^{\mathbb{Q}_s}(t,u)
    }{
        \sigma_r^{\mathbb{Q}_s}(t,u)
    },
    \\
    Y(u)
    :=&
    \frac{
        y(u)-\mu_y^{\mathbb{Q}_s}(t,u)
    }{
        \sigma_y^{\mathbb{Q}_s}(t,u)
    }.
\end{align*}
Note that $\sigma_r^{\mathbb{Q}_s}(t,t)=\sigma_y^{\mathbb{Q}_s}(t,t)=0$ at $u=t$.
Therefore, the above standardization is defined for $u\in(t,s]$.
When values at $u=t$ are required, they are treated as limits as $u\downarrow t$.
Under the $s$-forward measure $\mathbb{Q}_s$, the following relation holds:
\begin{align}
    \left.
        \begin{pmatrix}
            X(u)
            \\
            Y(u)
        \end{pmatrix}
    \right|_{\mathcal{F}_t}
    \sim&
    \mathcal{N}
    \left(
        \begin{pmatrix}
            0
            \\
            0
        \end{pmatrix}
        ,
        \begin{pmatrix}
            1 & \rho_{xy}(t,u)
            \\
            \rho_{xy}(t,u) & 1
        \end{pmatrix}
    \right),
    \nonumber
    \\
    \text{where} \quad
    \rho_{xy}(t,u)
    :=&
    \frac{
        \operatorname{Cov}^{\mathbb{Q}_s}
        \left[
            \left.
            r(u),y(u)
            \right|
            \mathcal{F}_t
        \right]
    }{
        \sigma_r^{\mathbb{Q}_s}(t,u)
        \sigma_y^{\mathbb{Q}_s}(t,u)
    },
    \nonumber
    \\
    \operatorname{Cov}^{\mathbb{Q}_s}
    \left[
        \left.
        r(u),y(u)
        \right|
        \mathcal{F}_t
    \right]
    =&
    \rho\sigma_r\sigma_y
    \int _t ^u e^{- (a_r + \kappa _y) (u-z)} \mathrm{d}z
    =
    \rho\sigma_r\sigma_y
    \frac{
        1-e^{-(a_r+\kappa_y)(u-t)}
    }{
        a_r+\kappa_y
    }.
    \label{eq:cov_r_y}
\end{align}
The indicator functions appearing on the right-hand side of Equation~\eqref{eq:lmd_br_Qs_analytic} satisfy the following relation:
\begin{align*}
    &
    r(u)
    \leq
    r_{\mathrm{flr}}(u)
    \quad
    \Leftrightarrow
    \quad
    X(u)
    \leq
    d(t,u),
    \\
    &
    \text{where} \quad
    d(t,u)
    :=
    \frac{
        r_{\mathrm{flr}}(u)
        -
        \mu_r^{\mathbb{Q}_s}(t,u)
    }{
        \sigma_r^{\mathbb{Q}_s}(t,u)
    }.
\end{align*}
First, the expectation in the first term on the right-hand side of Equation~\eqref{eq:lmd_br_Qs_analytic} can be evaluated as follows:
\begin{align*}
    &
    \mathbb{E}^{\mathbb{Q}_s}
    \left[
        \left.
        y(u)
        \mathbf{1}_{\left\{
            r(u)\leq r_{\mathrm{flr}}(u)
        \right\}}
        \right|
        \mathcal{F}_t
    \right]
    \\
    =&
    \mathbb{E}^{\mathbb{Q}_s}
    \left[
        \left.
        \left(
            \mu_y^{\mathbb{Q}_s}(t,u)
            +
            \sigma_y^{\mathbb{Q}_s}(t,u)Y(u)
        \right)
        \mathbf{1}_{\left\{
            X(u)\leq d(t,u)
        \right\}}
        \right|
        \mathcal{F}_t
    \right]
    \\
    =&
    \Phi\bigl(d(t,u)\bigr)
    \mu_y^{\mathbb{Q}_s}(t,u)
    -
    \rho_{xy}(t,u)
    \varphi\bigl(d(t,u)\bigr)
    \sigma_y^{\mathbb{Q}_s}(t,u).
\end{align*}
Similarly, the expectation in the second term on the right-hand side of Equation~\eqref{eq:lmd_br_Qs_analytic} can be evaluated as follows:
\begin{align*}
    &
    \mathbb{E}^{\mathbb{Q}_s}
    \left[
        \left.
        y(u)r(u)
        \mathbf{1}_{\left\{
            r(u)\leq r_{\mathrm{flr}}(u)
        \right\}}
        \right|
        \mathcal{F}_t
    \right]
    \\
    =&
    \mathbb{E}^{\mathbb{Q}_s}
    \left[
        \left.
        \left(
            \mu_y^{\mathbb{Q}_s}(t,u)
            +
            \sigma_y^{\mathbb{Q}_s}(t,u)Y(u)
        \right)
        \left(
            \mu_r^{\mathbb{Q}_s}(t,u)
            +
            \sigma_r^{\mathbb{Q}_s}(t,u)X(u)
        \right)
        \mathbf{1}_{\left\{
            X(u)\leq d(t,u)
        \right\}}
        \right|
        \mathcal{F}_t
    \right]
    \\
    =&
    \mu_y^{\mathbb{Q}_s}(t,u)
    \mu_r^{\mathbb{Q}_s}(t,u)
    \mathbb{E}^{\mathbb{Q}_s}
    \left[
        \left.
        \mathbf{1}_{\left\{
            X(u)\leq d(t,u)
        \right\}}
        \right|
        \mathcal{F}_t
    \right]
    \\
    &+
    \mu_y^{\mathbb{Q}_s}(t,u)
    \sigma_r^{\mathbb{Q}_s}(t,u)
    \mathbb{E}^{\mathbb{Q}_s}
    \left[
        \left.
        X(u)
        \mathbf{1}_{\left\{
            X(u)\leq d(t,u)
        \right\}}
        \right|
        \mathcal{F}_t
    \right]
    \\
    &+
    \mu_r^{\mathbb{Q}_s}(t,u)
    \sigma_y^{\mathbb{Q}_s}(t,u)
    \mathbb{E}^{\mathbb{Q}_s}
    \left[
        \left.
        Y(u)
        \mathbf{1}_{\left\{
            X(u)\leq d(t,u)
        \right\}}
        \right|
        \mathcal{F}_t
    \right]
    \\
    &+
    \sigma_y^{\mathbb{Q}_s}(t,u)
    \sigma_r^{\mathbb{Q}_s}(t,u)
    \mathbb{E}^{\mathbb{Q}_s}
    \left[
        \left.
        X(u)Y(u)
        \mathbf{1}_{\left\{
            X(u)\leq d(t,u)
        \right\}}
        \right|
        \mathcal{F}_t
    \right]
    \\
    =&
    \Phi\bigl(d(t,u)\bigr)
    \mu_y^{\mathbb{Q}_s}(t,u)
    \mu_r^{\mathbb{Q}_s}(t,u)
    \\
    &-
    \varphi\bigl(d(t,u)\bigr)
    \left(
        \mu_y^{\mathbb{Q}_s}(t,u)
        \sigma_r^{\mathbb{Q}_s}(t,u)
        +
        \rho_{xy}(t,u)
        \sigma_y^{\mathbb{Q}_s}(t,u)
        \mu_r^{\mathbb{Q}_s}(t,u)
    \right)
    \\
    &+
    \rho_{xy}(t,u)
    \left(
        \Phi\bigl(d(t,u)\bigr)
        -
        d(t,u)
        \varphi\bigl(d(t,u)\bigr)
    \right)
    \sigma_y^{\mathbb{Q}_s}(t,u)
    \sigma_r^{\mathbb{Q}_s}(t,u).
\end{align*}
Combining these results, we obtain the following analytical expression for $\bar{\lambda}_{\mathrm{r},1}^{\mathbb{Q}_s}(u)$:
\begin{align}
    \bar{\lambda}_{\mathrm{r},1}^{\mathbb{Q}_s}(u)
    ={}&\frac{bB_{\mathrm{r},10}}{10}
    \Big[
    \mu_y^{\mathbb{Q}_s}(t,u)
    \big\{
    \Delta_{\mathrm{r}}^{\mathbb{Q}_s}(t,u)\Phi(d(t,u))
    +\sigma_{\mathrm{r}}^{\mathbb{Q}_s}(t,u)\varphi(d(t,u))
    \big\}
    \nonumber\\
    &\quad-
    \rho _{xy} (t,u) \sigma ^{\mathbb{Q}_s} _r (t,u) \sigma ^{\mathbb{Q}_s} _y (t,u)
    \Phi(d(t,u))
    \Big].
    \label{eq:first_order_lambda_final_expression}
\end{align}
Here, we define $\Delta_r^{\mathbb{Q}_s}(t,v)$ as follows:
\begin{align*}
    \Delta_r^{\mathbb{Q}_s}(t,v)
    &:=
    r_{\mathrm{flr}}(v)
    -
    \mu_r^{\mathbb{Q}_s}(t,v).
\end{align*}
Equation~\eqref{eq:first_order_lambda_final_expression} can also be expressed as follows:
\begin{align*}
    \bar{\lambda}_{\mathrm{r},1}^{\mathbb{Q}_s}(u)
    ={}&\frac{bB_{\mathrm{r},10}}{10}
    \Big\{
    \mu_y^{\mathbb{Q}_s}(t,u)
    \mathbb{E}^{\mathbb{Q}_s}
    \left[
    \left(r_{\mathrm{flr}}(u)-r(u)\right)^+
    \mid \mathcal{F}_t
    \right]
    \nonumber\\
    &\quad-
    \mathrm{Cov}^{\mathbb{Q}_s}
    [r(u),y(u)\mid\mathcal{F}_t]
    \Phi(d(t,u))
    \Big\}.
\end{align*}
The first term on the right-hand side of the above expression corresponds to the conditional expectation of
$y(u)(r_{\mathrm{flr}}(u)-r(u))^+$
under $\mathbb{Q}_s$ when $r(u)$ and $y(u)$ are conditionally independent given $\mathcal{F}_t$.
In contrast, the second term represents a correction arising from the conditional covariance between $r(u)$ and $y(u)$.
However, since $\mu_y^{\mathbb{Q}_s}(t,u)$ itself also depends on the correlation coefficient $\rho$, the second term alone does not capture the entire effect of correlation.

Next, we derive a closed-form expression for $\bar{\lambda}_{\mathrm{r},2}^{\mathbb{Q}_s}(u,\eta)$.
Since the derivation follows essentially the same approach as that for $\bar{\lambda}_{\mathrm{r},1}^{\mathbb{Q}_s}(u)$, we provide only an outline here and refer the reader to Appendix~\ref{app:lambda_bar_2_prove} for details.
For $t\leq u\leq\eta\leq s$, $\bar{\lambda}_{\mathrm{r},2}^{\mathbb{Q}_s}(u,\eta)$ is expressed as follows:
\begin{align}
    \bar{\lambda}_{\mathrm{r},2}^{\mathbb{Q}_s}(u,\eta)
    =&
    \left(
        \frac{bB_{r,10}}{10}
    \right)^2
    \mathbb{E}^{\mathbb{Q}_s}
    \left[
        \left.
        y(u)y(\eta)
        \left(
            r_{\mathrm{flr}}(u)-r(u)
        \right)^+
        \left(
            r_{\mathrm{flr}}(\eta)-r(\eta)
        \right)^+
        \right|
        \mathcal{F}_t
    \right].
    \label{eq:second_lambda_br_exp}
\end{align}
In the following derivation using standardized variables, we assume $t<u<\eta\leq s$.
Values on the boundaries corresponding to $u=t$ and $u=\eta$ are treated as continuous limits when necessary.
The expectation on the right-hand side of the above equation can be expressed as follows:
\begin{align}
    &
    \mathbb{E}^{\mathbb{Q}_s}
    \left[
        \left.
        y(u)y(\eta)
        \left(
            r_{\mathrm{flr}}(u)-r(u)
        \right)^+
        \left(
            r_{\mathrm{flr}}(\eta)-r(\eta)
        \right)^+
        \right|
        \mathcal{F}_t
    \right]
    \nonumber
    \\
    =&
    \mathbb{E}^{\mathbb{Q}_s}
    \Bigg[
        \left\{
            \mu_y^{\mathbb{Q}_s}(t,u)
            +
            \sigma_y^{\mathbb{Q}_s}(t,u)Y(u)
        \right\}
        \left\{
            \mu_y^{\mathbb{Q}_s}(t,\eta)
            +
            \sigma_y^{\mathbb{Q}_s}(t,\eta)Y(\eta)
        \right\}
    \nonumber
    \\
    &\qquad\qquad\times
        \left\{
            \Delta_r^{\mathbb{Q}_s}(t,u)
            -
            \sigma_r^{\mathbb{Q}_s}(t,u)X(u)
        \right\}
        \left\{
            \Delta_r^{\mathbb{Q}_s}(t,\eta)
            -
            \sigma_r^{\mathbb{Q}_s}(t,\eta)X(\eta)
        \right\}
    \nonumber
    \\
    &\qquad\qquad\times
        \mathbf{1}_{\left\{
            X(u)\leq d(t,u)
        \right\}}
        \mathbf{1}_{\left\{
            X(\eta)\leq d(t,\eta)
        \right\}}
        \left|
        \mathcal{F}_t
        \right.
    \Bigg].
    \label{eq:second_order_expectation}
\end{align}
The random vector $\left.(X(u),Y(u),X(\eta),Y(\eta))^{\top}\right|_{\mathcal{F}_t}$ follows the following four-dimensional normal distribution:
\begin{align*}
    &
    \left.
        \begin{pmatrix}
            X(u)
            \\
            Y(u)
            \\
            X(\eta)
            \\
            Y(\eta)
        \end{pmatrix}
    \right|_{\mathcal{F}_t}
    \sim
    \mathcal{N}
    \left(
        \begin{pmatrix}
            0
            \\
            0
            \\
            0
            \\
            0
        \end{pmatrix}
        ,
        \bm{\Sigma}
    \right),
    \\
    \text{where} \quad
    \bm{\Sigma}
    :=&
    \begin{pmatrix}
        1
        &
        \rho_{xy}(t,u)
        &
        \rho_{xx}(t,u,\eta)
        &
        \rho_{xy}(t,u)\rho_{yy}(t,u,\eta)
        \\
        \rho_{xy}(t,u)
        &
        1
        &
        \rho_{xy}(t,u)\rho_{xx}(t,u,\eta)
        &
        \rho_{yy}(t,u,\eta)
        \\
        \rho_{xx}(t,u,\eta)
        &
        \rho_{xy}(t,u)\rho_{xx}(t,u,\eta)
        &
        1
        &
        \rho_{xy}(t,\eta)
        \\
        \rho_{xy}(t,u)\rho_{yy}(t,u,\eta)
        &
        \rho_{yy}(t,u,\eta)
        &
        \rho_{xy}(t,\eta)
        &
        1
    \end{pmatrix},
    \\
    \rho_{xx}(t,u,\eta)
    :=&
    \operatorname{Corr}^{\mathbb{Q}_s}
    \left[
        \left.
        X(u),X(\eta)
        \right|
        \mathcal{F}_t
    \right]
    =
    e^{-a_r(\eta-u)}
    \frac{
        \sigma_r^{\mathbb{Q}_s}(t,u)
    }{
        \sigma_r^{\mathbb{Q}_s}(t,\eta)
    },
    \\
    \rho_{yy}(t,u,\eta)
    :=&
    \operatorname{Corr}^{\mathbb{Q}_s}
    \left[
        \left.
        Y(u),Y(\eta)
        \right|
        \mathcal{F}_t
    \right]
    =
    e^{-\kappa_y(\eta-u)}
    \frac{
        \sigma_y^{\mathbb{Q}_s}(t,u)
    }{
        \sigma_y^{\mathbb{Q}_s}(t,\eta)
    }.
\end{align*}
To express the expectation in Equation~\eqref{eq:second_order_expectation} more compactly, we define the following coefficients:
\begin{align*}
    C_{00}(u,\eta)
    :=&
    \mu_y^{\mathbb{Q}_s}(t,u)
    \mu_y^{\mathbb{Q}_s}(t,\eta)
    +
    \sigma_y^{\mathbb{Q}_s}(t,u)
    \sigma_y^{\mathbb{Q}_s}(t,\eta)
    \gamma_y(t,u,\eta),
    \\
    C_{10}(u,\eta)
    :=&
    \mu_y^{\mathbb{Q}_s}(t,u)
    \sigma_y^{\mathbb{Q}_s}(t,\eta)
    \alpha_y(t,u,\eta)
    +
    \sigma_y^{\mathbb{Q}_s}(t,u)
    \mu_y^{\mathbb{Q}_s}(t,\eta)
    \rho_{xy}(t,u),
    \\
    C_{01}(u,\eta)
    :=&
    \mu_y^{\mathbb{Q}_s}(t,u)
    \sigma_y^{\mathbb{Q}_s}(t,\eta)
    \beta_y(t,u,\eta),
    \\
    C_{20}(u,\eta)
    :=&
    \sigma_y^{\mathbb{Q}_s}(t,u)
    \sigma_y^{\mathbb{Q}_s}(t,\eta)
    \rho_{xy}(t,u)
    \alpha_y(t,u,\eta),
    \\
    C_{11}(u,\eta)
    :=&
    \sigma_y^{\mathbb{Q}_s}(t,u)
    \sigma_y^{\mathbb{Q}_s}(t,\eta)
    \rho_{xy}(t,u)
    \beta_y(t,u,\eta).
\end{align*}
Here, $\alpha_y(t,u,\eta)$, $\beta_y(t,u,\eta)$, and $\gamma_y(t,u,\eta)$ are defined as follows:
\begin{align*}
    \alpha_y(t,u,\eta)
    :=&
    \frac{
        \rho_{xy}(t,u)\rho_{yy}(t,u,\eta)
        -
        \rho_{xx}(t,u,\eta)\rho_{xy}(t,\eta)
    }{
        1-\rho_{xx}^2(t,u,\eta)
    },
    \\
    \beta_y(t,u,\eta)
    :=&
    \frac{
        \rho_{xy}(t,\eta)
        -
        \rho_{xx}(t,u,\eta)
        \rho_{xy}(t,u)
        \rho_{yy}(t,u,\eta)
    }{
        1-\rho_{xx}^2(t,u,\eta)
    },
    \\
    \gamma_y(t,u,\eta)
    :=&
    \left(
        1-\rho_{xy}^2(t,u)
    \right)
    \rho_{yy}(t,u,\eta).
\end{align*}
Furthermore, we define the truncated moments $\mathcal{M}_{pq}(u,\eta)$ of $X(u)$ and $X(\eta)$ as follows:
\begin{align*}
    \mathcal{M}_{pq}(u,\eta)
    :=&
    \mathbb{E}^{\mathbb{Q}_s}
    \left[
        \left.
        X(u)^p
        X(\eta)^q
        \mathbf{1}_{\left\{
            X(u)\leq d(t,u)
        \right\}}
        \mathbf{1}_{\left\{
            X(\eta)\leq d(t,\eta)
        \right\}}
        \right|
        \mathcal{F}_t
    \right],
    \qquad
    p,q=0,1,2,\ldots.
\end{align*}
Analytical expressions are available for the truncated moments $\mathcal{M}_{pq}(u,\eta)$ required below.
Using these expressions, we can analytically evaluate the expectation in Equation~\eqref{eq:second_order_expectation} and ultimately obtain the following closed-form expression for $\bar{\lambda}_{\mathrm{r},2}^{\mathbb{Q}_s}(u,\eta)$:
\begin{align}
    \bar{\lambda}_{\mathrm{r},2}^{\mathbb{Q}_s}(u,\eta)
    ={}&
    \left(\frac{bB_{\mathrm{r},10}}{10}\right)^2
    \sum_{(p,q)\in\mathcal I}
    C_{pq}(u,\eta)
    \sum_{i=0}^{1}\sum_{j=0}^{1}
    (-1)^{i+j}
    \Delta_u^{1-i}\sigma_u^i
    \Delta_\eta^{1-j}\sigma_\eta^j
    M_{p+i,q+j}(u,\eta).
    \label{eq:second_order_lambda_final_expression}
\end{align}
Here, for notational simplicity, we have introduced
\begin{align*}
    \mathcal I
    &:=\{(0,0),(1,0),(0,1),(2,0),(1,1)\},\\
    \Delta_v
    &:=\Delta_{\mathrm{r}}^{\mathbb{Q}_s}(t,v),\\
    \sigma_v
    &:=\sigma_{\mathrm{r}}^{\mathbb{Q}_s}(t,v).
\end{align*}
For a comparison between the analytical expressions for $\bar{\lambda}_{\mathrm{r},1}^{\mathbb{Q}_s}(u)$ and $\bar{\lambda}_{\mathrm{r},2}^{\mathbb{Q}_s}(u,\eta)$ and their Monte Carlo estimates, see Appendix~\ref{app:lambda_expectations_numerical_validation}.

The above results yield the following approximate representation of the MBS price:
\begin{align}
    V(t,T)
    \simeq
    V^{(2)}(t,T)
    &:=
    V_0(t,T)
    +
    V_{\mathrm{r}}^{(2)}(t,T)
    -
    V_{\mathrm{op}}^{(2)}(t,T),
    \label{eq:mbs_closed_form_02}
    \\
    V_{\mathrm{r}}^{(2)}(t,T)
    &:=
    \int_t^T
    \bar{c}_{\mathrm{r}}^{(2)}(t,s)
    D(t,s)
    Z_{\mathrm{c}}(t,s)
    Z_{\mathrm{r}}^{(2)}(t,s)
    \mathrm{d}s,
    \label{eq:mbs_closed_form_02_01}
    \\
    V_{\mathrm{op}}^{(2)}(t,T)
    &:=
    \int_t^T
    \bar{c}_0(t,s)
    Z_{\mathrm{op}}^{(2)}(t,s)
    \mathrm{d}s,
    \label{eq:mbs_closed_form_03}
    \\
    \bar{c}^{(2)}(t,s)
    &:=
    \bar{c}_0(t,s)
    +
    \bar{c}_{\mathrm{r}}^{(2)}(t,s),
    \label{eq:mbs_closed_form_03_01}
    \\
    \bar{c}_{\mathrm{r}}^{(2)}(t,s)
    :=&
    e^{-\int_0^t\lambda(u)\mathrm{d}u}
    \tilde{P}(s)
    \left.
    \left(
        \frac{\partial}{\partial v}
        \kappa_1^{(s)}(t,v)
        -
        \frac{1}{2}
        \frac{\partial}{\partial v}
        \kappa_2^{(s)}(t,v)
    \right)
    \right|_{v=s},
    \label{eq:mbs_closed_form_03_02}
    \\
    Z_{\mathrm{r}}^{(2)}(t,s)
    &:=
    e^{-\kappa_1(t,s)+\kappa_2(t,s)/2},
    \label{eq:Z_r_cumulant_2}
    \\
    Z_{\mathrm{op}}^{(2)}(t,s)
    &:=
    D(t,s)
    Z_{\mathrm{c}}(t,s)
    \left(
        1
        -
        Z_{\mathrm{r}}^{(2)}(t,s)
    \right).
    \label{eq:mbs_closed_form_05}
\end{align}
Here,
$\kappa_n(t,s)=\kappa_n^{(s)}(t,s)$
denotes the diagonal value of the cumulant defined under the fixed $s$-forward measure $\mathbb{Q}_s$.
In contrast, the derivatives appearing in Equation~\eqref{eq:mbs_closed_form_03_02} are taken with respect to the auxiliary variable $v$, while holding $s$ and the measure $\mathbb{Q}_s$ fixed.
The superscript $(2)$ in quantities such as $V^{(2)}(t,T)$ indicates that they are approximated using the cumulant expansion truncated at the second cumulant.
Analytical expressions for
$\bar{\lambda}_{\mathrm{r},1}^{\mathbb{Q}_s}(u)$
and
$\bar{\lambda}_{\mathrm{r},2}^{\mathbb{Q}_s}(u,\eta)$,
which enter $\kappa_1(t,s)$ and $\kappa_2(t,s)$, are given by Equations~\eqref{eq:first_order_lambda_final_expression} and~\eqref{eq:second_order_lambda_final_expression}, respectively.
Therefore, the expectations required to evaluate $V^{(2)}(t,T)$ can be obtained analytically, and the remaining time integrals can be evaluated as deterministic integrals.

Rearranging this decomposition yields the following product form, which is consistent with the original MBS valuation formula:
\begin{align*}
    V^{(2)}(t,T)
    =\int_t^T\bar{c}^{(2)}(t,s)D(t,s)Z_{\mathrm{c}}(t,s)Z_{\mathrm{r}}^{(2)}(t,s)\mathrm{d}s.
\end{align*}
Note that although $Z_{\mathrm{r}}^{(2)}$ is strictly positive by construction as an exponential function, truncating the cumulant expansion at the second cumulant does not guarantee that $Z_{\mathrm{r}}^{(2)}\leq1$ or $\bar{c}_{\mathrm{r}}^{(2)}\geq0$.
Therefore, when interpreting the approximated quantities as nonnegative price components, these conditions must be verified numerically.
When the cumulant expansion is truncated at the first cumulant, $\kappa_2^{(s)}(t,v)$ and its derivative with respect to $v$ are set to zero, and $Z_{\mathrm{r}}^{(1)}(t,s)=e^{-\kappa_1(t,s)}$ is used.
In the following, we denote the MBS price obtained under this approximation by $V^{(1)}(t,T)$.

\section{Numerical Analysis of MBS Prices and Interest Rate Sensitivities}
\label{sec:mbs_numerical_analysis}

In this section, we numerically analyze MBS prices and their interest rate sensitivities using the closed-form approximation for MBS prices derived in Section~\ref{sec:mbs_approximation}.
First, we evaluate the accuracy of the closed-form approximation by comparing MBS prices obtained from the closed-form approximation with those obtained from Monte Carlo simulation.
Next, we decompose MBS prices into the price components introduced in the preceding section and examine how each component responds to changes in the interest rate level and refinancing sensitivity.
Furthermore, by decomposing the duration and convexity of MBS prices into the contributions of individual price components, we quantitatively investigate the effects of interest-rate-dependent prepayments on MBS interest rate sensitivities.
In particular, we clarify how duration shortening and negative convexity in declining interest rate environments are related to the interest rate sensitivities of the individual price components.

\subsection{Numerical Setup, Interest-Rate Sensitivities, and Approximation Accuracy}

In this subsection, we first present the baseline model parameters and computational settings used in the numerical analysis, and then assess the accuracy of the approximate closed-form MBS valuation formulas derived in the preceding sections.
Specifically, we compare MBS prices evaluated using cumulant expansions truncated after the first and second cumulants with those obtained by Monte Carlo simulation.
Unless otherwise stated, we use the model parameters and computational settings reported in Table~\ref{tab:base_parameters} throughout this section.
\begin{table}[!htbp]
    \centering
    \caption{
        Baseline Model Parameters and Computational Settings for Numerical Analysis
    }
    \label{tab:base_parameters}
    \begin{threeparttable}
        \small
        \setlength{\tabcolsep}{8pt}
        \renewcommand{\arraystretch}{1.15}
        \begin{tabular}{
            @{}
            >{\centering\arraybackslash}m{0.13\textwidth}
            >{\centering\arraybackslash}m{0.20\textwidth}
            >{\centering\arraybackslash}m{0.18\textwidth}
            @{}
        }
            \toprule
            Category
            & Parameter / Setting
            & Value
            \\
            \midrule

            \multirow{2}{*}{Term Structure}
            & $R$
            & $0.03$ (flat)
            \\
            & $D(0,t)$
            & $\exp(-Rt)$
            \\
            \midrule

            \multirow{4}{*}{$r(t)$}
            & $\vartheta(t)$
            & Consistent with the term structure
            \\
            & $a_r$
            & $0.03$
            \\
            & $\sigma_r$
            & $0.01$
            \\
            & $r(0)$
            & \makecell{
                $\left.
                    -\dfrac{\partial}{\partial t}\ln D(0,t)
                \right|_{t\to 0}$
                \\
                $=R=0.03$
            }
            \\
            \midrule

            \multirow{5}{*}{$y(t)$}
            & $\kappa_y$
            & $0.5$
            \\
            & $\bar{y}$
            & $1.0$
            \\
            & $\sigma_y$
            & $0.03$
            \\
            & $y(0)$
            & $0.00$
            \\
            & $\rho$
            & $-0.6$
            \\
            \midrule

            \multirow{4}{*}{$\lambda_{\mathrm{c}}(t)$}
            & $\kappa_{\lambda}$
            & $0.15$
            \\
            & $\bar{\lambda}$
            & $0.04$
            \\
            & $\sigma_{\lambda}$
            & $0.04$
            \\
            & $\lambda_{\mathrm{c}}(0)$
            & $0.02$
            \\
            \midrule

            \multirow{5}{*}{MBS}
            & $m$
            & $0.035$
            \\
            & $T$
            & $30.0$
            \\
            & $a$
            & $0.0$
            \\
            & $b$
            & $1.0$
            \\
            & Outstanding Principal Fraction
            & $1.0$
            \\
            & Cash-Flow Discretization Interval
            & $1/12$ year
            \\
            \midrule

            \multirow{2}{*}{MC1}
            & Number of Paths
            & $100{,}000$
            \\
            & Method
            & Multivariate Normal Sampling
            \\
            \midrule

            \multirow{3}{*}{MC2}
            & Number of Paths
            & $500{,}000$
            \\
            & $\Delta t$
            & $1/48$ year
            \\
            & Method
            & Euler--Maruyama Method
            \\
            \bottomrule
        \end{tabular}
    \end{threeparttable}
\end{table}

Note that $y(t)$, defined by Equation~\eqref{eq:SDE_y}, follows a Gaussian mean-reverting process and can therefore theoretically take negative values.
Under $y(0)=0$, the conditional mean and standard deviation of $y(t)$ under $\mathbb{Q}$ are given by
$\mu_y(t)=\bar{y}\left(1-e^{-\kappa_y t}\right)$
and
$\sigma_y(t)=\sigma_y\sqrt{(1-e^{-2\kappa_y t})/(2\kappa_y)}$,
respectively.
Thus, we obtain
$\mathbb{Q}\left(y(t)<0\right)=\Phi\left(-\mu_y(t)/\sigma_y(t)\right)$.
Using the parameters in Table~\ref{tab:base_parameters}, at the first monthly evaluation date $t=1/12$,
the probability is
$\mathbb{Q}(y(1/12)<0)\simeq7.5\times10^{-7}$
for $\bar{y}=1.0$ and
$\mathbb{Q}(y(1/12)<0)\simeq3.7\times10^{-128}$
for $\bar{y}=5.0$.
Therefore, under the parameter settings and at the monthly evaluation dates considered in the numerical analysis, the probability that $y(t)$ takes negative values is negligibly small.

\begin{figure}[bhtp]
    \begin{center}
        \includegraphics[width=17cm]{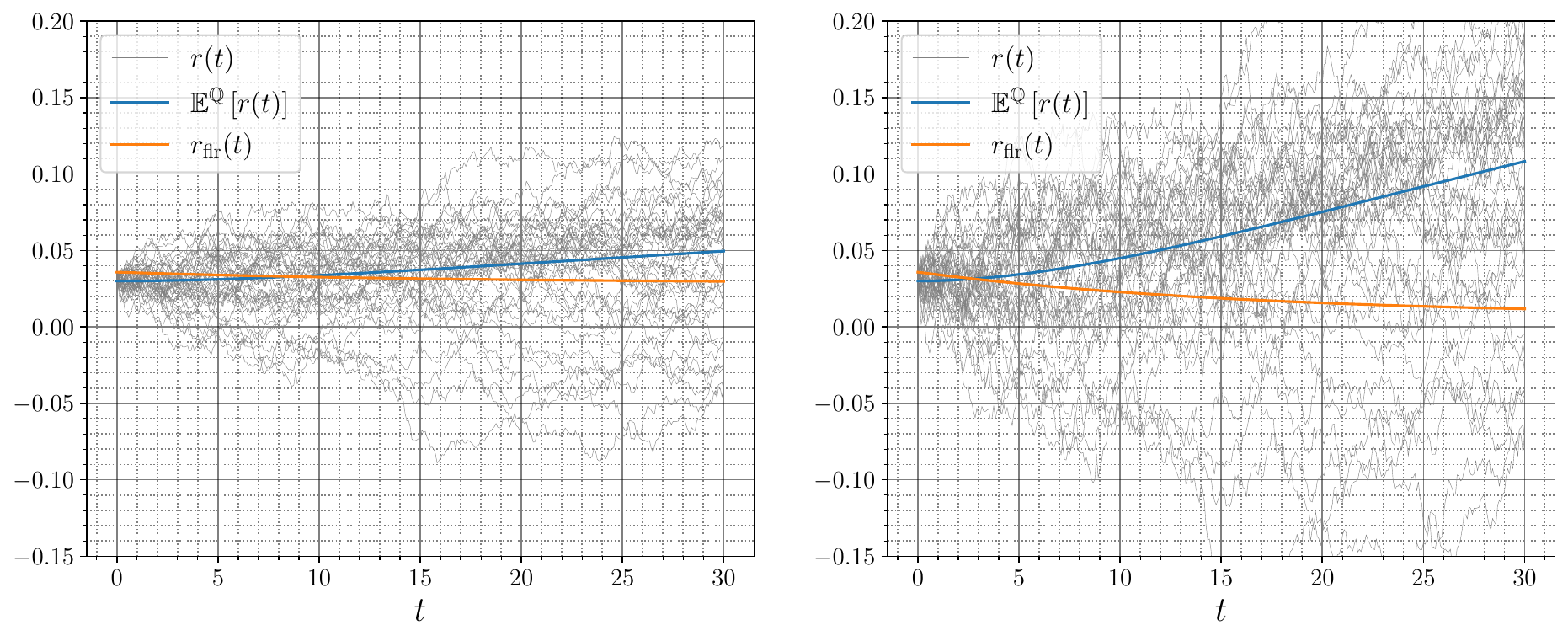}
        \caption{
            Evolution of several sample paths of $r(t)$, $\mathbb{E}^{\mathbb{Q}} \left[ r(t) \right]$, and $r_{\mathrm{flr}}(t)$.
            The left and right panels correspond to $\sigma_r = 0.01$ and $\sigma_r = 0.02$, respectively.
            The solid gray lines represent $40$ sample paths of $r(t)$, the solid blue line represents $\mathbb{E}^{\mathbb{Q}} \left[ r(t) \right]$, and the solid orange line represents $r_{\mathrm{flr}}(t)$.
            The other parameters and computational settings are given in Table~\ref{tab:base_parameters}.
            }
        \label{fig:r_t_r_flr}
    \end{center}
\end{figure}

Figure~\ref{fig:r_t_r_flr} shows several sample paths of $r(t)$, together with the evolution of $\mathbb{E}^{\mathbb{Q}} \left[ r(t) \right]$ and $r_{\mathrm{flr}}(t)$.
For $u \geq t$, $\mathbb{E}^{\mathbb{Q}}\left[ \left. r(u) \right| \mathcal{F}_t \right]$ is given by the following expression (see Appendix~\ref{app:expectation_r_under_Q_derivation} for the derivation):
\begin{align}
    \mathbb{E}^{\mathbb{Q}}\left[ \left. r(u) \right| \mathcal{F}_t \right]
    =&
    e^{- a_r (u - t)} r(t) - \frac{\partial}{\partial u} \ln D(0, u) + e^{- a_r (u-t)} \frac{\partial}{\partial t} \ln D(0, t)
    \nonumber
    \\
    &+
    \frac{\sigma _r ^2}{2 a_r ^2} \left( 1 - e^{- a_r (u-t)} \right) \left( 1 - e^{- a_r (u+t)} \right).
    \label{eq:expectation_r_t_under_Q}
\end{align}
In particular, when the zero-yield curve is flat, $\mathbb{E}^{\mathbb{Q}}\left[r(t)\right] \to R+\frac{\sigma_r^2}{2a_r^2}$ as $t \to +\infty$.
Thus, in the long-run limit, the expected short rate exceeds the zero yield $R$ by $\frac{\sigma_r^2}{2a_r^2}$, reflecting the uncertainty arising from interest rate fluctuations.
Figure~\ref{fig:r_t_r_flr} shows that the timing of positive prepayments arising from the interest-rate-dependent component varies across sample paths, depending on when the short rate $r(t)$ falls below the threshold $r_{\mathrm{flr}}(t)$.
Furthermore, the relative levels of $\mathbb{E}^{\mathbb{Q}}\left[r(t)\right]$ and $r_{\mathrm{flr}}(t)$ suggest that, under the parameter settings used in the figure, positive prepayments arising from the interest-rate-dependent component tend to occur more frequently at shorter horizons and less frequently at longer horizons.
Although we adopt the one-factor Hull--White model in this paper, under more general affine term structure models, the option can also be interpreted as an option whose payoff is given by the time integral of put-type payoffs written on the short rate.

In the following numerical analysis, as shown in Table~\ref{tab:base_parameters}, we assume a flat initial zero-yield curve at time $0$, with zero yields constant across maturities, so that $D(0,t)=\exp(-Rt)$.
Changes in $R$ are interpreted as parallel shifts of the entire initial zero-yield curve, while holding all other model parameters fixed.
Therefore, changes in $R$ affect not only the discount factor $D(0,t)$ but also $r(0)=R$, the drift term $\vartheta(t)$ of the Hull--White model, the distribution of the 10-year zero yield $R_{10}(t)$, and $r_{\mathrm{flr}}(t)$, in a manner consistent with the initial term structure.
Consequently, the derivatives of MBS prices with respect to $R$ considered below do not merely capture the effects of changes in discount rates, as in conventional fixed-rate bonds with predetermined future cash flows.
In this paper, the interest-rate-dependent prepayment rate $\lambda_{\mathrm{r}}(t)$ depends on interest rates through $R_{10}(t)$.
Thus, changes in $R$ affect the prepayment rate and, consequently, the future cash flows of the MBS itself, including prepaid principal and subsequent principal and interest payments.
Therefore, the first and second derivatives of MBS prices with respect to $R$ capture not only the price effects arising from changes in discount factors but also those arising from changes in future cash flows through interest-rate-dependent prepayments.

Furthermore, in the following numerical analysis, we compare two different settings for the mean-reversion level $\bar{y}$ of the refinancing sensitivity $y(t)$ to examine how MBS prices and their interest rate sensitivities change with the magnitude of the effects of interest-rate-dependent prepayments.
Specifically, we use $\bar{y}=1.0$, as shown in Table~\ref{tab:base_parameters}, as the baseline case and additionally analyze the case of $\bar{y}=5.0$.
A higher $\bar{y}$ leads, on average, to a higher interest-rate-dependent prepayment rate $\lambda_{\mathrm{r}}(t)$ when the refinancing incentive is positive.
Therefore, the case of $\bar{y}=5.0$ represents a setting in which the effects of interest-rate-dependent prepayments are stronger than those in the case of $\bar{y}=1.0$.
By comparing these two settings, we examine how the relative contributions of the individual MBS price components change as the effects of interest-rate-dependent prepayments become stronger, and how these changes give rise to characteristic MBS interest rate sensitivities, such as duration shortening and negative convexity.

Under this setting, we define the duration $\mathcal{D}(t,T)$ and convexity $\mathcal{C}(t,T)$, which measure the sensitivities of the MBS price $V(t,T)$ to the zero yield $R$, as follows:
\begin{align*}
    \mathcal{D} (t,T)
    :=&
    - \frac{1}{V (t,T)} \frac{\partial V}{\partial R} (t,T),
    \\
    \mathcal{C} (t,T)
    :=&
    \frac{1}{V (t,T)} \frac{\partial ^2 V}{\partial R^2} (t,T).
\end{align*}
Using the price decomposition given in Equation~\eqref{eq:mbs_closed_form_02}, we define $\mathcal{D}^{(2)}(t,T)$ as an approximation to MBS duration, which can be decomposed into the contributions of the individual price components as follows:
\begin{align}
    \mathcal{D}^{(2)} (t,T)
    :=&
    - \frac{1}{V^{(2)} (t,T)} \frac{\partial V^{(2)}}{\partial R} (t,T)
    \nonumber
    \\
    =&
    \frac{V_0 (t,T)}{V^{(2)} (t,T)} \mathcal{D}_0 (t,T)
    + \frac{V_{\mathrm{r}}^{(2)} (t,T)}{V^{(2)} (t,T)} \mathcal{D}_{\mathrm{r}}^{(2)} (t,T)
    - \frac{V_{\mathrm{op}}^{(2)} (t,T)}{V^{(2)} (t,T)} \mathcal{D}_{\mathrm{op}}^{(2)} (t,T),
    \label{eq:duration_dcmp}
    \\
    \text{where} \quad
    \mathcal{D}_0 (t,T)
    :=&
    - \frac{1}{V_0 (t,T)} \frac{\partial V_0}{\partial R} (t,T),
    \nonumber
    \\
    \mathcal{D}_{\mathrm{r}}^{(2)} (t,T)
    :=&
    - \frac{1}{V_{\mathrm{r}}^{(2)} (t,T)}
    \frac{\partial V_{\mathrm{r}}^{(2)}}{\partial R} (t,T),
    \nonumber
    \\
    \mathcal{D}_{\mathrm{op}}^{(2)} (t,T)
    :=&
    - \frac{1}{V_{\mathrm{op}}^{(2)} (t,T)}
    \frac{\partial V_{\mathrm{op}}^{(2)}}{\partial R} (t,T).
    \nonumber
\end{align}
Each term on the right-hand side of Equation~\eqref{eq:duration_dcmp} is expressed as the product of the duration of an individual price component and its relative weight in the total MBS price.
Therefore, changes in each term reflect not only changes in the duration of the corresponding price component itself but also changes in its price weight relative to the total MBS price.
In the following, we interpret these terms as the contributions of the individual price components to MBS duration and examine how they vary with the interest rate level.

Similarly, we define $\mathcal{C}^{(2)}(t,T)$ as an approximation to MBS convexity as follows:
\begin{align}
    \mathcal{C}^{(2)} (t,T)
    :=&
    \frac{1}{V^{(2)} (t,T)}
    \frac{\partial^2 V^{(2)}}{\partial R^2} (t,T)
    \nonumber
    \\
    =&
    \frac{V_0 (t,T)}{V^{(2)} (t,T)} \mathcal{C}_0 (t,T)
    + \frac{V_{\mathrm{r}}^{(2)} (t,T)}{V^{(2)} (t,T)} \mathcal{C}_{\mathrm{r}}^{(2)} (t,T)
    - \frac{V_{\mathrm{op}}^{(2)} (t,T)}{V^{(2)} (t,T)} \mathcal{C}_{\mathrm{op}}^{(2)} (t,T),
    \label{eq:convexity_dcmp}
    \\
    \text{where} \quad
    \mathcal{C}_0 (t,T)
    :=&
    \frac{1}{V_0 (t,T)}
    \frac{\partial^2 V_0}{\partial R^2} (t,T),
    \nonumber
    \\
    \mathcal{C}_{\mathrm{r}}^{(2)} (t,T)
    :=&
    \frac{1}{V_{\mathrm{r}}^{(2)} (t,T)}
    \frac{\partial^2 V_{\mathrm{r}}^{(2)}}{\partial R^2} (t,T),
    \nonumber
    \\
    \mathcal{C}_{\mathrm{op}}^{(2)} (t,T)
    :=&
    \frac{1}{V_{\mathrm{op}}^{(2)} (t,T)}
    \frac{\partial^2 V_{\mathrm{op}}^{(2)}}{\partial R^2} (t,T).
    \nonumber
\end{align}
Similarly, each term on the right-hand side of Equation~\eqref{eq:convexity_dcmp} is expressed as the product of the convexity of an individual price component and its price weight, reflecting changes in both.
In the following, we interpret these terms as the contributions of the individual price components to the overall MBS convexity and examine how they vary with the interest rate level.
Note that the duration and convexity of each individual price component are defined only when the corresponding price is nonzero.
Even when the price of a component is zero, its contribution can be evaluated directly using the following expressions, provided that $V^{(2)}(t,T)>0$:
\begin{align}
    \mathcal{D}^{(2)}
    =&-\frac{\partial_R V_0}{V^{(2)}}
       -\frac{\partial_R V_{\mathrm{r}}^{(2)}}{V^{(2)}}
       +\frac{\partial_R V_{\mathrm{op}}^{(2)}}{V^{(2)}},
       \label{eq:duration_dcmp_direct}
    \\
    \mathcal{C}^{(2)}
    =&\frac{\partial_R^2 V_0}{V^{(2)}}
       +\frac{\partial_R^2 V_{\mathrm{r}}^{(2)}}{V^{(2)}}
       -\frac{\partial_R^2 V_{\mathrm{op}}^{(2)}}{V^{(2)}}.
       \nonumber
\end{align}
Here, the arguments $(t,T)$ have been omitted for notational simplicity.
In particular, a necessary and sufficient condition for negative convexity is
\begin{align}
    \partial_R^2 V_{\mathrm{op}}^{(2)}(t,T)
    >\partial_R^2 V_0(t,T)+\partial_R^2 V_{\mathrm{r}}^{(2)}(t,T)
    \label{eq:negative_convexity_condition}
\end{align}
This condition compares the second-order sensitivities of the price components rather than their price levels or first-order sensitivities.
Since the second component also incorporates the reduction in outstanding principal through $Z_{\mathrm{r}}^{(2)}$, the signs of its contributions to duration and convexity are not necessarily fixed.
In the following numerical analysis, we use the Pathwise method to evaluate duration and convexity by Monte Carlo simulation.
Details of this method are provided in Appendix~\ref{app:pathwise_greeks}.
For the expressions used to evaluate duration and convexity analytically in an approximate form,
see Appendix~\ref{app:mbs_duration_convexity_closed_form}.

Table~\ref{tab:comparison_V_y_br} compares MBS prices obtained from the closed-form approximation and Monte Carlo simulation.
\begin{table}[!htbp]
    \centering
    \caption{
        Comparison of MBS prices obtained from the closed-form approximations using up to the first and second cumulants and from Monte Carlo simulation.
        $V^{(1)}$ and $V^{(2)}$ denote MBS prices obtained from the closed-form approximations using up to the first and second cumulants, respectively, while $V_{\mathrm{MC}}$ denotes the MBS price evaluated by Monte Carlo simulation.
        Relative deviations $1$ and $2$ are defined as $100(V^{(1)}-V_{\mathrm{MC}})/V_{\mathrm{MC}}$ and $100(V^{(2)}-V_{\mathrm{MC}})/V_{\mathrm{MC}}$, respectively.
        We set $R \in \left\{0.000, 0.005, \cdots, 0.070\right\}$, and use the other parameters given in Table~\ref{tab:base_parameters}.
    }
    \label{tab:comparison_V_y_br}
    \small
    \setlength{\tabcolsep}{7pt}
    \renewcommand{\arraystretch}{1.15}
    \begin{tabular}{
        c
        ccc
        cc
    }
        \toprule
        $R-m$
        & $V^{(1)}$
        & $V^{(2)}$
        & $V_{\mathrm{MC}}$
        & Relative deviation $1$ $(\%)$
        & Relative deviation $2$ $(\%)$
        \\
        \midrule

        \multicolumn{6}{c}{$\bar{y}=1$}
        \\
        \midrule

        $-0.035$ & 1.16927 & 1.16942 & 1.17063 & $-0.11609$ & $-0.10323$ \\
        $-0.030$ & 1.14359 & 1.14373 & 1.14475 & $-0.10118$ & $-0.08978$ \\
        $-0.025$ & 1.11827 & 1.11838 & 1.11924 & $-0.08639$ & $-0.07645$ \\
        $-0.020$ & 1.09330 & 1.09339 & 1.09409 & $-0.07196$ & $-0.06348$ \\
        $-0.015$ & 1.06870 & 1.06877 & 1.06932 & $-0.05826$ & $-0.05123$ \\
        $-0.010$ & 1.04448 & 1.04454 & 1.04495 & $-0.04563$ & $-0.04000$ \\
        $-0.005$ & 1.02067 & 1.02072 & 1.02102 & $-0.03448$ & $-0.03018$ \\
        $\phantom{-}0.000$ & 0.99732 & 0.99735 & 0.99757 & $-0.02478$ & $-0.02168$ \\
        $\phantom{-}0.005$ & 0.97448 & 0.97450 & 0.97465 & $-0.01702$ & $-0.01495$ \\
        $\phantom{-}0.010$ & 0.95219 & 0.95221 & 0.95230 & $-0.01132$ & $-0.01004$ \\
        $\phantom{-}0.015$ & 0.93050 & 0.93051 & 0.93057 & $-0.00729$ & $-0.00654$ \\
        $\phantom{-}0.020$ & 0.90941 & 0.90942 & 0.90946 & $-0.00451$ & $-0.00411$ \\
        $\phantom{-}0.025$ & 0.88895 & 0.88895 & 0.88897 & $-0.00258$ & $-0.00237$ \\
        $\phantom{-}0.030$ & 0.86910 & 0.86910 & 0.86911 & $-0.00135$ & $-0.00124$ \\
        $\phantom{-}0.035$ & 0.84985 & 0.84985 & 0.84985 & $-0.00070$ & $-0.00065$ \\

        \midrule
        \multicolumn{6}{c}{$\bar{y}=5$}
        \\
        \midrule

        $-0.035$ & 1.12638 & 1.12793 & 1.13346 & $-0.62403$ & $-0.48781$ \\
        $-0.030$ & 1.10958 & 1.11105 & 1.11578 & $-0.55580$ & $-0.42461$ \\
        $-0.025$ & 1.09202 & 1.09339 & 1.09735 & $-0.48588$ & $-0.36135$ \\
        $-0.020$ & 1.07367 & 1.07492 & 1.07815 & $-0.41514$ & $-0.29913$ \\
        $-0.015$ & 1.05454 & 1.05565 & 1.05819 & $-0.34511$ & $-0.23980$ \\
        $-0.010$ & 1.03466 & 1.03562 & 1.03754 & $-0.27748$ & $-0.18540$ \\
        $-0.005$ & 1.01415 & 1.01493 & 1.01633 & $-0.21462$ & $-0.13829$ \\
        $\phantom{-}0.000$ & 0.99317 & 0.99375 & 0.99473 & $-0.15729$ & $-0.09832$ \\
        $\phantom{-}0.005$ & 0.97194 & 0.97234 & 0.97300 & $-0.10923$ & $-0.06748$ \\
        $\phantom{-}0.010$ & 0.95070 & 0.95095 & 0.95139 & $-0.07251$ & $-0.04539$ \\
        $\phantom{-}0.015$ & 0.92965 & 0.92980 & 0.93008 & $-0.04610$ & $-0.02985$ \\
        $\phantom{-}0.020$ & 0.90895 & 0.90903 & 0.90921 & $-0.02801$ & $-0.01895$ \\
        $\phantom{-}0.025$ & 0.88870 & 0.88875 & 0.88884 & $-0.01578$ & $-0.01105$ \\
        $\phantom{-}0.030$ & 0.86897 & 0.86899 & 0.86904 & $-0.00818$ & $-0.00584$ \\
        $\phantom{-}0.035$ & 0.84978 & 0.84979 & 0.84982 & $-0.00419$ & $-0.00309$ \\
        \bottomrule
    \end{tabular}
\end{table}
Table~\ref{tab:comparison_V_y_br} shows that, for both values of $\bar{y}$, the relative deviations of the approximations using up to the first and second cumulants decrease as $R-m$ increases.
This is likely because a higher $R-m$ reduces the contribution of the interest-rate-dependent component of the prepayment rate, thereby limiting the effects of the approximation.
Furthermore, for both values of $\bar{y}$ and across all values of $R-m$, the absolute relative deviations of the approximation using up to the second cumulant are smaller than those of the approximation using only the first cumulant.
These results indicate that incorporating the second cumulant improves the approximation accuracy, consistent with the expected improvement from including higher-order terms.

\subsection{Decomposition of MBS Prices}

In this subsection, we numerically examine the contributions of the individual price components to the overall MBS price, based on the price decomposition given in Equation~\eqref{eq:lmd_c_nonzero_V_decomp} and its second-order cumulant approximation in Equation~\eqref{eq:mbs_closed_form_02}.
In particular, we investigate how the component corresponding to interest-rate-independent cash flows, the component representing the present value of principal repaid through interest-rate-dependent prepayments, and the option component reflecting the reduction in future cash flows due to prepayments vary with the interest rate level and the magnitude of interest-rate-dependent prepayments.
This allows us to distinguish the different effects of increased prepayments resulting from declining interest rates on MBS prices.

\begin{figure}[bhtp]
    \begin{center}
        \includegraphics[width=15cm]{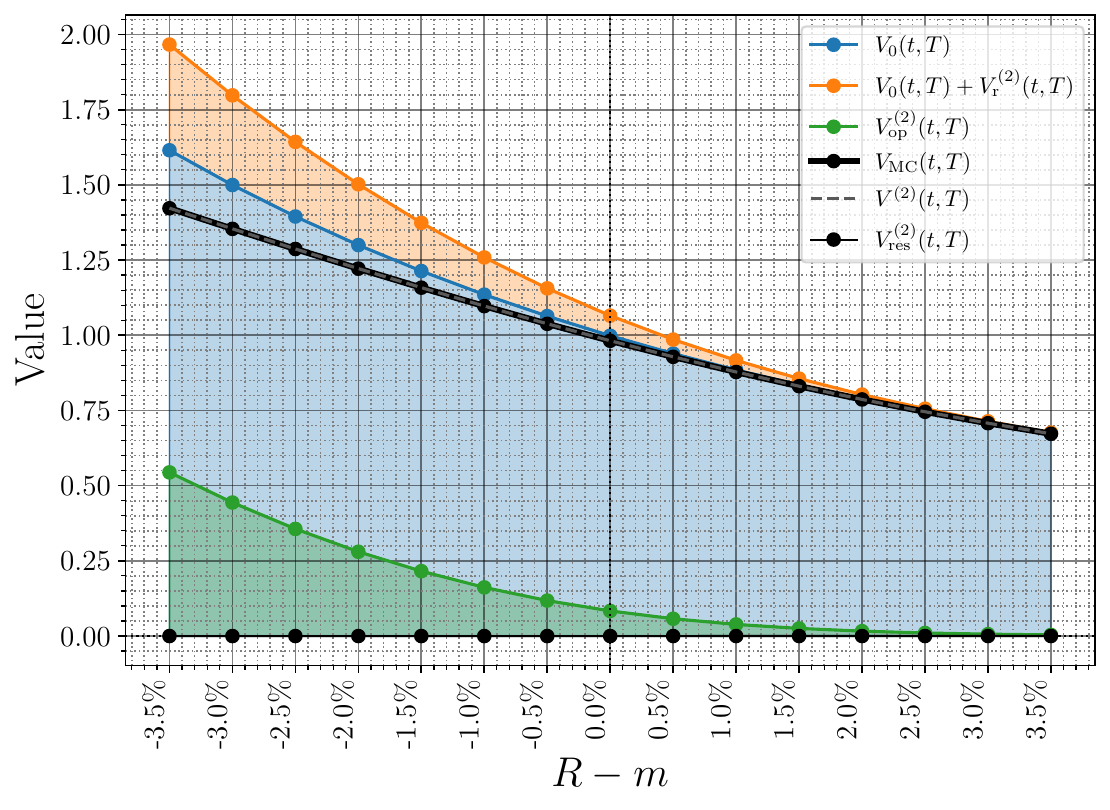}
        \caption{
            Decomposition of MBS prices based on Equation~\eqref{eq:lmd_c_nonzero_V_decomp} for $\bar{y}=1.0$.
            We set $R \in \left\{0.000, 0.005, \cdots, 0.070\right\}$.
            The other parameters are based on those in Table~\ref{tab:base_parameters}, except that we set $Z_{\mathrm{c}}(t,v)\equiv 1$ to focus on the interest rate dependence of MBS prices.
            The thick black solid line and black markers represent $V(t,T)$ evaluated using Equations~\eqref{eq:Z_r_mc_01}--\eqref{eq:Z_r_mc_05}, with the expectations on the right-hand sides of Equations~\eqref{eq:Z_closed_form_03} and \eqref{eq:Z_r_mc_05} evaluated by Monte Carlo simulation.
            The blue, orange, and green shaded areas represent
            $V_0(t,T)$, $V_{\mathrm{r}}^{(2)}(t,T)$,
            and $V_{\mathrm{op}}^{(2)}(t,T)$, respectively,
            evaluated using the approximate closed-form expressions given in
            Equations~\eqref{eq:mbs_closed_form_02}--\eqref{eq:mbs_closed_form_05}.
            The orange markers represent $V_0(t,T)+V_{\mathrm{r}}^{(2)}(t,T)$.
            The gray dashed line represents the second-order cumulant approximation of the MBS price constructed from these components,
            $V^{(2)}(t,T)
            :=V_0(t,T)+V_{\mathrm{r}}^{(2)}(t,T)-V_{\mathrm{op}}^{(2)}(t,T)$.
            The thin black solid line and black markers represent the residual calculated as
            $V_{\mathrm{res}}^{(2)}(t,T):=V_{\mathrm{MC}} (t,T)-V^{(2)}(t,T)$.
            This residual includes sampling errors and time discretization errors associated with Monte Carlo evaluation, in addition to approximation errors.
            }
        \label{fig:moneyness_vs_V_decomposition_fixed_m_0.035_y_br_1.0_T_30}
    \end{center}
\end{figure}

\begin{figure}[bhtp]
    \begin{center}
        \includegraphics[width=15cm]{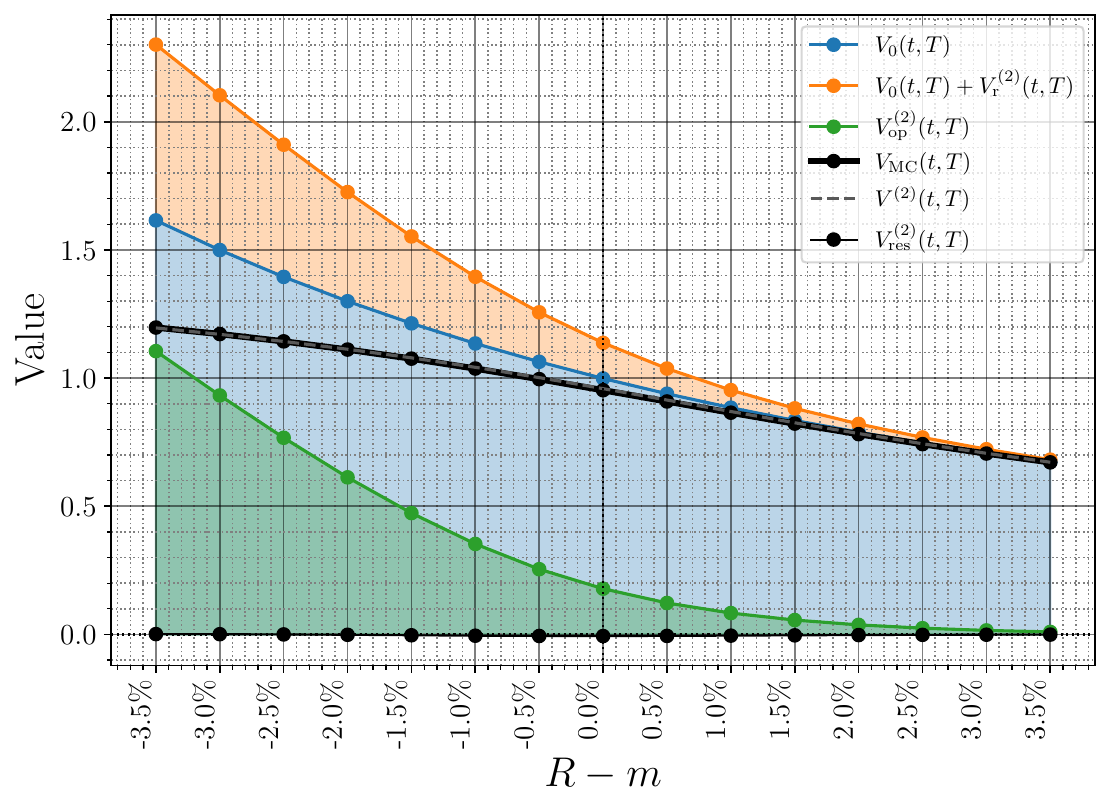}
        \caption{
            Decomposition of MBS prices based on Equation~\eqref{eq:lmd_c_nonzero_V_decomp} for $\bar{y}=5.0$.
            All other conditions are the same as those in Figure~\ref{fig:moneyness_vs_V_decomposition_fixed_m_0.035_y_br_1.0_T_30}.
            }
        \label{fig:moneyness_vs_V_decomposition_fixed_m_0.035_y_br_5.0_T_30}
    \end{center}
\end{figure}

Figures~\ref{fig:moneyness_vs_V_decomposition_fixed_m_0.035_y_br_1.0_T_30} and \ref{fig:moneyness_vs_V_decomposition_fixed_m_0.035_y_br_5.0_T_30} show the decomposition of MBS prices based on Equation~\eqref{eq:lmd_c_nonzero_V_decomp} for $\bar{y}=1.0$ and $\bar{y}=5.0$, respectively.
Here, we define the residual price component $V_{\mathrm{res}} (t,T)$, which is not captured by the approximation, as $V_{\mathrm{res}} (t,T) := V_{\mathrm{MC}} (t,T) - V^{(2)} (t,T)$.

Figures~\ref{fig:moneyness_vs_V_decomposition_fixed_m_0.035_y_br_1.0_T_30} and
\ref{fig:moneyness_vs_V_decomposition_fixed_m_0.035_y_br_5.0_T_30}
show that, for both values of $\bar{y}$,
the sum of the first and second components,
$V_0(t,T)+V_{\mathrm{r}}^{(2)}(t,T)$,
and the third component, $V_{\mathrm{op}}^{(2)}(t,T)$,
both increase as the interest rate level $R$ decreases.
Since the MBS price is represented as the net amount obtained by subtracting the latter from the former,
the increase in the sum of the first and second components resulting from declining interest rates
is partially offset by the increase in the third component,
which reflects the reduction in future cash flows due to prepayments.
Furthermore, as the interest rate level increases,
the contributions of both interest-rate-dependent prepayment components diminish,
and the MBS price approaches the first component, $V_0(t,T)$.

Next, focusing on the differences in $\bar{y}$,
we observe that, compared with $\bar{y}=1.0$,
the positive contribution of the second component
and the negative contribution of the third component
are both larger in the low-interest-rate region when $\bar{y}=5.0$.
In particular, at the same low interest rate level,
the increase in the third component associated with a higher $\bar{y}$
exceeds the increase in the sum of the first and second components.
Consequently, the MBS price is lower for $\bar{y}=5.0$
than for $\bar{y}=1.0$.
This indicates that an increase in interest-rate-dependent prepayments
not only increases the second component, which corresponds to accelerated principal repayments,
but also increases the third component, which reflects the reduction in future cash flows.

Furthermore, when $\bar{y}=5.0$,
the increase in the MBS price resulting from declining interest rates
gradually diminishes in the low-interest-rate region,
and the black curve representing the MBS price exhibits a shape
consistent with a negative second derivative with respect to the interest rate level $R$.
This suggests that the interest rate sensitivity of the option value corresponding to the third component
has a substantial influence on the curvature of the MBS price,
giving rise to negative convexity.
However, the occurrence of negative convexity cannot be determined
solely by comparing the levels or first-order sensitivities of the individual price components.
Therefore, a quantitative analysis of the sign of convexity
and the second-order sensitivities of the individual components
is presented in Section~\ref{subsec:convexity_decomposition}.

\subsection{Decomposition of MBS Duration}

In this subsection, we decompose MBS duration into the contributions of the individual price components, based on the MBS price decomposition presented in the preceding subsection.
In particular, we examine how the component corresponding to interest-rate-independent cash flows, the component representing the present value of principal repaid through interest-rate-dependent prepayments, and the option component reflecting the reduction in future cash flows due to prepayments each contribute to the overall MBS duration.
This allows us to investigate which price components and their interest rate sensitivities are responsible for the characteristic shortening of MBS duration as interest rates decline.

\begin{figure}[bhtp]
    \begin{center}
        \includegraphics[width=15cm]{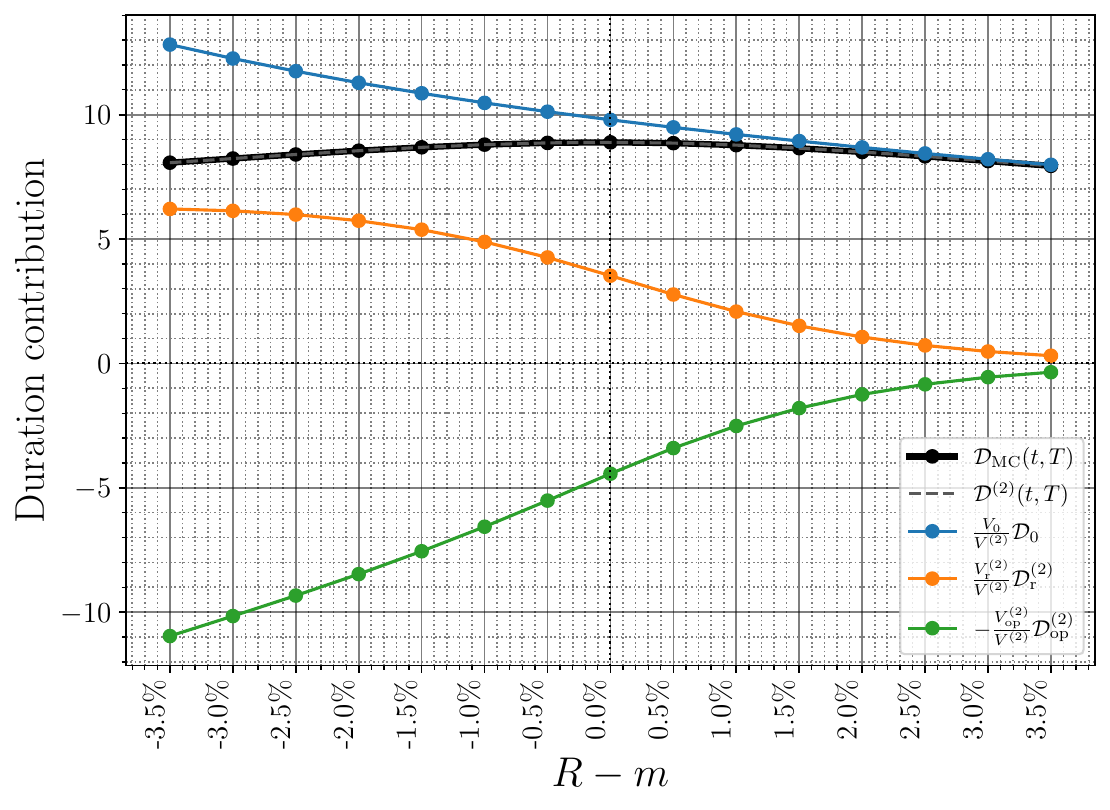}
        \caption{
            Decomposition of MBS duration into individual contributions based on Equation~\eqref{eq:duration_dcmp} and comparison with Monte Carlo results for $\bar{y}=1.0$.
            The black solid line (black markers) represents the duration $\mathcal{D}_{\mathrm{MC}}(t,T)$ evaluated using the Pathwise Monte Carlo method with exact conditional Gaussian sampling (see Appendix~\ref{app:pathwise_duration}), while the gray dashed line represents the duration $\mathcal{D}^{(2)}(t,T)$ obtained from the second-order cumulant approximation.
            The blue, orange, and green markers represent the contributions of the respective terms on the far right-hand side of Equation~\eqref{eq:duration_dcmp}, evaluated using analytical expressions based on the second-order cumulant approximation.
            We set $R \in \left\{ 0.000, 0.005, \cdots, 0.070 \right\}$, and use the other parameters given in Table~\ref{tab:base_parameters}.
            }
        \label{fig:moneyness_vs_duration_contribution_cumulant_2_fixed_m_0.035_y_br_1.0_T_30}
    \end{center}
\end{figure}

\begin{figure}[bhtp]
    \begin{center}
        \includegraphics[width=15cm]{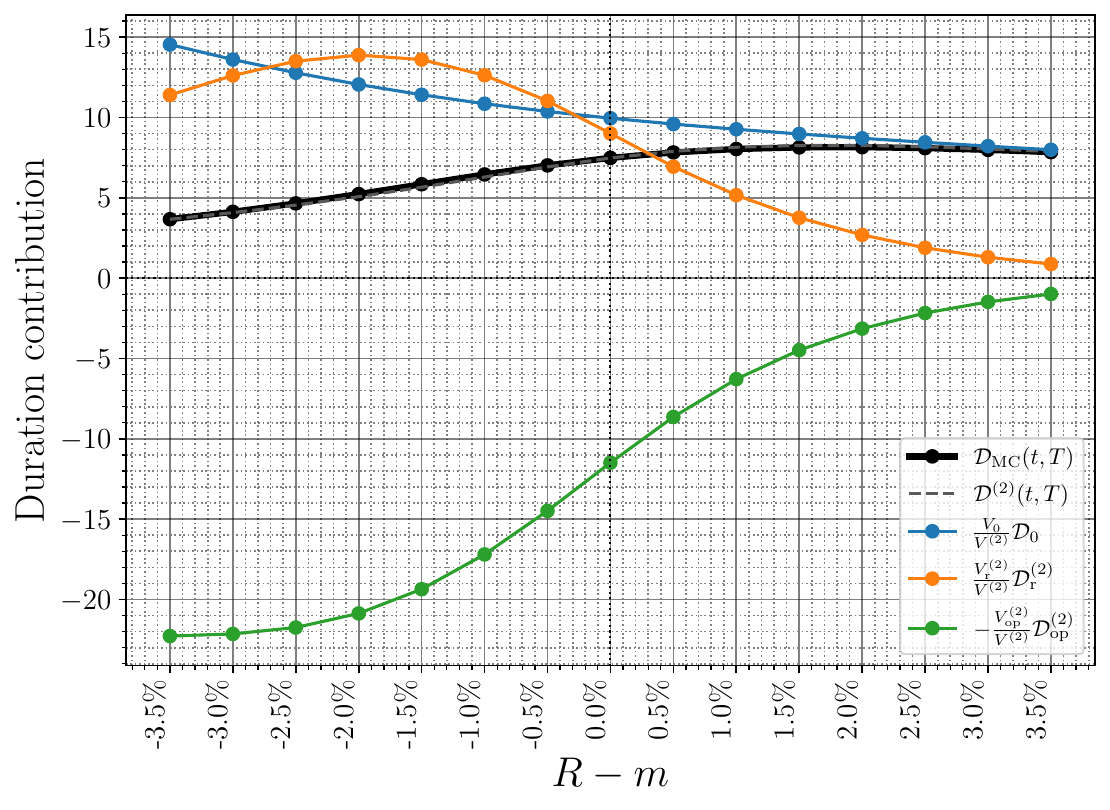}
        \caption{
            Decomposition of MBS duration into individual contributions based on Equation~\eqref{eq:duration_dcmp} and comparison with Monte Carlo results for $\bar{y}=5.0$.
            All other conditions, including the line styles and color conventions, are the same as those in Figure~\ref{fig:moneyness_vs_duration_contribution_cumulant_2_fixed_m_0.035_y_br_1.0_T_30}.
            }
        \label{fig:moneyness_vs_duration_contribution_cumulant_2_fixed_m_0.035_y_br_5.0_T_30}
    \end{center}
\end{figure}

Figures~\ref{fig:moneyness_vs_duration_contribution_cumulant_2_fixed_m_0.035_y_br_1.0_T_30} and \ref{fig:moneyness_vs_duration_contribution_cumulant_2_fixed_m_0.035_y_br_5.0_T_30} show the decomposition of the approximate MBS duration $\mathcal{D}^{(2)}(t,T)$, obtained using the cumulant expansion up to the second order, into individual contributions based on Equation~\eqref{eq:duration_dcmp}, for $\bar{y}=1.0$ and $\bar{y}=5.0$, respectively.

Figures~\ref{fig:moneyness_vs_duration_contribution_cumulant_2_fixed_m_0.035_y_br_1.0_T_30} and
\ref{fig:moneyness_vs_duration_contribution_cumulant_2_fixed_m_0.035_y_br_5.0_T_30}
show that, for both values of $\bar{y}$,
the first and second components make positive contributions to MBS duration,
whereas the third component makes a negative contribution.
In particular, as the interest rate level $R$ decreases,
both the positive contribution of the first component
and the absolute magnitude of the negative contribution of the third component increase.
Thus, the change in overall MBS duration resulting from declining interest rates
is determined by the sum of these contributions,
with the negative contribution of the third component
acting to offset the increase in duration.

Next, focusing on the differences in $\bar{y}$,
we observe a nonmonotonic relationship when $\bar{y}=1.0$,
with MBS duration initially increasing as interest rates decline
and then decreasing in the low-interest-rate region.
In contrast, when $\bar{y}=5.0$,
the shortening of MBS duration as interest rates decline is more pronounced.
In particular, in the low-interest-rate region,
although the positive contribution of the second component is larger than that for $\bar{y}=1.0$,
the absolute magnitude of the negative contribution of the third component increases by an even greater amount.
Consequently, the overall MBS duration in the low-interest-rate region
is substantially shorter for $\bar{y}=5.0$ than for $\bar{y}=1.0$.
This indicates that, when refinancing sensitivity is high,
the interest rate sensitivity of the option component,
which reflects the reduction in future cash flows due to prepayments,
contributes substantially to the shortening of overall MBS duration.

Furthermore, for both values of $\bar{y}$,
the duration $\mathcal{D}^{(2)}(t,T)$ obtained from the second-order cumulant approximation
closely agrees with the duration $\mathcal{D}_{\mathrm{MC}}(t,T)$
evaluated using the Pathwise Monte Carlo method.
These results indicate that, under the present parameter settings,
the second-order cumulant approximation accurately captures
the first-order interest rate sensitivity of MBS prices.

The first term in Equation~\eqref{eq:duration_dcmp} corresponds to the interest rate sensitivity of $V_0(t,T)$, whose underlying cash flows are independent of interest rates.
The second term corresponds to the sensitivity of principal payments arising from interest-rate-dependent prepayments, while the third term corresponds to the sensitivity of the option component reflecting the reduction in future cash flows due to prepayments.
From Equation~\eqref{eq:duration_dcmp_direct}, the second term makes a positive contribution to duration when $V_{\mathrm{r}}^{(2)}(t,T)$ decreases as interest rates rise, whereas the third term makes a negative contribution when $V_{\mathrm{op}}^{(2)}(t,T)$ decreases as interest rates rise.
However, since the second term incorporates both changes in the prepayment rate and reductions in outstanding principal resulting from earlier prepayments, the sign of its duration contribution is not determined in general.
Furthermore, the duration contribution of each component is the product of its corresponding price weight and its own duration, with a negative sign applied to the third component.
Therefore, changes in these contributions cannot be inferred solely from changes in their price weights.
This decomposition allows us to interpret the shortening of overall MBS duration as interest rates decline in terms of the interest rate sensitivities associated with both accelerated principal repayments and the reduction in future cash flows.
This result is also consistent with the qualitative explanation that prepayments accelerate principal repayments and shorten the average time to receive cash flows~\parencite{Hanson:2014}.

\subsection{Decomposition of MBS Convexity}
\label{subsec:convexity_decomposition}

In this subsection, we decompose MBS convexity into the contributions of the individual price components, based on the MBS price decomposition.
In particular, we examine how the component corresponding to interest-rate-independent cash flows, the component representing the present value of principal repaid through interest-rate-dependent prepayments, and the option component reflecting the reduction in future cash flows due to prepayments each contribute to the overall MBS convexity.
This allows us to investigate which price components and their second-order interest rate sensitivities are responsible for the negative convexity characteristic of MBSs.

\begin{figure}[bhtp]
    \begin{center}
        \includegraphics[width=15cm]{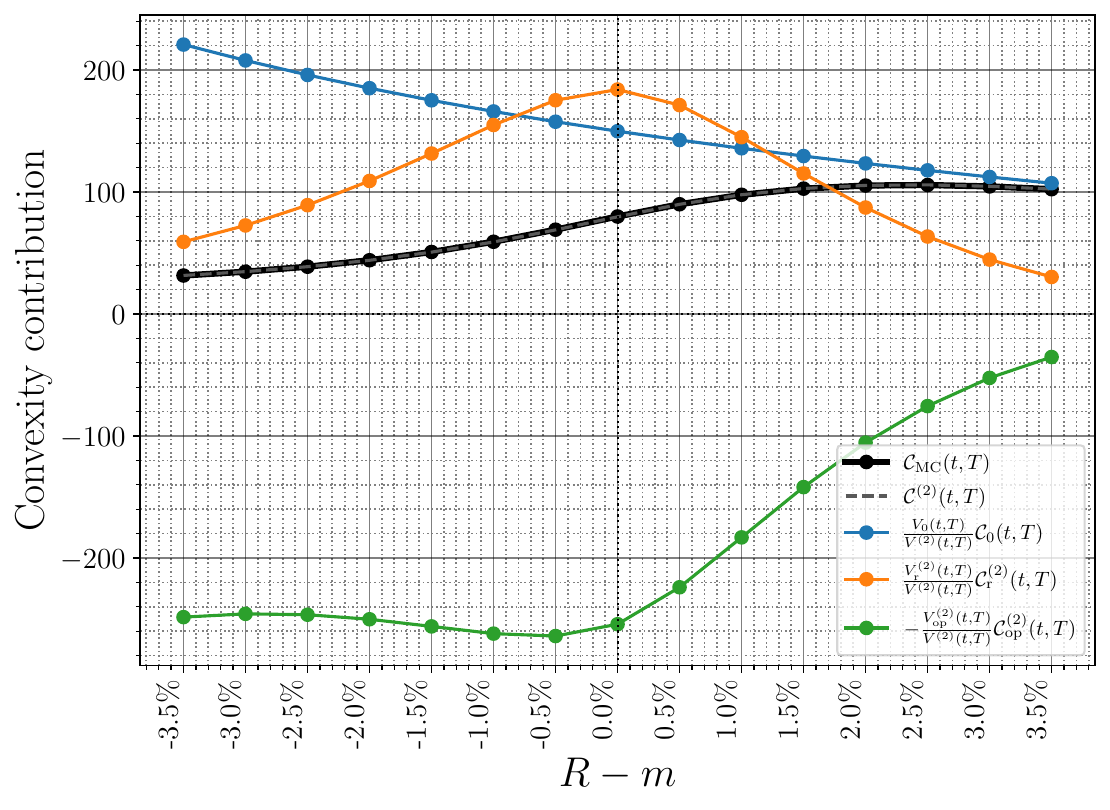}
        \caption{
            Decomposition of MBS convexity into individual contributions based on Equation~\eqref{eq:convexity_dcmp} and comparison with Monte Carlo results for $\bar{y}=1.0$.
            The black solid line (black markers) represents the convexity $\mathcal{C}_{\mathrm{MC}}(t,T)$ evaluated using the Pathwise Monte Carlo method with exact conditional Gaussian sampling (see Appendix~\ref{app:pathwise_convexity}), while the gray dashed line represents the convexity $\mathcal{C}^{(2)}(t,T)$ obtained from the second-order cumulant approximation.
            The blue, orange, and green markers represent the contributions of the respective terms on the far right-hand side of Equation~\eqref{eq:convexity_dcmp}, evaluated using analytical expressions based on the second-order cumulant approximation.
            We set $R \in \left\{ 0.000, 0.005, \cdots, 0.070 \right\}$, and use the other parameters given in Table~\ref{tab:base_parameters}.
            }
        \label{fig:moneyness_vs_convexity_contribution_cumulant_2_fixed_m_0.035_y_br_1.0_T_30}
    \end{center}
\end{figure}

\begin{figure}[bhtp]
    \begin{center}
        \includegraphics[width=15cm]{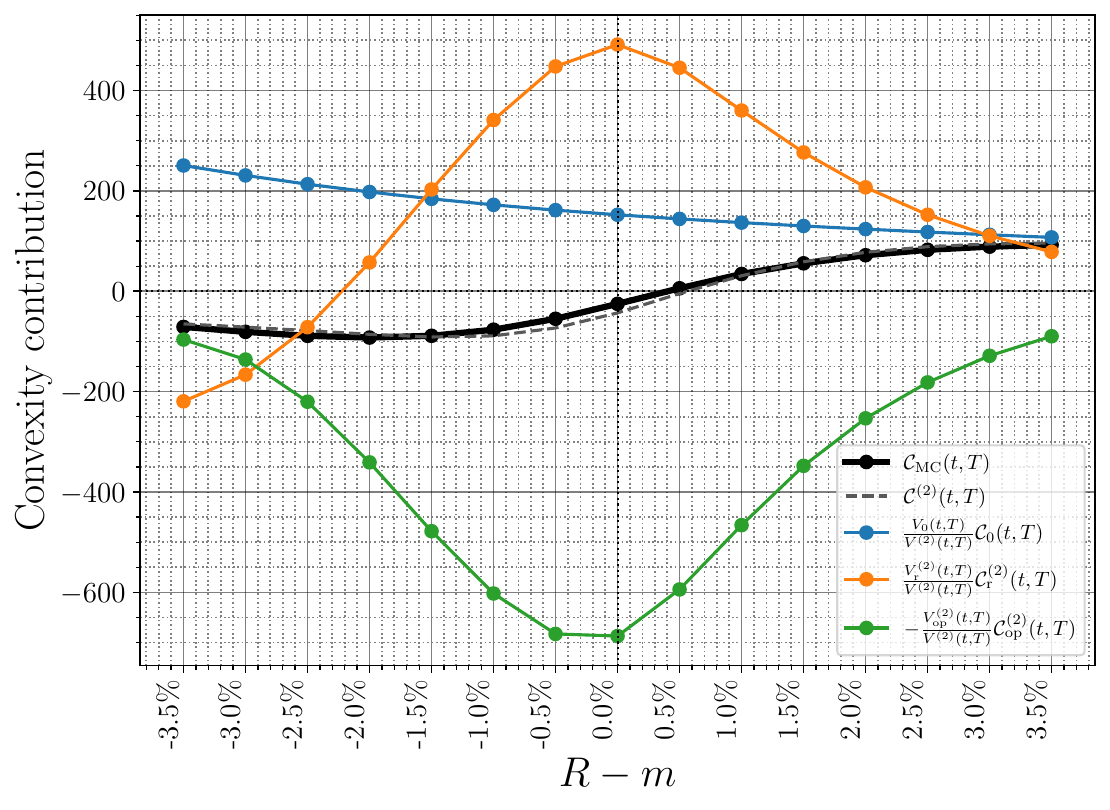}
        \caption{
            Decomposition of MBS convexity into individual contributions based on Equation~\eqref{eq:convexity_dcmp} and comparison with Monte Carlo results for $\bar{y}=5.0$.
            All other conditions, including the line styles and color conventions, are the same as those in Figure~\ref{fig:moneyness_vs_convexity_contribution_cumulant_2_fixed_m_0.035_y_br_1.0_T_30}.
            }
        \label{fig:moneyness_vs_convexity_contribution_cumulant_2_fixed_m_0.035_y_br_5.0_T_30}
    \end{center}
\end{figure}

Figures~\ref{fig:moneyness_vs_convexity_contribution_cumulant_2_fixed_m_0.035_y_br_1.0_T_30} and \ref{fig:moneyness_vs_convexity_contribution_cumulant_2_fixed_m_0.035_y_br_5.0_T_30} show the decomposition of the approximate MBS convexity $\mathcal{C}^{(2)}(t,T)$, obtained using the cumulant expansion up to the second order, into individual contributions based on Equation~\eqref{eq:convexity_dcmp}, for $\bar{y}=1.0$ and $\bar{y}=5.0$, respectively.

Figures~\ref{fig:moneyness_vs_convexity_contribution_cumulant_2_fixed_m_0.035_y_br_1.0_T_30} and
\ref{fig:moneyness_vs_convexity_contribution_cumulant_2_fixed_m_0.035_y_br_5.0_T_30}
show that, for both values of $\bar{y}$,
the first component makes a positive contribution to MBS convexity,
whereas the third component makes a negative contribution.
In contrast, the contribution of the second component depends on the interest rate level and $\bar{y}$.
While it is positive across all interest rate levels when $\bar{y}=1.0$,
it becomes negative in part of the low-interest-rate region when $\bar{y}=5.0$.
Furthermore, as the interest rate level increases,
the absolute magnitudes of the contributions from the second and third components decrease,
and the overall MBS convexity approaches the contribution of the first component.

Next, focusing on the differences in $\bar{y}$,
we observe that, when $\bar{y}=1.0$,
the overall MBS convexity remains positive across all interest rate levels,
because the sum of the positive contributions from the first and second components
exceeds the absolute magnitude of the negative contribution from the third component.
In contrast, when $\bar{y}=5.0$,
the overall MBS convexity becomes negative in the low-interest-rate region.
In particular, for $R-m$ ranging approximately from $-1.0\%$ to $0.0\%$,
the absolute magnitude of the negative contribution from the third component
exceeds the sum of the positive contributions from the first and second components.
This corresponds to the condition expressed in Equation~\eqref{eq:negative_convexity_condition},
under which the second-order interest rate sensitivity of the option component
exceeds the sum of the second-order sensitivities of the other components.
Therefore, in this region,
the second-order interest rate sensitivity of the option component,
which reflects the reduction in future cash flows due to prepayments,
plays a crucial role in generating negative convexity.

Furthermore, when $\bar{y}=5.0$,
the convexity contribution of the second component itself becomes negative
in the extremely low-interest-rate region.
In contrast, in this region,
the absolute magnitude of the negative contribution from the third component
is smaller than at higher interest rate levels.
Therefore, the occurrence of negative convexity cannot be explained
solely by an increase in the negative contribution from the third component.
Since the second component incorporates both changes in the prepayment rate
and reductions in outstanding principal resulting from earlier prepayments,
the sign of its second-order interest rate sensitivity is not determined in general.
Moreover, the negative contribution from the third component
varies nonmonotonically with the interest rate level,
and its absolute magnitude does not necessarily increase as interest rates decline.
These results indicate that the mechanism underlying negative convexity
depends on the interest rate level and refinancing sensitivity.

Furthermore, when $\bar{y}=1.0$,
the convexity $\mathcal{C}^{(2)}(t,T)$ obtained from the second-order cumulant approximation
closely agrees with the convexity $\mathcal{C}_{\mathrm{MC}}(t,T)$
evaluated using the Pathwise Monte Carlo method.
In contrast, when $\bar{y}=5.0$,
although the two methods yield broadly consistent results
regarding the interest rate dependence and sign changes of convexity,
noticeable discrepancies are observed in the low-interest-rate region.

The first term in Equation~\eqref{eq:convexity_dcmp}
represents the second-order interest rate sensitivity of $V_0(t,T)$,
which corresponds to interest-rate-independent cash flows,
and makes a positive contribution to convexity, as in the case of a conventional bond.
The second term represents the second-order interest rate sensitivity of
$V_{\mathrm{r}}^{(2)}(t,T)$,
which corresponds to the present value of principal repaid through interest-rate-dependent prepayments.
This component incorporates both changes in the prepayment rate
and reductions in outstanding principal resulting from earlier prepayments.
On the other hand, the third term represents the negative of the
second-order interest rate sensitivity of the option component
$V_{\mathrm{op}}^{(2)}(t,T)$,
which reflects the reduction in future cash flows due to prepayments.
Since the signs of the convexity contributions from the second and third components
are not determined in general by the decomposition alone,
the occurrence of negative convexity must be assessed
by evaluating these contributions jointly
and applying the condition given in Equation~\eqref{eq:negative_convexity_condition}.

Finally, we examine the relationship between the shortening of duration and negative convexity.
From the definitions of duration and convexity, the following relationship holds:
\begin{align*}
    \frac{\partial\mathcal{D}^{(2)}(t,T)}{\partial R}
    =\left(\mathcal{D}^{(2)}(t,T)\right)^2-\mathcal{C}^{(2)}(t,T).
\end{align*}
Therefore, when $\mathcal{C}^{(2)}(t,T)<0$,
we have $\partial_R\mathcal{D}^{(2)}(t,T)>0$,
implying that duration shortens as interest rates decline.
More generally, however, duration shortens whenever
$\mathcal{C}^{(2)}(t,T)<(\mathcal{D}^{(2)}(t,T))^2$.
Thus, negative convexity is a sufficient but not a necessary condition for the shortening of duration.
This relationship connects the shortening of duration observed in the preceding subsection
with the sign of convexity examined in this subsection
and is important for understanding the characteristic interest rate sensitivities of MBSs arising from prepayments
~\parencite[e.g.,][]{Anderson_et_al:1993,Hanson:2014,Malkhozov_et_al:2016}.

Note that discrepancies between the convexity obtained from the second-order cumulant approximation
and that evaluated using the Pathwise Monte Carlo method
are observed, particularly in the low-interest-rate region when $\bar{y}=5.0$.
Therefore, in the subsequent numerical analysis of the parameter regions
where the sign of convexity changes,
we use the results obtained from the Pathwise Monte Carlo method
to avoid the effects of approximation errors.

\section{Phase Diagrams and Nonmonotonic Interest-Rate Dependence of MBS Convexity}
\label{sec:convexity_phase_diagram}

In this section, we use phase diagrams of MBS convexity with respect to the interest rate level and refinancing sensitivity
to identify the parameter regions in which negative convexity occurs
and examine the nonmonotonic dependence of convexity on the interest rate level.
In Section~\ref{subsec:convexity_phase_diagram},
we construct phase diagrams of convexity with respect to the interest rate level $R$
and the initial refinancing sensitivity $y(0)$
to numerically identify the boundaries at which the sign of convexity changes.
We also investigate how the long-run mean $\bar{y}$ of refinancing sensitivity
affects the regions in which negative convexity occurs.
In Section~\ref{subsec:convexity_nonmonotonicity},
we focus on the fact that the absolute magnitude of negative convexity
does not necessarily increase monotonically as interest rates decline
and examine the mechanism underlying this nonmonotonicity
using a simplified model incorporating prepayments.

\subsection{Convexity Phase Diagrams and Sign Boundaries}
\label{subsec:convexity_phase_diagram}

In this subsection, we use phase diagrams with respect to the interest rate level and refinancing sensitivity
to identify the boundaries at which the sign of MBS convexity changes
and examine how these boundaries depend on the initial value and long-run mean of refinancing sensitivity.

\begin{figure}[bhtp]
    \begin{center}
        \includegraphics[width=15cm]{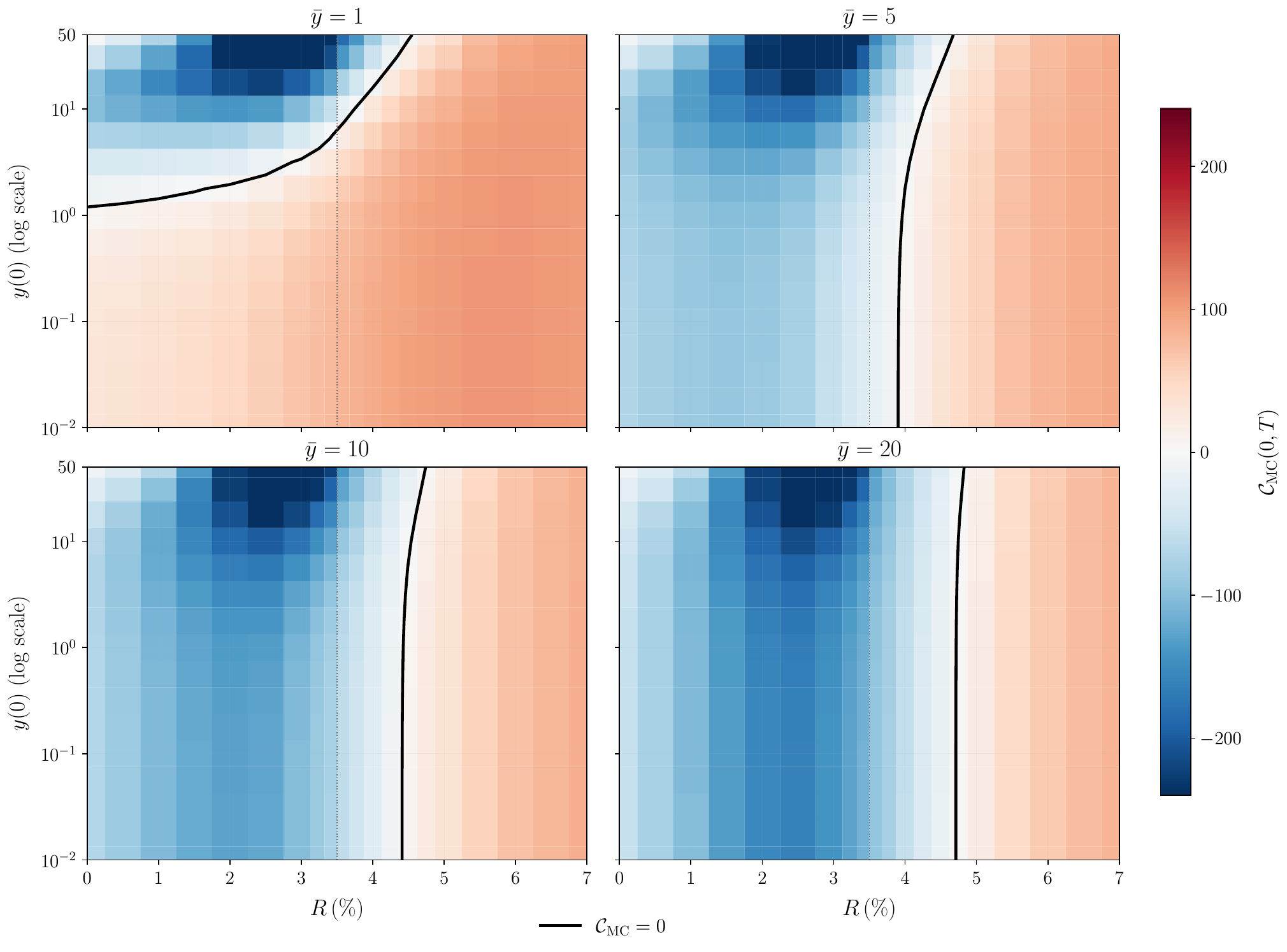}
        \caption{
            Phase diagrams of MBS convexity with respect to
            the interest rate level $R$ and the initial refinancing sensitivity $y(0)$.
            The four panels correspond to the long-run means of refinancing sensitivity
            $\bar{y}=1,5,10,20$.
            The horizontal axis represents the interest rate level $R$,
            while the vertical axis represents $y(0)$ on a logarithmic scale,
            with $R\in[0,0.07]$ and $y(0)\in[10^{-2},50]$.
            Convexity $\mathcal{C}_{\mathrm{MC}}(0,T)$ is evaluated using
            the Pathwise Monte Carlo method.
            Red indicates positive convexity and blue indicates negative convexity,
            with color intensity reflecting the absolute magnitude of convexity.
            The solid black lines indicate the boundaries at which
            $\mathcal{C}_{\mathrm{MC}}(0,T)=0$.
            We set $m=0.035$, and all other parameters are
            specified as in Table~\ref{tab:base_parameters}.
            The vertical gray dotted line indicates $R=m$.
        }
        \label{fig:convexity_phase_diagram_pathwise_mc}
    \end{center}
\end{figure}

Figure~\ref{fig:convexity_phase_diagram_pathwise_mc}
presents phase diagrams of MBS convexity with respect to
the interest rate level $R$ and the initial refinancing sensitivity $y(0)$.
By comparing the four cases with
$\bar{y}=1,5,10,20$,
we examine the parameter regions in which the sign of convexity changes
and how these regions depend on the long-run mean of refinancing sensitivity.

First, for all values of $\bar{y}$,
negative convexity tends to occur more frequently as the interest rate level $R$ declines.
This tendency can be attributed to the acceleration of prepayments
as declining interest rates increase refinancing incentives.
For straight bonds, a decline in interest rates raises prices
through a reduction in discount rates.
In contrast, for MBSs, prepayments reduce future cash flows,
thereby limiting the increase in MBS prices.
Consequently, as the interest rate level declines,
the negative contribution to convexity arising from prepayments
becomes relatively more pronounced,
causing overall convexity to turn from positive to negative.

Next, we observe that the parameter region in which negative convexity occurs
expands as the long-run mean of refinancing sensitivity $\bar{y}$ increases.
In particular, the shape of the solid black lines,
which indicate the boundaries where $\mathcal{C}_{\mathrm{MC}}(0,T)=0$,
shows that these boundaries generally shift toward higher interest rate levels
as $\bar{y}$ increases.
When $\bar{y}=1$,
positive convexity is maintained even at low interest rate levels
provided that $y(0)$ is sufficiently small.
In contrast, when $\bar{y}=5,10,20$,
negative convexity occurs even for small values of $y(0)$,
and the interest rate level at which the sign of convexity changes
increases with $\bar{y}$.

Furthermore, the shape of the zero-convexity boundary
also varies with $\bar{y}$.
When $\bar{y}=1$,
the zero-convexity boundary shifts substantially
toward higher interest rate levels as $y(0)$ increases.
In contrast, when $\bar{y}=10,20$,
the zero-convexity boundary is nearly vertical
for relatively small values of $y(0)$,
and its location is relatively insensitive to $y(0)$.
This difference can be explained by the mean-reversion level
of refinancing sensitivity $y(t)$.
In our model,
$\mathbb{E}^{\mathbb{Q}}[y(u)\mid\mathcal{F}_t]
=\bar{y}+(y(t)-\bar{y})e^{-\kappa_y(u-t)}$
holds.
Therefore, even when the initial value $y(0)$ is small,
a large long-run mean $\bar{y}$ causes future refinancing sensitivity
to revert toward a high level.
As a result, the reduction in future cash flows due to prepayments
has a substantial impact on convexity.
These results suggest that the sign of MBS convexity
depends strongly not only on the current refinancing sensitivity
but also on its long-run mean.

Furthermore, negative convexity occurs even at interest rate levels above $R=m$.
Since we set $a=0$ and $b=1$ in this study,
and the initial zero-yield curve is assumed to be flat,
the refinancing incentive at the initial time is absent when $R>m$.
However, even in this case,
future declines in interest rates may generate refinancing incentives.
Therefore, even when the refinancing option is out of the money
at the initial time,
prepayment risk arising from future interest rate fluctuations
can affect MBS convexity.
Indeed, Figure~\ref{fig:convexity_phase_diagram_pathwise_mc}
shows that negative convexity occurs even in the region $R>m$,
particularly when $\bar{y}$ is large.

\subsection{Nonmonotonic Interest-Rate Dependence of MBS Convexity}
\label{subsec:convexity_nonmonotonicity}

In the previous subsection, we confirmed that the sign of MBS convexity
varies with the interest rate level and refinancing sensitivity.
In this subsection, we focus on the fact that
the magnitude of negative convexity is not necessarily monotonic
with respect to the interest rate level.

When refinancing sensitivity is high,
convexity turns from positive to negative as interest rates decline.
However, as interest rates decline further,
the absolute magnitude of negative convexity tends to decrease.
This can be interpreted as follows:
when the refinancing option becomes deep in the money,
prepayments occur substantially earlier,
leaving less room for further declines in interest rates
to accelerate future cash flows.

This property can be explained using a simplified continuous-time model.
Consider a perpetual bond with a constant interest rate level $R$
and a constant prepayment intensity $\lambda$,
which pays a coupon at a constant rate $m$
until prepayment occurs, at which time the principal amount of $1$ is repaid.
The bond price is then given by
$V(R)=(m+\lambda)/(R+\lambda)$.
Suppose that refinancing sensitivity is a constant $y>0$
and that $\lambda=y(m-R)^+$.
For $R<m$, the bond price becomes
$V(R)=\{m+y(m-R)\}/\{R+y(m-R)\}$,
and its second derivative with respect to the interest rate level
is given by
\begin{equation}
    \frac{\partial^2 V(R)}{\partial R^2}
    =
    -\frac{2m(y-1)}
    {\{R+y(m-R)\}^3}.
    \label{eq:simplified_mbs_convexity}
\end{equation}
Equation~\eqref{eq:simplified_mbs_convexity} shows that
convexity is negative when $y>1$.
However, as $R$ declines, the denominator increases,
and hence the absolute magnitude of negative convexity decreases.
Since the bond price $V(R)$ is positive,
the same property holds for normalized convexity.
Therefore, when refinancing sensitivity is high,
the phenomenon in which negative convexity emerges as interest rates decline,
but subsequently decreases in absolute magnitude as rates decline further,
can be understood as a reduction in interest rate sensitivity
resulting from the acceleration of prepayments.
However, because this simplified model does not account for
stochastic interest rate fluctuations
or the mean-reverting behavior of refinancing sensitivity,
it does not provide an exact derivation of the numerical results
shown in Figure~\ref{fig:convexity_phase_diagram_pathwise_mc}.

These results demonstrate that negative MBS convexity
is not determined solely by the presence of a prepayment option,
but rather by the interaction among the interest rate level,
the initial value of refinancing sensitivity,
and its long-run mean.
In particular, the phase diagrams presented in this study
identify the parameter regions in which negative convexity occurs
and numerically characterize the corresponding boundaries
and the nonlinear dependence of convexity on the interest rate level.

\section{Conclusion and Discussion}
\label{sec:conclusion_discussion}

This paper investigates why negative convexity arises in pass-through MBSs
backed by fixed-rate mortgages and under what conditions it occurs.
Specifically, we examine the mechanism underlying negative convexity
by decomposing MBS prices and interest rate sensitivities into their constituent components.
We also identify the conditions under which negative convexity occurs
and characterize its dependence on the interest rate level
using phase diagrams of convexity with respect to the interest rate level
and refinancing sensitivity.

First, under a prepayment process that responds nonlinearly to interest rates,
we decomposed the MBS price into three components:
the present value of interest-rate-independent cash flows,
the present value of principal repaid early through interest-rate-dependent prepayments,
and an interest rate option component reflecting
the reduction in future cash flows due to prepayments.
Furthermore, we showed that, under a first-order approximation,
the second component can be interpreted as a portfolio of interest rate options
resembling floorlets,
while the third component can be interpreted
as a portfolio of Asian-style interest rate options.
This decomposition allows us to distinguish
the effects of interest-rate-dependent prepayments on MBS prices
between the present value of principal repaid early
and the present value of future cash flows lost as a result of those prepayments.

Next, to analytically evaluate the interest rate sensitivities
of the individual price components,
we applied a cumulant expansion to the expectations arising in MBS valuation
and derived closed-form approximations under the one-factor Hull--White model.
Our valuation method is also applicable when the sensitivity of prepayment rates
to refinancing incentives is itself stochastic.
The numerical analysis confirmed that including the second cumulant
improves the accuracy of the MBS price approximation
relative to the approximation based only on the first cumulant.
These results enable an approximate analytical evaluation
of the interest rate sensitivities of the individual price components,
even under a nonlinear prepayment process.

Based on this price decomposition,
we decomposed MBS duration and convexity
into the contributions of the individual price components.
In particular, under the second-order cumulant approximation,
we expressed the necessary and sufficient condition for negative convexity
as an inequality requiring that the second-order interest rate sensitivity
of the option component reflecting the reduction in future cash flows
due to prepayments exceeds the sum of the corresponding sensitivities
of the other price components.
We also confirmed that the sign of the convexity contribution
from the component corresponding to the present value of principal
repaid through interest-rate-dependent prepayments
is not necessarily fixed and can become negative
depending on the interest rate level and refinancing sensitivity.
Thus, negative MBS convexity cannot be explained solely
by the negative contribution of the prepayment option,
but is determined by the relative magnitudes
of the second-order interest rate sensitivities
of the individual price components.
This result provides an analytical framework for understanding
the mechanism underlying negative convexity
in terms of the constituent components of MBS prices.

Furthermore, using phase diagrams of convexity
with respect to the interest rate level $R$
and the initial refinancing sensitivity $y(0)$,
we numerically identified the parameter regions
in which negative convexity occurs and the corresponding sign boundaries.
The results confirmed that negative convexity becomes more likely
as interest rates decline
and that the region of negative convexity expands
as the long-run mean of refinancing sensitivity $\bar{y}$ increases.
In particular, when $\bar{y}$ is large,
negative convexity occurs even for small initial values of $y(0)$,
and the sign boundary becomes relatively insensitive to $y(0)$.
This can be attributed to the mean-reverting behavior
of refinancing sensitivity,
which allows future prepayments to have a substantial impact
on MBS interest rate sensitivities
even when refinancing sensitivity is initially low.
These findings suggest that the sign of MBS convexity
depends strongly not only on current refinancing sensitivity
but also on its long-run mean.

We also confirmed that negative convexity can occur
even when the initial interest rate level $R$ exceeds the mortgage rate $m$
and no refinancing incentive exists at the initial time.
This can be interpreted as reflecting the possibility
that future declines in interest rates generate refinancing incentives,
which affect MBS interest rate sensitivities even at the initial time.
Therefore, the fact that the refinancing option is out of the money
at the initial time does not necessarily imply
that MBS convexity is positive.
This result highlights the importance of considering
not only the initial refinancing incentive
but also the possibility of future prepayments
arising from subsequent interest rate movements
when examining the conditions under which negative convexity occurs.

Moreover, when refinancing sensitivity is high,
we observed that convexity turns from positive to negative
as interest rates decline,
but the absolute magnitude of negative convexity tends to decrease
as interest rates decline further.
Using a simplified model with a constant interest rate level
and a constant prepayment intensity,
we analytically demonstrated that,
when refinancing sensitivity is sufficiently high,
the absolute magnitude of negative convexity decreases
as interest rates decline.
This can be interpreted as follows:
when refinancing incentives increase and prepayments occur substantially earlier,
there is less room for further declines in interest rates
to accelerate future cash flows.
Thus, even after negative convexity emerges as interest rates decline,
its absolute magnitude does not necessarily increase monotonically
with further declines in interest rates.
However, because this simplified model does not account for
stochastic interest rate fluctuations
or the mean-reverting behavior of refinancing sensitivity,
it does not provide an exact derivation
of the numerical results shown in the phase diagrams.

Overall, this paper elucidated the mechanism underlying negative MBS convexity
in terms of the second-order interest rate sensitivities
of the individual price components
through a decomposition of MBS prices and interest rate sensitivities.
Furthermore, using phase diagrams of convexity,
we demonstrated that the regions in which negative convexity occurs
depend on the interest rate level,
the initial value of refinancing sensitivity,
and its long-run mean.
We also confirmed that the absolute magnitude of negative convexity
can exhibit nonmonotonic behavior as interest rates decline.
These results demonstrate that negative MBS convexity
is not uniformly determined by the mere presence of a prepayment option,
but arises from the interaction among multiple price effects associated
with prepayments, the interest rate level, and refinancing sensitivity.

We next discuss several limitations of this study and directions for future research.

First, the parameters used in our numerical analysis
were chosen to clarify the mechanisms and conditions
underlying negative convexity,
and the model was not calibrated to observed MBS market prices.
Since the prepayment process in this paper is formulated
under the risk-neutral probability measure,
practical valuation requires calibrating the interest rate model parameters
to the interest rate derivatives market
and estimating the prepayment process parameters
to be consistent with observed MBS market prices.
For example, the cross-section of TBA prices with different coupons
and the market prices of interest-only (IO) and principal-only (PO) securities
corresponding to the same mortgage pool
may provide useful information for identifying
the interest-rate-independent prepayment component
and the sensitivity of prepayments to refinancing incentives.
An important direction for future research is to calibrate the model
using such market data,
evaluate the accuracy of the closed-form approximations
derived in this paper,
and examine the conditions under which negative convexity occurs
under parameters consistent with market prices.

Second, we adopted the one-factor Hull--White model
to facilitate analytical tractability.
However, actual yield curve movements are characterized
by multiple factors, such as level, slope, and curvature.
Furthermore, while the prepayment intensity in our model
depends on refinancing incentives through the 10-year zero yield,
the discounting of MBS cash flows depends on the entire yield curve.
Therefore, extending our framework to the two-factor Hull--White model
or other multifactor Gaussian interest rate models
would allow us to analyze negative MBS convexity
under more realistic interest rate dynamics.
Even under multifactor Gaussian interest rate models,
if zero yields can be expressed as affine functions of the state variables,
the nonlinearity of the prepayment intensity considered in this paper
can be represented as the positive part
of a linear combination of Gaussian state variables.
Consequently, the analytical evaluation of the first and second cumulants
derived in this paper may be extended to multifactor models
using moments of truncated multivariate normal distributions.

Furthermore, extending the interest rate model to multiple factors
is also important for generalizing the decomposition
of interest rate sensitivities presented in this paper
to nonparallel yield curve movements.
In our numerical analysis,
we defined duration and convexity with respect to the parameter $R$,
which shifts the entire initial zero-yield curve in parallel,
and decomposed these sensitivities
into the contributions of the individual price components.
Under a multifactor interest rate model,
one could instead define first- and second-order price sensitivities
with respect to interest rate factors
corresponding to the level, slope, and curvature of the yield curve,
and decompose each sensitivity
into the contributions of the individual price components.
This would make it possible to quantitatively examine
how interest rate sensitivities arising from prepayments
differ between changes in the level and slope of the yield curve.
In particular, identifying how the conditions for negative convexity
depend on the direction of yield curve movements
would be an important extension of our analysis
to a more general framework of interest rate risk.

Third, the prepayment model employed in this paper
consists of an interest-rate-independent prepayment component
and a component that responds nonlinearly to refinancing incentives.
However, actual prepayment behavior is also influenced
by factors such as loan age, seasonality, burnout,
and borrower heterogeneity.
In particular, when refinancing incentives remain high
for an extended period,
borrowers who are more likely to refinance
may prepay their mortgages earlier,
thereby reducing the refinancing sensitivity
of the remaining mortgage pool.
Accounting for such burnout effects and borrower heterogeneity
is important not only for understanding changes in prepayments
as interest rates decline,
but also for explaining the magnitude of negative convexity
and its nonmonotonic dependence on the interest rate level.

Incorporating these factors into the prepayment model
would allow for more realistic MBS valuation
and interest rate sensitivity analysis.
However, the resulting increase in model complexity
may make it difficult to retain the closed-form approximations
and the decomposition of interest rate sensitivities
that characterize our framework.
Therefore, another direction for future research
is to examine the extent to which analytical tractability
can be preserved while incorporating more realistic prepayment behavior,
and to investigate how these additional factors affect
the regions in which negative convexity occurs
and the corresponding sign boundaries.

\small
\printbibliography

\appendix

\section{Supplementary Derivations for MBS Valuation}

In this section, we provide supplementary derivations for MBS valuation, including a proof of Equation~\eqref{eq:MBS_value_result_01}, which was omitted from the main text, and the derivation of an analytical expression for $Z_{\mathrm{c}}(t,s)$.
In Section~\ref{app:derivation_mbs_value}, we derive Equations~\eqref{eq:MBS_value_result_05}--\eqref{eq:MBS_value_result_01} using a change of measure to the $s$-forward measure and differentiation of conditional expectations.
In Section~\ref{subsec:analytic_Zc}, we derive an analytical expression for $Z_{\mathrm{c}}(t,s)$, corresponding to the interest-rate-independent prepayment process, and use it to obtain an analytical expression for the effective prepayment rate $\bar{\lambda}_{\mathrm{c}}(t,s)$.

\subsection{Proof of \texorpdfstring{Equations~\eqref{eq:MBS_value_result_05}--\eqref{eq:MBS_value_result_01}}{the MBS Valuation Equations}}
\label{app:derivation_mbs_value}

In this subsection, we derive Equations~\eqref{eq:MBS_value_result_05}--\eqref{eq:MBS_value_result_01}.
Using the $s$-forward probability measure defined in Equation~\eqref{eq:def_RND}, the expectation on the right-hand side of Equation~\eqref{eq:MBS_value} can be evaluated as follows:
\begin{align}
    &
    \mathbb{E} ^{\mathbb{Q}} \left[ \left. e^{- \int _t ^s ( r(u) + \lambda (u) ) \mathrm{d}u} \left( c + \tilde{P} (s) \lambda (s) \right) \right| \mathcal{F}_t \right]
    \nonumber
    \\
    =&
    \mathbb{E} ^{\mathbb{Q}_s} \left[ \left. \frac{L(t)}{L(s)} e^{- \int _t ^s ( r(u) + \lambda (u) ) \mathrm{d}u} \left( c + \tilde{P} (s) \lambda (s) \right) \right| \mathcal{F}_t \right]
    \nonumber
    \\
    =&
    \mathbb{E} ^{\mathbb{Q}_s} \left[ \left. D(t,s) e^{\int _t ^s r(u) \mathrm{d}u} e^{- \int _t ^s ( r(u) + \lambda (u) ) \mathrm{d}u} \left( c + \tilde{P} (s) \lambda (s) \right) \right| \mathcal{F}_t \right]
    \nonumber
    \\
    =&
    D(t,s) \,
    \mathbb{E} ^{\mathbb{Q}_s} \left[ \left. e^{- \int _t ^s \lambda (u) \mathrm{d}u} \left( c + \tilde{P} (s) \lambda (s) \right) \right| \mathcal{F}_t \right]
    \nonumber
    \\
    =&
    D(t,s) \left( c \, \mathbb{E}^{\mathbb{Q}_s} \left[ \left. e^{- \int _t ^s \lambda (u) \mathrm{d}u} \right| \mathcal{F}_t \right] + \tilde{P} (s) \mathbb{E}^{\mathbb{Q}_s} \left[ \left. e^{- \int _t ^s \lambda (u) \mathrm{d}u} \lambda (s) \right| \mathcal{F}_t \right] \right).
    \label{eq:calc_mbs_price}
\end{align}
Here, we consider $Z^{(s)}(t,v)$, defined in Equation~\eqref{eq:MBS_value_result_05} for a fixed cash flow date $s$.
Differentiating with respect to $v$ while holding $s$ and $\mathbb{Q}_s$ fixed yields
\begin{align*}
    \frac{\partial}{\partial v}
    Z^{(s)}(t,v)
    =
    -
    \mathbb{E}^{\mathbb{Q}_s}
    \left[
        \left.
        \lambda(v)
        e^{-\int_t^v\lambda(u)\mathrm{d}u}
        \right|
        \mathcal{F}_t
    \right].
\end{align*}
Therefore,
\begin{align*}
    -
    \frac{\partial}{\partial v}
    \ln Z^{(s)}(t,v)
    =
    \frac{
        \mathbb{E}^{\mathbb{Q}_s}
        \left[
            \left.
            e^{-\int_t^v\lambda(u)\mathrm{d}u}
            \lambda(v)
            \right|
            \mathcal{F}_t
        \right]
    }{
        \mathbb{E}^{\mathbb{Q}_s}
        \left[
            \left.
            e^{-\int_t^v\lambda(u)\mathrm{d}u}
            \right|
            \mathcal{F}_t
        \right]
    }.
\end{align*}
Setting $v=s$ yields Equation~\eqref{eq:MBS_value_result_04}.
That is,
\begin{align*}
    \mathbb{E}^{\mathbb{Q}_s}
    \left[
        \left.
        e^{-\int_t^s\lambda(u)\mathrm{d}u}
        \lambda(s)
        \right|
        \mathcal{F}_t
    \right]
    =
    \bar{\lambda}^{\mathbb{Q}_s}(t,s)
    Z(t,s).
\end{align*}
Substituting this expression into Equation~\eqref{eq:calc_mbs_price}, we obtain
\begin{align*}
    &
    D(t,s)
    \left(
        c Z(t,s)
        +
        \tilde{P}(s)
        \bar{\lambda}^{\mathbb{Q}_s}(t,s)
        Z(t,s)
    \right)
    \\
    =&
    D(t,s)
    Z(t,s)
    \left(
        c
        +
        \tilde{P}(s)
        \bar{\lambda}^{\mathbb{Q}_s}(t,s)
    \right)
\end{align*}
Thus, we obtain Equation~\eqref{eq:MBS_value_result_01}.

\subsection{Analytical Expression for \texorpdfstring{$Z_{\mathrm{c}}(t,s)$}{Zc(t,s)}}
\label{subsec:analytic_Zc}

$Z_{\mathrm{c}}(t,s)$ admits the following analytical expression:
\begin{align*}
    Z_{\mathrm{c}} (t,s)
    =&
    e^{- (A_{\mathrm{c}}(\tau) \lambda _{\mathrm{c}}(t) + B_{\mathrm{c}} (\tau))},
    \\
    \text{where} \quad
    A_{\mathrm{c}} (\tau)
    :=&
    \frac{\sinh (\xi \tau)}{g(\tau)},
    \\
    B_{\mathrm{c}} (\tau)
    :=&
    \frac{2 \kappa _{\lambda} \bar{\lambda}}{\sigma _{\lambda} ^2} \ln \left( \frac{g(\tau)}{\xi \exp (\kappa _{\lambda} \tau / 2)} \right),
    \\
    g(\tau)
    :=&
    \xi \cosh (\xi \tau) + \frac{\kappa _{\lambda}}{2} \sinh (\xi \tau),
    \\
    \xi
    :=&
    \frac{1}{2} \sqrt{\kappa _{\lambda} ^2 + 2 \sigma _{\lambda} ^2}.
\end{align*}
Here, $\tau := s-t$.
Furthermore, for a fixed time $s$, the survival-weighted conditional mean of $\lambda_{\mathrm{c}}(s)$ under $\mathbb{Q}_s$, given by $-\left.\frac{\partial}{\partial v}\ln Z_{\mathrm{c}}(t,v)\right|_{v=s}$, also admits the following analytical expression:
\begin{align*}
    -
    \left.
        \frac{\partial}{\partial v}
        \ln Z_{\mathrm{c}}(t,v)
    \right|_{v=s}
    =&
    \frac{
        \mathbb{E}^{\mathbb{Q}_s}
        \left[
            \left.
            e^{-\int_t^s\lambda_{\mathrm{c}}(u)\mathrm{d}u}
            \lambda_{\mathrm{c}}(s)
            \right|
            \mathcal{F}_t
        \right]
    }{
        \mathbb{E}^{\mathbb{Q}_s}
        \left[
            \left.
            e^{-\int_t^s\lambda_{\mathrm{c}}(u)\mathrm{d}u}
            \right|
            \mathcal{F}_t
        \right]
    }
    \\
    =&
    A_{\mathrm{c}}^{\prime}(s-t)
    \lambda_{\mathrm{c}}(t)
    +
    B_{\mathrm{c}}^{\prime}(s-t),
    \\
    \text{where} \quad
    A_{\mathrm{c}}^{\prime}(\tau)
    =&
    1
    -
    \kappa_{\lambda}
    A_{\mathrm{c}}(\tau)
    -
    \frac{1}{2}
    \sigma_{\lambda}^2
    A_{\mathrm{c}}^2(\tau),
    \\
    B_{\mathrm{c}}^{\prime}(\tau)
    =&
    \kappa_{\lambda}
    \bar{\lambda}
    A_{\mathrm{c}}(\tau).
\end{align*}

\section{Supplementary Results for the Hull--White Model}

In this section, we present supplementary results for the Hull--White model.
First, we derive an expression for $\vartheta(t)$ that ensures that the theoretical discount curve derived from the model matches the market-observed discount curve.
Next, we derive an analytical expression for zero-coupon bond prices and their stochastic differential equation.
Finally, we derive the conditional expectation of the short rate under the risk-neutral measure.

\subsection{Market-Consistent \texorpdfstring{$\vartheta(t)$}{vartheta(t)}}
\label{app:market_consistent_vartheta}

In this subsection, we show that if the theoretical discount curve derived from the short-rate process in Equation~\eqref{eq:HW} matches the market-observed discount curve, then $\vartheta(t)$ must satisfy
\begin{align}
    \vartheta(t)
    =
    -
    \frac{\partial^2}{\partial t^2}
    \ln D(0,t)
    -
    a_r
    \frac{\partial}{\partial t}
    \ln D(0,t)
    +
    \frac{\sigma_r^2}{2a_r}
    \left(
        1-e^{-2a_rt}
    \right).
    \label{eq:vartheta}
\end{align}
We consider the following ansatz for $D(t,T)$.
\begin{align}
    D(t,T)
    =
    \mathbb{E}^{\mathbb{Q}} \left[ \left. e^{- \int _t ^T r(u) \mathrm{d}u} \right| \mathcal{F}_t \right]
    =
    e^{A_r (\tau) - B_r (\tau) r(t)}.
    \label{eq:D_ansatz}
\end{align}
Here, $\tau := T-t$.
The following process is a $\mathbb{Q}$-martingale.
\begin{align*}
    e^{- \int _0 ^t r(u) \mathrm{d}u} D(t,T)
    =
    \mathbb{E}^{\mathbb{Q}} \left[ \left. e^{- \int _0 ^T r(u) \mathrm{d}u} \right| \mathcal{F}_t \right].
\end{align*}
Applying Ito's formula to $D = D(t, r(t))$ yields
\begin{align}
    \mathrm{d} D
    =&
    \partial _t D \mathrm{d}t + (\partial _r D) \mathrm{d}r + \frac{1}{2} (\partial _r ^2 D) (\mathrm{d}r)^2
    \nonumber
    \\
    =&
    \left( \partial _t D + (\vartheta (t) - a_r r) \partial _r D + \frac{1}{2} \sigma _r ^2 \partial _r ^2 D \right) \mathrm{d}t + \sigma _r (\partial _r D) \mathrm{d} W^{\mathbb{Q}} _r (t).
    \label{eq:Ito_D}
\end{align}
Similarly, applying Ito's formula to $e^{- \int _0 ^t r(u) \mathrm{d}u} D(t,T)$ gives
\begin{align*}
    &
    \mathrm{d} \left( e^{- \int _0 ^t r(u) \mathrm{d}u} D(t, r(t)) \right)
    \\
    =&
    e^{- \int _0 ^t r(u) \mathrm{d}u} \left( \partial _t D + (\vartheta (t) - a_r r) \partial _r D + \frac{1}{2} \sigma _r ^2 \partial _r ^2 D - rD \right) \mathrm{d}t + (\text{diffusion term}).
\end{align*}
Since $e^{- \int _0 ^t r(u) \mathrm{d}u} D(t, r(t))$ is a $\mathbb{Q}$-martingale, the above result implies that $D(t,r)$ satisfies the following partial differential equation.
\begin{align}
    &
    \partial _t D(t,r) + (\vartheta (t) - a_r r) \partial _r D(t,r) + \frac{1}{2} \sigma _r ^2 \partial _r ^2 D(t,r)
    =
    rD(t,r),
    \label{eq:PDE_D}
    \\
    &
    D(T,r)
    =
    1.
    \nonumber
\end{align}
Substituting the ansatz for $D(t,T)$ in Equation~\eqref{eq:D_ansatz} into the above partial differential equation yields the following ordinary differential equations.
\begin{align*}
    &
    B_r ^{\prime} (\tau) + a_r B_r (\tau) - 1 = 0,
    \quad
    B_r (0)
    =
    0,
    \\
    &
    A_r ^{\prime} (\tau) + \vartheta (t) B_r (\tau) - \frac{1}{2} \sigma _r ^2 B_r ^2 (\tau)
    =
    0,
    \quad 
    A_r (0)
    =
    0.
\end{align*}
The equation for $B_r(\tau)$ can be solved directly, yielding
\begin{align}
    B_r (\tau)
    =
    \frac{1 - e^{-a_r \tau}}{a_r}.
    \label{eq:B_r_sol}
\end{align}
The equation for $A_r(\tau)$ can also be solved as follows.
\begin{align}
    A_r(\tau)
    =
    \int _0 ^{\tau} \left( - \vartheta (T- \tau^{\prime}) B_r (\tau ^{\prime}) + \frac{1}{2} \sigma _r ^2 B_r ^2 (\tau ^{\prime}) \right) \mathrm{d} \tau ^{\prime}
    =
    \int _t ^T \left( - \vartheta (v) B_r (T-v) + \frac{1}{2} \sigma _r ^2 B_r ^2 (T-v) \right) \mathrm{d}v.
    \label{eq:A_r_sol}
\end{align}
In the second equality, we applied the change of variables $v=T-\tau^{\prime}$.
In particular, setting $t=0$ in Equation~\eqref{eq:A_r_sol} yields
\begin{align*}
    A_r (T)
    =
    \int _0 ^T \left( - \vartheta (v) B_r (T-v) + \frac{1}{2} \sigma _r ^2 B_r ^2 (T-v) \right) \mathrm{d}v.
\end{align*}
The first derivative of $A_r(T)$ with respect to $T$ is given by
\begin{align}
    A_r ^{\prime} (T)
    =&
    \left( - \vartheta (T) B_r (0) + \frac{1}{2} \sigma _r ^2 B_r ^2 (0) \right)
    \nonumber
    \\
    &+
    \int _0 ^T \left( - \vartheta (v) B_r ^{\prime} (T-v) + \frac{1}{2} \sigma _r ^2 \cdot 2 B_r (T-v) B_r ^{\prime} (T-v) \right) \mathrm{d}v
    \label{eq:A_r_prime_calc}
    \\
    =&
    \int _0 ^T \left( - \vartheta (v) e^{- a_r (T-v)} + \sigma _r ^2 e^{-a_r (T-v)} B_r (T-v) \right) \mathrm{d}v.
    \nonumber
\end{align}
Here, the first line on the right-hand side of Equation~\eqref{eq:A_r_prime_calc} represents the boundary term arising from the dependence of the upper integration limit on $T$, while the integral term in the second line represents the contribution arising from the dependence of the integrand on $T$.
Since $B_r(0)=0$, the boundary term vanishes.
Similarly, the second derivative of $A_r(T)$ with respect to $T$ is given by
\begin{align*}
    A^{\prime \prime}_r (\tau)
    =&
    \left( - \vartheta (T) + \sigma _r ^2 B_r (0) \right)
    \\
    &+
    \int _0 ^T \bigg\{ - \vartheta (v) (- a_r) e^{- a_r (T-v)} \bigg.
    \\
    &+
    \bigg. \sigma _r ^2 \left[ (-a_r) e^{-a_r (T-v)} B_r (T-v) + e^{-a_r (T-v)} e^{-a_r (T-v)} \right] \bigg\}
    \\
    =&
    - \vartheta (T)
    +
    \int _0 ^T \left[ a_r e^{-a_r (T-v)} \vartheta (v) + \sigma _r ^2 e^{-a_r (T-v)} \left( -1 + 2 e^{-a_r (T-v)} \right) \right] \mathrm{d}v.
\end{align*}
Using the relation $\ln D(0,T) = A_r (T) - B_r (T) r (0)$, we differentiate both sides with respect to $T$ to obtain the first and second derivatives as follows.
\begin{align*}
    a_r \frac{\partial}{\partial T} \ln D(0,T)
    =&
    a_r A_r ^{\prime} (T) - a_r e^{-a_r T} r(0),
    \\
    \frac{\partial ^2}{\partial T^2} \ln D(0,T)
    =&
    A_r ^{\prime \prime} (T) + a_r e^{-a_r T} r(0).
\end{align*}
Adding these two equations yields
\begin{align*}
    &
    a_r \frac{\partial}{\partial T} \ln D(0,T) + \frac{\partial ^2}{\partial T^2}\ln D(0,T)
    \\
    =&
    a_r A_r ^{\prime} (T) + A_r ^{\prime \prime} (T)
    \\
    =&
    a_r \int _0 ^T \left( - \vartheta (v) e^{- a_r (T-v)} + \sigma _r ^2 e^{-a_r (T-v)} B_r (T-v) \right) \mathrm{d}v
    \\
    &
    - \vartheta (T)
    +
    \int _0 ^T \left[ a_r e^{-a_r (T-v)} \vartheta (v) + \sigma _r ^2 e^{-a_r (T-v)} \left( -1 + 2 e^{-a_r (T-v)} \right) \right] \mathrm{d}v
    \\
    =&
    - \vartheta (T) + \int _0 ^T \sigma _r ^2 e^{-2 a_r (T-v)} \mathrm{d}v
    \\
    =&
    - \vartheta (T) + \frac{\sigma _r ^2}{2 a_r} \left( 1 - e^{-2a_r T} \right).
\end{align*}
Rearranging the above expression yields Equation~\eqref{eq:vartheta}.

\subsection{Zero-Coupon Bond Prices}
\label{app:hw_bond_price_derivation}

In this subsection, we show that, when the Hull--White model is consistent with the market-observed discount curve, the zero-coupon bond price $D(t,T)$ satisfies the following stochastic differential equation.
\begin{align}
    \frac{\mathrm{d}D(t,T)}{D(t,T)}
    =
    r(t)\mathrm{d}t
    -
    \sigma_r B_r(T-t)
    \mathrm{d}W_r^{\mathbb{Q}}(t).
    \label{eq:SDE_D}
\end{align}
We also show that, if the theoretical discount curve derived from the short-rate process in Equation~\eqref{eq:HW} matches the market-observed discount curve, the zero-coupon bond price $D(t,T)$ under the Hull--White model is given by Equations~\eqref{eq:D_sol_HW_01}--\eqref{eq:D_sol_HW_03}.

First, we establish Equation~\eqref{eq:SDE_D}.
Substituting Equation~\eqref{eq:PDE_D} into Equation~\eqref{eq:Ito_D}, we find that $D(t,T)$ satisfies
\begin{align*}
    \frac{\mathrm{d} D(t,T)}{D(t,T)}
    =
    r(t) \mathrm{d}t + \sigma _r (\partial _r \ln D(t,r)) \mathrm{d} W^{\mathbb{Q}} _r(t).
\end{align*}
From Equations~\eqref{eq:D_ansatz} and \eqref{eq:B_r_sol}, we have $\partial _r \ln D(t,r) = - B_r (T-t)$, which establishes Equation~\eqref{eq:SDE_D}.

Next, we establish Equations~\eqref{eq:D_sol_HW_01}--\eqref{eq:D_sol_HW_03}.
Since Equations~\eqref{eq:D_sol_HW_01} and \eqref{eq:D_sol_HW_03} have already been derived in Appendix~\ref{app:market_consistent_vartheta}, we focus on deriving Equation~\eqref{eq:D_sol_HW_02}.
Specifically, we establish this expression by substituting Equation~\eqref{eq:vartheta} into Equation~\eqref{eq:A_r_sol}.
Let $\ell(v):=\ln D(0,v)$. Then, we obtain
\begin{align}
    \int _t ^T \vartheta (v) B_r (T-v) \mathrm{d}v
    =
    \int _t ^T \left[ - \frac{\partial ^2 \ell}{\partial v^2} (v) - a_r \frac{\partial \ell}{\partial v} (v) + \frac{\sigma _r ^2}{2 a_r} (1 - e^{-2a_r v}) \right] B_r (T-v) \mathrm{d}v.
    \label{eq:A_r_calc_tmp}
\end{align}
Straightforward calculations yield the following three expressions.
\begin{align*}
    &
    \int _t ^T \frac{\partial \ell}{\partial v}(v) B_r (T-v) \mathrm{d}v
    \\
    =&
    - \ell (t) B_r (T-t) + \int _t ^T \ell (v) e^{-a_r (T-v)} \mathrm{d}v,
    \\
    &
    \int _t ^T \frac{\partial ^2 \ell}{\partial v^2} (v) B_r (T-v) \mathrm{d}v
    \\
    =&
    - \frac{\partial \ell}{\partial v} (v=t) B_r (T-t) + \ell(T) - \ell(t) e^{-a_r (T-t)} + a_r \int _t ^T \ell (v) e^{-a_r (T-v)} \mathrm{d}v,
    \\
    &
    - \int _t ^T \frac{\sigma _r ^2}{2 a_r} (1 - e^{-2 a_r v}) B_r (T-v) \mathrm{d}v + \int _t ^T \frac{1}{2} \sigma _r ^2 B_r ^2 (T-v) \mathrm{d}v
    \\
    =&
    - \frac{\sigma _r ^2}{4a_r} (1 - e^{-2a_r t}) B_r ^2 (T-t).
\end{align*}
Substituting these three expressions into Equation~\eqref{eq:A_r_calc_tmp} and using Equation~\eqref{eq:A_r_sol}, we obtain Equation~\eqref{eq:D_sol_HW_02}.

\subsection{Derivation of Equation~(\ref{eq:expectation_r_t_under_Q})}
\label{app:expectation_r_under_Q_derivation}

In this subsection, we derive Equation~\eqref{eq:expectation_r_t_under_Q}.
First, under the risk-neutral measure $\mathbb{Q}$, the short-rate process $r(t)$ admits the following solution.
\begin{align*}
    r(u)
    =&
    e^{-a_r (u-t)} r(t) + \int _t ^u e^{-a_r (u-v)} \vartheta (v) \mathrm{d}v + \sigma _r \int _t ^u e^{-a_r (u-v)} \mathrm{d} W^{\mathbb{Q}} _r (v).
\end{align*}
Using Equation~\eqref{eq:vartheta}, the second term on the right-hand side can be evaluated as follows.
\begin{align*}
    &
    \int _t ^u e^{- a_r (u-v)} \vartheta (v) \mathrm{d}v
    \\
    =&
    - \frac{\partial}{\partial u} \ln D(0, u) + e^{- a_r (u-t)} \frac{\partial}{\partial t} \ln D(0, t)
    +
    \frac{\sigma_r ^2}{2 a_r} \int _t ^u e^{- a_r (u-v)} \left( 1 - e^{-2 a_r v} \right) \mathrm{d}v
    \\
    =&
    - \frac{\partial}{\partial u} \ln D(0, u) + e^{- a_r (u-t)} \frac{\partial}{\partial t} \ln D(0, t)
    +
    \frac{\sigma _r ^2}{2 a_r ^2} \left( 1 - e^{- a_r (u-t)} \right) \left( 1 - e^{- a_r (u+t)} \right).
\end{align*}
Combining these results yields Equation~\eqref{eq:expectation_r_t_under_Q}.

\section{Change of Measure and Stochastic Process Dynamics}

In this section, we provide supplementary details on the change of measure from the risk-neutral measure $\mathbb{Q}$ to the $s$-forward measure $\mathbb{Q}_s$ and the resulting dynamics of the stochastic processes.
In Section~\ref{app:change_of_measure_to_forward}, we derive an explicit expression for the Radon--Nikodym derivative used in the change of measure.
In Section~\ref{app:stochastic_processes_forward_measure}, using Girsanov's theorem, we establish the relationship between the Brownian motions under the two measures and derive the dynamics of the short rate $r(t)$ and the refinancing sensitivity $y(t)$ under $\mathbb{Q}_s$.
In Section~\ref{app:conditional_means_variances}, we derive analytical expressions for the conditional means and variances of these stochastic processes, which are used in the closed-form approximation presented in the main text.

\subsection{Change of Measure to the Forward Measure}
\label{app:change_of_measure_to_forward}

In this subsection, we derive an explicit expression for $L(t)$ defined in Equation~\eqref{eq:def_RND}.
Using Equation~\eqref{eq:SDE_D}, we obtain
\begin{align*}
    &
    \left.
        \mathrm{d}
        \left(
            \frac{D(t,s)}{M(t)}
        \right)
        \middle/
        \frac{D(t,s)}{M(t)}
    \right.
    \\
    =&
    \frac{\mathrm{d}D(t,s)}{D(t,s)}
    -
    \frac{\mathrm{d}M(t)}{M(t)}
    +
    \left(
        \frac{\mathrm{d}M(t)}{M(t)}
    \right)^2
    -
    \frac{\mathrm{d}D(t,s)}{D(t,s)}
    \frac{\mathrm{d}M(t)}{M(t)}
    \\
    =&
    -
    \sigma_r B_r(s-t)
    \mathrm{d}W_r^{\mathbb{Q}}(t).
\end{align*}
Therefore, $L(t)$ satisfies
\begin{align*}
    \frac{\mathrm{d}L(t)}{L(t)}
    =
    -
    \sigma_r B_r(s-t)
    \mathrm{d}W_r^{\mathbb{Q}}(t).
\end{align*}
Solving this stochastic differential equation yields
\begin{align}
    L(t)
    =
    \left.
        \frac{\mathrm{d}\mathbb{Q}_s}
        {\mathrm{d}\mathbb{Q}}
    \right|_{\mathcal{F}_t}
    =
    e^{
        -\int_0^t
        \sigma_r B_r(s-v)
        \mathrm{d}W_r^{\mathbb{Q}}(v)
        -
        \frac{1}{2}
        \int_0^t
        \sigma_r^2 B_r^2(s-v)
        \mathrm{d}v
    }.
    \label{eq:RND_concrete_expression}
\end{align}

\subsection{Stochastic Processes under the Forward Measure}
\label{app:stochastic_processes_forward_measure}

In this subsection, we derive the relationship between the Brownian motions under the risk-neutral measure $\mathbb{Q}$ and those under the $s$-forward measure $\mathbb{Q}_s$ resulting from the change of measure.
Furthermore, using this relationship, we derive the stochastic differential equations for the short rate $r(t)$ and the refinancing sensitivity $y(t)$ under $\mathbb{Q}_s$.

Consider standard Brownian motions $W_r^{\mathbb{Q}_s}$, $W_y^{\mathbb{Q}_s}$, and $W_{\lambda}^{\mathbb{Q}_s}$ under $\mathbb{Q}_s$ satisfying
\begin{align}
    \mathrm{d}W_r^{\mathbb{Q}_s}(t)
    \mathrm{d}W_y^{\mathbb{Q}_s}(t)
    =&
    \rho\mathrm{d}t,
    \nonumber
    \\
    \mathrm{d}W_r^{\mathbb{Q}_s}(t)
    \mathrm{d}W_{\lambda}^{\mathbb{Q}_s}(t)
    =&
    \mathrm{d}W_y^{\mathbb{Q}_s}(t)
    \mathrm{d}W_{\lambda}^{\mathbb{Q}_s}(t)
    =
    0.
    \label{eq:Qs_W_correlation}
\end{align}
By Equation~\eqref{eq:RND_concrete_expression} and Girsanov's theorem, the standard Brownian motions $W_r^{\mathbb{Q}_s}$, $W_y^{\mathbb{Q}_s}$, and $W_{\lambda}^{\mathbb{Q}_s}$ under $\mathbb{Q}_s$ are related to the standard Brownian motions $W_r^{\mathbb{Q}}$, $W_y^{\mathbb{Q}}$, and $W_{\lambda}^{\mathbb{Q}}$ under $\mathbb{Q}$ as follows.
\begin{align}
    \mathrm{d}W_r^{\mathbb{Q}_s}(t)
    =&
    \mathrm{d}W_r^{\mathbb{Q}}(t)
    +
    \sigma_r B_r(s-t)\mathrm{d}t,
    \label{eq:relation_W_Q_Qs_01}
    \\
    \mathrm{d}W_y^{\mathbb{Q}_s}(t)
    =&
    \mathrm{d}W_y^{\mathbb{Q}}(t)
    +
    \rho\sigma_r B_r(s-t)\mathrm{d}t,
    \label{eq:relation_W_Q_Qs_02}
    \\
    \mathrm{d}W_{\lambda}^{\mathbb{Q}_s}(t)
    =&
    \mathrm{d}W_{\lambda}^{\mathbb{Q}}(t).
    \label{eq:relation_W_Q_Qs_03}
\end{align}
We establish these relationships below.

First, using mutually independent standard Brownian motions under $\mathbb{Q}$, we can represent $W^{\mathbb{Q}}_r$, $W^{\mathbb{Q}}_y$, and $W^{\mathbb{Q}}_{\lambda}$ as follows.
\begin{align}
    \mathrm{d} W^{\mathbb{Q}}_r (t)
    =&
    \mathrm{d} W^{\mathbb{Q}} _1 (t),
    \label{eq:Q_rylmd_123_relation_01}
    \\
    \mathrm{d} W^{\mathbb{Q}}_y (t)
    =&
    \rho \mathrm{d} W^{\mathbb{Q}} _1 (t) + \sqrt{1 - \rho ^2} \mathrm{d} W^{\mathbb{Q}} _2 (t),
    \label{eq:Q_rylmd_123_relation_02}
    \\
    \mathrm{d} W^{\mathbb{Q}} _{\lambda} (t)
    =&
    \mathrm{d} W^{\mathbb{Q}} _3 (t).
    \label{eq:Q_rylmd_123_relation_03}
\end{align}
By Equation~\eqref{eq:RND_concrete_expression} and Girsanov's theorem, the processes $W^{\mathbb{Q}_s} _1$, $W^{\mathbb{Q}_s} _2$, and $W^{\mathbb{Q}_s} _3$ defined below are mutually independent standard Brownian motions under $\mathbb{Q}_s$.
\begin{align}
    \mathrm{d} W^{\mathbb{Q}_s}_1 (t)
    :=&
    \mathrm{d} W^{\mathbb{Q}} _1 (t) + \sigma _r B_r (s-t) \mathrm{d}t,
    \label{eq:def_Qs_123_01}
    \\
    \mathrm{d} W^{\mathbb{Q}_s}_2 (t)
    :=&
    \mathrm{d} W^{\mathbb{Q}} _2 (t),
    \label{eq:def_Qs_123_02}
    \\
    \mathrm{d} W^{\mathbb{Q}_s} _3 (t)
    :=&
    \mathrm{d} W^{\mathbb{Q}} _3 (t).
    \label{eq:def_Qs_123_03}
\end{align}
Using $W^{\mathbb{Q}_s} _1$, $W^{\mathbb{Q}_s} _2$, and $W^{\mathbb{Q}_s} _3$ defined above, we can express $W^{\mathbb{Q}}_r$, $W^{\mathbb{Q}}_y$, and $W^{\mathbb{Q}}_{\lambda}$ as follows.
\begin{align*}
    &
    \begin{pmatrix}
        \mathrm{d} W^{\mathbb{Q}} _r (t)
        \\
        \mathrm{d} W^{\mathbb{Q}} _y (t)
        \\
        \mathrm{d} W^{\mathbb{Q}} _{\lambda} (t)
    \end{pmatrix}
    \\
    =&
    \begin{pmatrix}
        1 & 0 & 0
        \\
        \rho & \sqrt{1 - \rho ^2} & 0
        \\
        0 & 0 & 1
    \end{pmatrix}
    \begin{pmatrix}
        \mathrm{d} W^{\mathbb{Q}} _1 (t)
        \\
        \mathrm{d} W^{\mathbb{Q}} _2 (t)
        \\
        \mathrm{d} W^{\mathbb{Q}} _3 (t)
    \end{pmatrix}
    \\
    =&
    \begin{pmatrix}
        1 & 0 & 0
        \\
        \rho & \sqrt{1 - \rho ^2} & 0
        \\
        0 & 0 & 1
    \end{pmatrix}
    \left[
        \begin{pmatrix}
            \mathrm{d} W^{\mathbb{Q}_s} _1 (t)
            \\
            \mathrm{d} W^{\mathbb{Q}_s} _2 (t)
            \\
            \mathrm{d} W^{\mathbb{Q}_s} _3 (t)
        \end{pmatrix}
        -
        \begin{pmatrix}
            \sigma _r B_r (s-t)
            \\
            0
            \\
            0
        \end{pmatrix}
        \mathrm{d}t
    \right]
    \\
    =&
    \begin{pmatrix}
        1 & 0 & 0
        \\
        \rho & \sqrt{1 - \rho ^2} & 0
        \\
        0 & 0 & 1
    \end{pmatrix}   
    \begin{pmatrix}
        \mathrm{d} W^{\mathbb{Q}_s} _1 (t)
        \\
        \mathrm{d} W^{\mathbb{Q}_s} _2 (t)
        \\
        \mathrm{d} W^{\mathbb{Q}_s} _3 (t)
    \end{pmatrix}
    -
    \begin{pmatrix}
        1 & 0 & 0
        \\
        \rho & \sqrt{1 - \rho ^2} & 0
        \\
        0 & 0 & 1
    \end{pmatrix}   
    \begin{pmatrix}
        \sigma _r B_r (s-t)
        \\
        0
        \\
        0
    \end{pmatrix}
    \mathrm{d}t.
\end{align*}
Here, the first equality follows from Equations~\eqref{eq:Q_rylmd_123_relation_01}--\eqref{eq:Q_rylmd_123_relation_03}, while the second equality follows from Equations~\eqref{eq:def_Qs_123_01}--\eqref{eq:def_Qs_123_03}.
Therefore, we obtain
\begin{align}
    \begin{pmatrix}
        1 & 0 & 0
        \\
        \rho & \sqrt{1 - \rho ^2} & 0
        \\
        0 & 0 & 1
    \end{pmatrix}   
    \begin{pmatrix}
        \mathrm{d} W^{\mathbb{Q}_s} _1 (t)
        \\
        \mathrm{d} W^{\mathbb{Q}_s} _2 (t)
        \\
        \mathrm{d} W^{\mathbb{Q}_s} _3 (t)
    \end{pmatrix}
    =
    \begin{pmatrix}
        \mathrm{d} W^{\mathbb{Q}} _r (t)
        \\
        \mathrm{d} W^{\mathbb{Q}} _y (t)
        \\
        \mathrm{d} W^{\mathbb{Q}} _{\lambda} (t)
    \end{pmatrix}
    +
    \begin{pmatrix}
        1 & 0 & 0
        \\
        \rho & \sqrt{1 - \rho ^2} & 0
        \\
        0 & 0 & 1
    \end{pmatrix}   
    \begin{pmatrix}
        \sigma _r B_r (s-t)
        \\
        0
        \\
        0
    \end{pmatrix}
    \mathrm{d}t
    .
    \label{eq:realtion_Qs_123_Wrylmd}
\end{align}
Using the mutually independent standard Brownian motions under $\mathbb{Q}_s$ defined in Equations~\eqref{eq:def_Qs_123_01}--\eqref{eq:def_Qs_123_03}, we define $W^{\mathbb{Q}_s} _r$, $W^{\mathbb{Q}_s} _y$, and $W^{\mathbb{Q}_s} _{\lambda}$ as follows.
\begin{align}
    \begin{pmatrix}
        \mathrm{d} W^{\mathbb{Q}_s} _r (t)
        \\
        \mathrm{d} W^{\mathbb{Q}_s} _y (t)
        \\
        \mathrm{d} W^{\mathbb{Q}_s} _{\lambda} (t)
    \end{pmatrix}
    =
    \begin{pmatrix}
        1 & 0 & 0
        \\
        \rho & \sqrt{1 - \rho ^2} & 0
        \\
        0 & 0 & 1
    \end{pmatrix}
    \begin{pmatrix}
        \mathrm{d} W^{\mathbb{Q}_s} _1 (t)
        \\
        \mathrm{d} W^{\mathbb{Q}_s} _2 (t)
        \\
        \mathrm{d} W^{\mathbb{Q}_s} _3 (t)
    \end{pmatrix}
    .
    \label{eq:def_Qs_rylmd}
\end{align}
It follows directly from their definitions that $W^{\mathbb{Q}_s} _r$, $W^{\mathbb{Q}_s} _y$, and $W^{\mathbb{Q}_s} _{\lambda}$ satisfy Equation~\eqref{eq:Qs_W_correlation}.
Substituting Equation~\eqref{eq:def_Qs_rylmd} into Equation~\eqref{eq:realtion_Qs_123_Wrylmd} yields
\begin{align*}
    \begin{pmatrix}
        \mathrm{d} W^{\mathbb{Q}_s} _r (t)
        \\
        \mathrm{d} W^{\mathbb{Q}_s} _y (t)
        \\
        \mathrm{d} W^{\mathbb{Q}_s} _{\lambda} (t)
    \end{pmatrix}
    =
    \begin{pmatrix}
        \mathrm{d} W^{\mathbb{Q}} _r (t)
        \\
        \mathrm{d} W^{\mathbb{Q}} _y (t)
        \\
        \mathrm{d} W^{\mathbb{Q}} _{\lambda} (t)
    \end{pmatrix}
    +
    \begin{pmatrix}
        1 & 0 & 0
        \\
        \rho & \sqrt{1 - \rho ^2} & 0
        \\
        0 & 0 & 1
    \end{pmatrix}   
    \begin{pmatrix}
        \sigma _r B_r (s-t)
        \\
        0
        \\
        0
    \end{pmatrix}
    \mathrm{d}t.
\end{align*}
Thus, we have established Equations~\eqref{eq:relation_W_Q_Qs_01}--\eqref{eq:relation_W_Q_Qs_03}.

We also derive the dynamics of $r(t)$ and $y(t)$ under the $s$-forward measure $\mathbb{Q}_s$.
Applying \eqref{eq:relation_W_Q_Qs_01}--\eqref{eq:relation_W_Q_Qs_03} to the stochastic differential equations for $r(t)$ and $y(t)$ under the risk-neutral measure $\mathbb{Q}$ yields the following stochastic differential equations.
\begin{align*}
    &
    \mathrm{d} r(t)
    =
    \big( \vartheta (t) - \sigma _r ^2 B_r (s-t) - a_r r(t) \big) \mathrm{d}t + \sigma _r \mathrm{d} W^{\mathbb{Q}_s} _r (t),
    \\
    &
    \mathrm{d} y(t)
    =
    \kappa _y \left( \bar{y} - \frac{\rho \sigma _r \sigma _y}{\kappa _y} B_r (s-t) - y(t) \right) \mathrm{d}t + \sigma _y \mathrm{d} W^{\mathbb{Q}_s} _y (t).
\end{align*}
Here, $B_r(x) := (1 - e^{- a_r x}) / a_r$.

\subsection{Analytical Expressions for Conditional Means and Variances}
\label{app:conditional_means_variances}

In this subsection, we show that the conditional means and variances of the processes $r$ and $y$ admit the following analytical expressions:
\begin{align}
    \mu_r^{\mathbb{Q}_s}(t,u)
    =&
    -
    \frac{\partial}{\partial u}
    \ln D(0,u)
    +
    e^{-a_r(u-t)}
    \left(
        r(t)
        +
        \frac{\partial}{\partial t}
        \ln D(0,t)
    \right)
    \nonumber
    \\
    &+
    \frac{\sigma_r^2}{2a_r}
    B_r(u-t)
    \left[
        e^{-a_r(s-u)}
        \left(
            1+e^{-a_r(u-t)}
        \right)
        -
        \left(
            1+e^{-a_r(u+t)}
        \right)
    \right],
    \label{eq:analytic_ry_01}
    \\
    \sigma_r^{\mathbb{Q}_s}(t,u)
    =&
    \sigma_r
    \sqrt{
        \frac{
            1-e^{-2a_r(u-t)}
        }{
            2a_r
        }
    },
    \label{eq:analytic_ry_02}
    \\
    \mu_y^{\mathbb{Q}_s}(t,u)
    =&
    e^{-\kappa_y(u-t)}y(t)
    +
    \bar{y}
    \left[
        1-e^{-\kappa_y(u-t)}
    \right]
    \nonumber
    \\
    &-
    \frac{\rho\sigma_r\sigma_y}{a_r}
    \left[
        \frac{
            1-e^{-\kappa_y(u-t)}
        }{
            \kappa_y
        }
        -
        e^{-a_r(s-u)}
        \frac{
            1-e^{-(\kappa_y+a_r)(u-t)}
        }{
            \kappa_y+a_r
        }
    \right],
    \label{eq:analytic_ry_03}
    \\
    \sigma_y^{\mathbb{Q}_s}(t,u)
    =&
    \sigma_y
    \sqrt{
        \frac{
            1-e^{-2\kappa_y(u-t)}
        }{
            2\kappa_y
        }
    }.
    \label{eq:analytic_ry_04}
\end{align}

First, we derive $\mu ^{\mathbb{Q}_s} _r (t, u)$.
The short-rate process $r(u)$ admits the following solution:
\begin{align}
    r(u)
    =&
    e^{-a_r (u-t)} r(t) + \int _t ^u e^{-a_r (u-v)} \left( \vartheta (v) - \sigma _r ^2 B_r (s-v) \right) \mathrm{d}v
    \nonumber
    \\
    & \hspace{3.5cm} +
    \sigma _r \int _t ^u e^{-a_r (u-v)} \mathrm{d} W^{\mathbb{Q}_s} _r (v).
    \label{eq:sol_for_r_u}
\end{align}
Taking conditional expectations of both sides under $\mathbb{Q}_s$ given $\mathcal{F}_t$, and defining $\ell (v) := \ln D(0, v)$, we can use Equation~\eqref{eq:vartheta} to evaluate the second term on the right-hand side of Equation~\eqref{eq:sol_for_r_u} as follows:
\begin{align*}
    &
    \int _t ^u e^{-a_r (u-v)} \vartheta (v) \mathrm{d}v
    \\
    =&
    \int _t ^u e^{-a_r (u-v)} \left[ - \ell ^{\prime \prime}(v) - a_r \ell ^{\prime} (v) + \frac{\sigma _r ^2}{2 a_r} (1 - e^{-2a_r v}) \right] \mathrm{d}v
    \\
    =&
    \int _t ^u \left( - e^{-a_r (u-v)} \ell ^{\prime} (v) \right)^{\prime} \mathrm{d}v + \frac{\sigma _r ^2}{2 a_r} \int _t ^u e^{-a_r (u-v)} (1 - e^{-2a_r v}) \mathrm{d}v
    \\
    =&
    - \frac{\partial}{\partial u} \ln D(0,u) + e^{-a_r (u-t)} \frac{\partial}{\partial t} \ln D(0,t) + \frac{\sigma _r ^2}{2 a_r} \int _t ^u e^{-a_r (u-v)} (1 - e^{-2a_r v}) \mathrm{d}v.
\end{align*}
Furthermore, the following calculation yields
\begin{align*}
    &
    \frac{\sigma _r ^2}{2 a_r} \int _t ^u e^{-a_r (u-v)} (1 - e^{-2a_r v}) \mathrm{d}v - \sigma _r ^2 \int _t ^u e^{-a_r (u-v)} B_r (s-v) \mathrm{d}v
    \\
    =&
    \frac{\sigma _r ^2}{2 a_r ^2} \left( -1 + e^{-a_r (u-t)} - e^{-a_r (u+t)} + e^{-2a_r u} + e^{-a_r (s-u)} - e^{-a_r (s+u-2t)} \right)
    \\
    =&
    \frac{\sigma_r^2}{2a_r}
    B_r(u-t)
    \left[
    e^{-a_r(s-u)}
    \left(1+e^{-a_r(u-t)}\right)
    -
    \left(1+e^{-a_r(u+t)}\right)
    \right].
\end{align*}
Combining these results, we obtain Equation~\eqref{eq:analytic_ry_01}.
Setting $t=0$ in Equation~\eqref{eq:analytic_ry_01} yields an expression that coincides with the formula given by \textcite{Brigo_Mercurio:2006}.

For $\sigma_r^{\mathbb{Q}_s}(t,u)$, Equation~\eqref{eq:sol_for_r_u} yields
\begin{align*}
    \mathbb{V}\mathrm{ar}^{\mathbb{Q}_s}
    \left[
        \left.
        r(u)
        \right|
        \mathcal{F}_t
    \right]
    =&
    \sigma_r^2
    \int_t^u
    e^{-2a_r(u-v)}
    \mathrm{d}v
    =
    \sigma_r^2
    \frac{1-e^{-2a_r(u-t)}}{2a_r}.
\end{align*}
This establishes Equation~\eqref{eq:analytic_ry_02}.

Next, we derive $\mu_y^{\mathbb{Q}_s}(t,u)$.
The process $y(u)$ admits the following solution:
\begin{align}
    y(u)
    =&
    e^{- \kappa _y (u-t)} y(t) + \int _t ^u e^{- \kappa _y (u-v)} \left( \kappa _y \bar{y} - \rho \sigma _r \sigma _y B_r (s-v) \right) \mathrm{d}v
    \nonumber
    \\
    &
    \hspace{4.5cm}
    + \sigma _y \int _t ^u e^{- \kappa _y (u-v)} \mathrm{d} W^{\mathbb{Q}_s} _y (v).
    \label{eq:sol_for_y}
\end{align}
Taking conditional expectations of both sides under $\mathbb{Q}_s$ given $\mathcal{F}_t$ and evaluating the integrals, we obtain Equation~\eqref{eq:analytic_ry_03}.

For $\sigma_y^{\mathbb{Q}_s}(t,u)$, Equation~\eqref{eq:sol_for_y} yields
\begin{align*}
    \mathbb{V} \mathrm{ar} ^{\mathbb{Q}_s} \left[ \left. y(u) \right| \mathcal{F}_t \right]
    =
    \sigma _y ^2 \int _t ^u e^{-2 \kappa _y (u-v)} \mathrm{d}v
    =
    \frac{\sigma _y ^2}{2 \kappa _y} \left[ 1 - e^{-2 \kappa _y (u-t)} \right].
\end{align*}
This establishes Equation~\eqref{eq:analytic_ry_04}.

\begin{figure}[bhtp]
    \begin{center}
        \includegraphics[width=17cm]{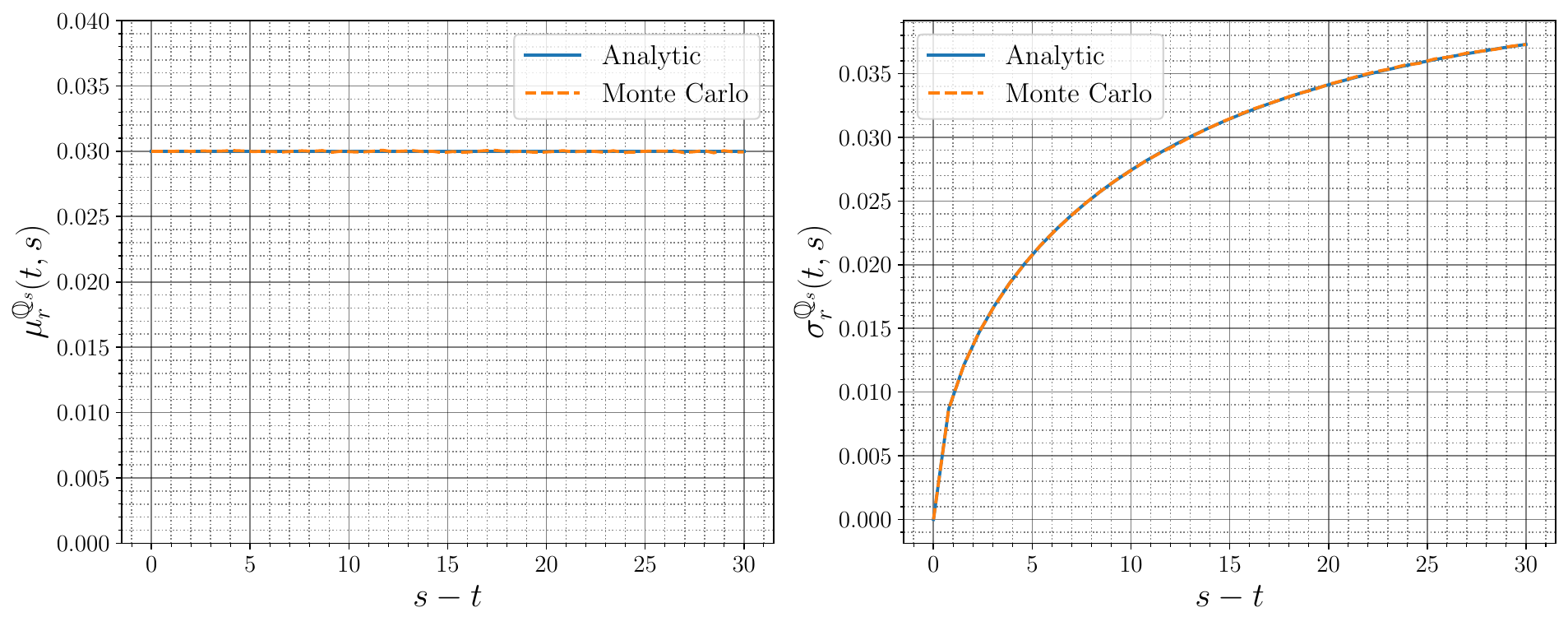}
        \caption{
            Comparison between the analytical and Monte Carlo evaluations of
            $\mu_r^{\mathbb{Q}_s}(t,s)$ and
            $\sigma_r^{\mathbb{Q}_s}(t,s)$.
            The left panel shows the conditional mean
            $\mu_r^{\mathbb{Q}_s}(t,s)$,
            and the right panel shows the conditional standard deviation
            $\sigma_r^{\mathbb{Q}_s}(t,s)$.
            Analytic (solid blue line) represents the results obtained by evaluating
            $\mu_r^{\mathbb{Q}_s}(t,s)$ and
            $\sigma_r^{\mathbb{Q}_s}(t,s)$
            analytically based on
            Equations~\eqref{eq:analytic_ry_01} and
            \eqref{eq:analytic_ry_02}.
            Monte Carlo (dashed orange line) represents the results obtained by evaluating the conditional mean and conditional standard deviation by Monte Carlo simulation, based on the definitions in
            Equations~\eqref{eq:def_r_mean_Qs} and
            \eqref{eq:def_r_sd_Qs},
            where paths of $r(s)$ were generated by discretizing
            Equation~\eqref{eq:r_t_under_Q_s},
            the stochastic differential equation under $\mathbb{Q}_s$,
            using the Euler--Maruyama method.
            The model parameters are given in Table~\ref{tab:base_parameters},
            and the Monte Carlo simulation uses the MC2 computational settings reported in the same table.
        }
        \label{fig:mu_r_Qs_sigma_r_Qs}
    \end{center}
\end{figure}

\begin{figure}[bhtp]
    \begin{center}
        \includegraphics[width=17cm]{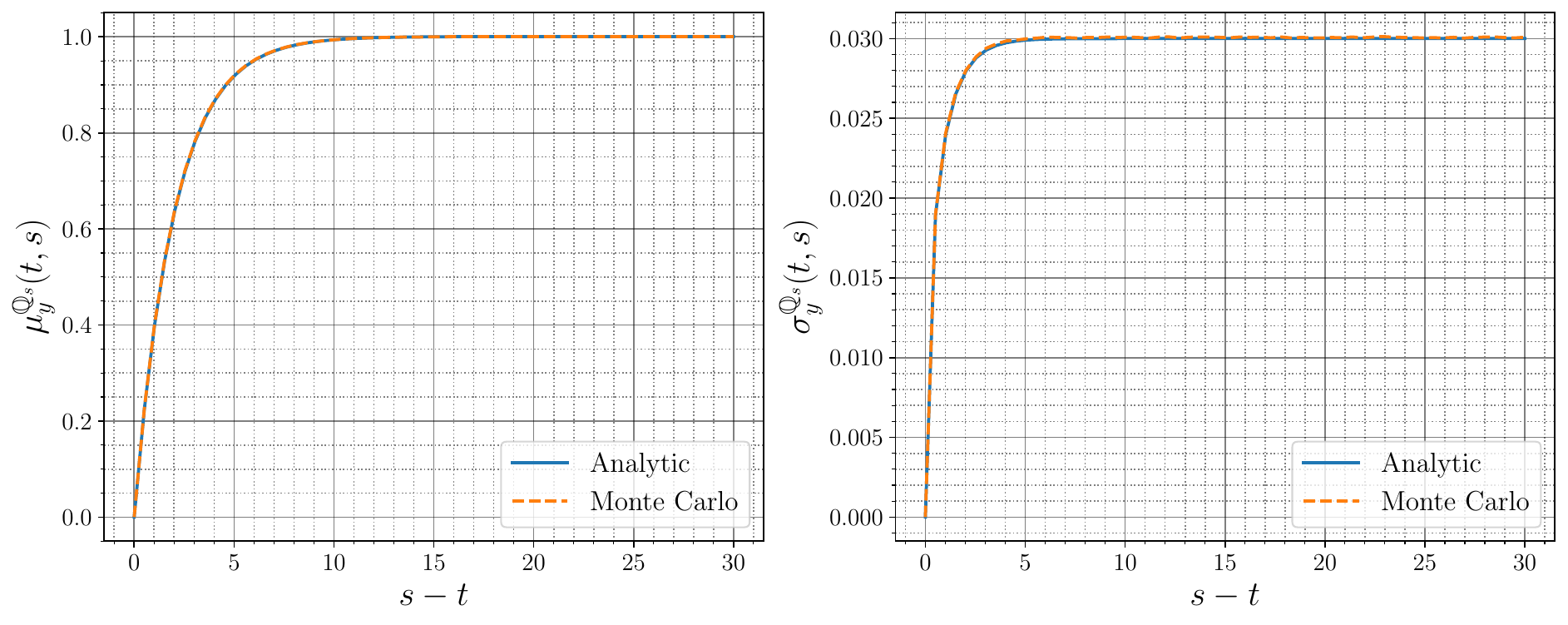}
        \caption{
            Comparison between the analytical and Monte Carlo evaluations of
            $\mu_y^{\mathbb{Q}_s}(t,s)$ and
            $\sigma_y^{\mathbb{Q}_s}(t,s)$.
            The left panel shows the conditional mean
            $\mu_y^{\mathbb{Q}_s}(t,s)$,
            and the right panel shows the conditional standard deviation
            $\sigma_y^{\mathbb{Q}_s}(t,s)$.
            Analytic (solid blue lines) represents the results obtained by analytically evaluating
            $\mu_y^{\mathbb{Q}_s}(t,s)$ and
            $\sigma_y^{\mathbb{Q}_s}(t,s)$
            based on Equations~\eqref{eq:analytic_ry_03} and~\eqref{eq:analytic_ry_04}.
            Monte Carlo (dashed orange lines) represents the results obtained by evaluating the conditional mean and conditional standard deviation by Monte Carlo simulation, based on the definitions in Equations~\eqref{eq:def_y_mean_Qs} and~\eqref{eq:def_y_sd_Qs}, where paths of $y(s)$ are generated by discretizing Equation~\eqref{eq:y_t_under_Q_s}, the stochastic differential equation under $\mathbb{Q}_s$, using the Euler--Maruyama method.
            The model parameters are given in Table~\ref{tab:base_parameters}, and the MC2 computational settings in the same table are used for the Monte Carlo simulation.
        }
        \label{fig:mu_y_Qs_sigma_y_Qs}
    \end{center}
\end{figure}

Figure~\ref{fig:mu_r_Qs_sigma_r_Qs} compares the analytical and Monte Carlo evaluations of
$\mu_r^{\mathbb{Q}_s}(t,s)$ and
$\sigma_r^{\mathbb{Q}_s}(t,s)$.
In addition, Figure~\ref{fig:mu_y_Qs_sigma_y_Qs} compares the analytical and Monte Carlo evaluations of
$\mu_y^{\mathbb{Q}_s}(t,s)$ and
$\sigma_y^{\mathbb{Q}_s}(t,s)$.

\section{Conditional Moments of Prepayment Intensity under the Hull--White Model}

In this section, we derive analytical expressions
for the conditional moments of interest-rate-dependent prepayment intensity
under the Hull--White model
and examine their numerical validity.
In Appendix~\ref{app:lambda_bar_2_prove},
we derive an analytical expression for
$\bar{\lambda}_{\mathrm{r},2}^{\mathbb{Q}_s}(u,\eta)$
using truncated moments of a bivariate normal distribution.
Next, in Appendix~\ref{app:lambda_expectations_numerical_validation},
we compare analytical evaluations with Monte Carlo simulation results
for the two conditional expectations
appearing in the derivation of
$\bar{\lambda}_{\mathrm{r},1}^{\mathbb{Q}_s}(u)$,
as well as for the first cumulant $\kappa_1(t,s)$
and the second cumulant $\kappa_2(t,s)$,
thereby numerically validating the analytical expressions.

\subsection{
    Derivation of an Analytical Expression for
    \texorpdfstring{
        $\bar{\lambda}^{\mathbb{Q}_s}_{\mathrm{r},2}(u,\eta)$
    }{
        lambda-bar-r,2 under Qs
    }
}
\label{app:lambda_bar_2_prove}

In this subsection, we derive an analytical expression for $\bar{\lambda}_{\mathrm{r},2}^{\mathbb{Q}_s}(u,\eta)$.
Specifically, exploiting the joint normality of $r(u)$, $y(u)$, $r(\eta)$, and $y(\eta)$, we evaluate the conditional expectation using truncated moments of a bivariate normal distribution.
As in the main text, we consider the case $t<u<\eta\leq s$.
The boundary cases $u=t$ and $u=\eta$ are treated as continuous limits of the expressions derived below.

First, from Equations~\eqref{eq:r_t_under_Q_s} and \eqref{eq:y_t_under_Q_s}, the deviations of $r(v)$ and $y(v)$ from their conditional means, for $v\in(t,s]$, can be expressed as follows:
\begin{align*}
    r(v)
    -
    \mu_r^{\mathbb{Q}_s}(t,v)
    =&
    \sigma_r
    \int_t^v
    e^{-a_r(v-z)}
    \mathrm{d}W_r^{\mathbb{Q}_s}(z),
    \\
    y(v)
    -
    \mu_y^{\mathbb{Q}_s}(t,v)
    =&
    \sigma_y
    \int_t^v
    e^{-\kappa_y(v-z)}
    \mathrm{d}W_y^{\mathbb{Q}_s}(z).
\end{align*}
Therefore, using the It\^o isometry and Equation~\eqref{eq:cov_r_y}, we obtain the following covariances for $t<u<\eta\leq s$:
\begin{align*}
    \operatorname{Cov}^{\mathbb{Q}_s}
    \left[
        \left.
        r(u),r(\eta)
        \right|
        \mathcal{F}_t
    \right]
    =&
    e^{-a_r(\eta-u)}
    \left\{
        \sigma_r^{\mathbb{Q}_s}(t,u)
    \right\}^2,
    \\
    \operatorname{Cov}^{\mathbb{Q}_s}
    \left[
        \left.
        y(u),y(\eta)
        \right|
        \mathcal{F}_t
    \right]
    =&
    e^{-\kappa_y(\eta-u)}
    \left\{
        \sigma_y^{\mathbb{Q}_s}(t,u)
    \right\}^2,
    \\
    \operatorname{Cov}^{\mathbb{Q}_s}
    \left[
        \left.
        r(u),y(\eta)
        \right|
        \mathcal{F}_t
    \right]
    =&
    e^{-\kappa_y(\eta-u)}
    \operatorname{Cov}^{\mathbb{Q}_s}
    \left[
        \left.
        r(u),y(u)
        \right|
        \mathcal{F}_t
    \right],
    \\
    \operatorname{Cov}^{\mathbb{Q}_s}
    \left[
        \left.
        y(u),r(\eta)
        \right|
        \mathcal{F}_t
    \right]
    =&
    e^{-a_r(\eta-u)}
    \operatorname{Cov}^{\mathbb{Q}_s}
    \left[
        \left.
        y(u),r(u)
        \right|
        \mathcal{F}_t
    \right].
\end{align*}
Using the correlation coefficients introduced in the main text, we thus obtain
\begin{align*}
    \operatorname{Corr}^{\mathbb{Q}_s}
    \left[
        \left.
        X(u),X(\eta)
        \right|
        \mathcal{F}_t
    \right]
    =&
    \rho_{xx}(t,u,\eta),
    \\
    \operatorname{Corr}^{\mathbb{Q}_s}
    \left[
        \left.
        Y(u),Y(\eta)
        \right|
        \mathcal{F}_t
    \right]
    =&
    \rho_{yy}(t,u,\eta),
    \\
    \operatorname{Corr}^{\mathbb{Q}_s}
    \left[
        \left.
        X(u),Y(\eta)
        \right|
        \mathcal{F}_t
    \right]
    =&
    \rho_{xy}(t,u)
    \rho_{yy}(t,u,\eta),
    \\
    \operatorname{Corr}^{\mathbb{Q}_s}
    \left[
        \left.
        Y(u),X(\eta)
        \right|
        \mathcal{F}_t
    \right]
    =&
    \rho_{xy}(t,u)
    \rho_{xx}(t,u,\eta).
\end{align*}
Consequently, the covariance matrix of
$\left.(X(u),Y(u),X(\eta),Y(\eta))^\top\right|_{\mathcal{F}_t}$
coincides with $\bm{\Sigma}$ given in the main text.
We now define
\begin{align*}
    \bm{X}
    &:=
    \begin{pmatrix}
        X(u)
        \\
        X(\eta)
    \end{pmatrix},
    \qquad
    \bm{Y}
    :=
    \begin{pmatrix}
        Y(u)
        \\
        Y(\eta)
    \end{pmatrix}.
\end{align*}
The blocks of the covariance matrix corresponding to $\bm{X}$ and $\bm{Y}$ are given by
\begin{align*}
    \bm{\Sigma}_{XX}
    =&
    \begin{pmatrix}
        1
        &
        \rho_{xx}(t,u,\eta)
        \\
        \rho_{xx}(t,u,\eta)
        &
        1
    \end{pmatrix},
    \\
    \bm{\Sigma}_{YX}
    =&
    \begin{pmatrix}
        \rho_{xy}(t,u)
        &
        \rho_{xy}(t,u)\rho_{xx}(t,u,\eta)
        \\
        \rho_{xy}(t,u)\rho_{yy}(t,u,\eta)
        &
        \rho_{xy}(t,\eta)
    \end{pmatrix}.
\end{align*}
By the formula for the conditional distribution of a multivariate normal distribution, we obtain
\footnote{
    The proof is as follows.
    Let $A := \bm{\Sigma}_{YX}\bm{\Sigma}_{XX}^{-1}$ and $\bm{\varepsilon} := \bm{Y}-A\bm{X}$.
    Then,
    $\mathrm{Cov}^{\mathbb{Q}_s}[\bm{\varepsilon},\bm{X}\mid\mathcal{F}_t]
    =
    \mathrm{Cov}^{\mathbb{Q}_s}[\bm{Y},\bm{X}\mid\mathcal{F}_t]
    -
    A\mathrm{Cov}^{\mathbb{Q}_s}[\bm{X},\bm{X}\mid\mathcal{F}_t]
    =
    \bm{\Sigma}_{YX}
    -
    \bm{\Sigma}_{YX}\bm{\Sigma}_{XX}^{-1}\bm{\Sigma}_{XX}
    =0$.
    Since uncorrelated jointly normally distributed random vectors are independent, $\bm{\varepsilon}$ and $\bm{X}$ are conditionally independent given $\mathcal{F}_t$.
    Therefore,
    $\mathbb{E}^{\mathbb{Q}_s}[\bm{\varepsilon}\mid\mathcal{G}_t]
    =
    \mathbb{E}^{\mathbb{Q}_s}[\bm{\varepsilon}\mid\mathcal{F}_t]
    =0$.
    Consequently,
    $\mathbb{E}^{\mathbb{Q}_s}[\bm{Y}\mid\mathcal{G}_t]
    =
    A\bm{X}
    +
    \mathbb{E}^{\mathbb{Q}_s}[\bm{\varepsilon}\mid\mathcal{G}_t]
    =
    \bm{\Sigma}_{YX}\bm{\Sigma}_{XX}^{-1}\bm{X}$.
}.
\begin{align}
    \mathbb{E}^{\mathbb{Q}_s}
    \left[
        \left.
        \bm{Y}
        \right|
        \mathcal{G}_t
    \right]
    =
    \bm{\Sigma}_{YX}
    \bm{\Sigma}_{XX}^{-1}
    \bm{X}.
    \label{eq:Y_Conditional_expectation}
\end{align}
Here, $\mathcal{G}_t := \mathcal{F}_t \vee \sigma (X(u), X(\eta))$.
Since
\begin{align*}
    \bm{\Sigma}_{XX}^{-1}
    =
    \frac{1}{
        1-\rho_{xx}^2(t,u,\eta)
    }
    \begin{pmatrix}
        1
        &
        -\rho_{xx}(t,u,\eta)
        \\
        -\rho_{xx}(t,u,\eta)
        &
        1
    \end{pmatrix}
\end{align*}
we obtain
\begin{align*}
    \mathbb{E}^{\mathbb{Q}_s}
    \left[
        \left.
        Y(u)
        \right|
        \mathcal{G}_t
    \right]
    =&
    \rho_{xy}(t,u)X(u),
    \\
    \mathbb{E}^{\mathbb{Q}_s}
    \left[
        \left.
        Y(\eta)
        \right|
        \mathcal{G}_t
    \right]
    =&
    \alpha_y(t,u,\eta)X(u)
    +
    \beta_y(t,u,\eta)X(\eta),
\end{align*}
Furthermore, using
$\mathrm{Cov}^{\mathbb{Q}_s}[\bm{Y},\bm{Y}\mid\mathcal{G}_t]
=
\bm{\Sigma}_{YY}
-
\bm{\Sigma}_{YX}\bm{\Sigma}_{XX}^{-1}\bm{\Sigma}_{XY}$
\footnote{
    The proof is as follows.
    We use the same notation as in the proof of Equation~\eqref{eq:Y_Conditional_expectation}.
    Noting that
    $\mathbb{E}^{\mathbb{Q}_s}[\bm{X}\mid\mathcal{G}_t]
    =
    \mathbb{E}^{\mathbb{Q}_s}[\bm{X}\mid\mathcal{F}_t\vee\sigma(X(u),X(\eta))]
    =
    \bm{X}$
    and that $\bm{X}$ and $\bm{\varepsilon}$ are conditionally independent, we use
    $\bm{Y}=A\bm{X}+\bm{\varepsilon}$ to obtain
    $\mathrm{Cov}^{\mathbb{Q}_s}[\bm{Y},\bm{Y}\mid\mathcal{G}_t]
    =
    \mathrm{Cov}^{\mathbb{Q}_s}[\bm{\varepsilon},\bm{\varepsilon}\mid\mathcal{G}_t]
    =
    \mathrm{Cov}^{\mathbb{Q}_s}[\bm{\varepsilon},\bm{\varepsilon}\mid\mathcal{F}_t]$.
    Noting that
    $A^{\top}=\Sigma_{XX}^{-1}\Sigma_{YX}^{\top}$
    and substituting
    $\bm{\varepsilon}=\bm{Y}-A\bm{X}$
    into the last expression yields the desired result.
},
we obtain
\begin{align*}
    &
    \operatorname{Cov}^{\mathbb{Q}_s}
    \left[
        \left.
        Y(u),Y(\eta)
        \right|
        \mathcal{G}_t
    \right]
    \\
    =&
    \rho_{yy}(t,u,\eta)
    -
    \rho_{xy}^2(t,u)
    \rho_{yy}(t,u,\eta)
    \\
    =&
    \left(
        1-\rho_{xy}^2(t,u)
    \right)
    \rho_{yy}(t,u,\eta)
    \\
    =&
    \gamma_y(t,u,\eta).
\end{align*}
Therefore,
\begin{align*}
    &
    \mathbb{E}^{\mathbb{Q}_s}
    \left[
        \left.
        Y(u)Y(\eta)
        \right|
        \bm{X},
        \mathcal{F}_t
    \right]
    \\
    =&
    \gamma_y(t,u,\eta)
    +
    \rho_{xy}(t,u)
    X(u)
    \left\{
        \alpha_y(t,u,\eta)X(u)
        +
        \beta_y(t,u,\eta)X(\eta)
    \right\}.
\end{align*}
Using these results, the conditional expectation of $y(u)y(\eta)$ given $\mathcal{G}_t$ can be expressed as follows:
\begin{align*}
    &
    \mathbb{E}^{\mathbb{Q}_s}
    \left[
        \left.
        y(u)y(\eta)
        \right|
        \mathcal{G}_t
    \right]
    \\
    =&
    C_{00}(u,\eta)
    +
    C_{10}(u,\eta)X(u)
    +
    C_{01}(u,\eta)X(\eta)
    \\
    &
    +
    C_{20}(u,\eta)X(u)^2
    +
    C_{11}(u,\eta)X(u)X(\eta),
\end{align*}
where $C_{00}(u,\eta)$, $C_{10}(u,\eta)$, $C_{01}(u,\eta)$,
$C_{20}(u,\eta)$, and $C_{11}(u,\eta)$ are the coefficients defined in the main text.

On the other hand, using
$\Delta_r^{\mathbb{Q}_s}(t,v)
=
r_{\mathrm{flr}}(v)-\mu_r^{\mathbb{Q}_s}(t,v)$,
as defined in the main text, we obtain
\begin{align*}
    &
    \left(
        r_{\mathrm{flr}}(v)-r(v)
    \right)^+
    \\
    =&
    \left\{
        \Delta_r^{\mathbb{Q}_s}(t,v)
        -
        \sigma_r^{\mathbb{Q}_s}(t,v)X(v)
    \right\}
    \mathbf{1}_{\left\{
        X(v)\leq d(t,v)
    \right\}}.
\end{align*}
Therefore, by the law of iterated expectations, the expectation on the right-hand side of Equation~\eqref{eq:second_order_expectation} can be expressed as follows:
\begin{align*}
    &
    \mathbb{E}^{\mathbb{Q}_s}
    \left[
        \left.
        y(u)y(\eta)
        \left(
            r_{\mathrm{flr}}(u)-r(u)
        \right)^+
        \left(
            r_{\mathrm{flr}}(\eta)-r(\eta)
        \right)^+
        \right|
        \mathcal{F}_t
    \right]
    \\
    =&
    \mathbb{E}^{\mathbb{Q}_s}
    \Bigg[
        \Big\{
            C_{00}(u,\eta)
            +
            C_{10}(u,\eta)X(u)
            +
            C_{01}(u,\eta)X(\eta)
    \\
    &\hspace{30mm}
            +
            C_{20}(u,\eta)X(u)^2
            +
            C_{11}(u,\eta)X(u)X(\eta)
        \Big\}
    \\
    &\hspace{15mm}\times
        \left\{
            \Delta_r^{\mathbb{Q}_s}(t,u)
            -
            \sigma_r^{\mathbb{Q}_s}(t,u)X(u)
        \right\}
    \\
    &\hspace{15mm}\times
        \left\{
            \Delta_r^{\mathbb{Q}_s}(t,\eta)
            -
            \sigma_r^{\mathbb{Q}_s}(t,\eta)X(\eta)
        \right\}
    \\
    &\hspace{15mm}\times
        \mathbf{1}_{\left\{
            X(u)\leq d(t,u)
        \right\}}
        \mathbf{1}_{\left\{
            X(\eta)\leq d(t,\eta)
        \right\}}
        \left|
        \mathcal{F}_t
        \right.
    \Bigg].
\end{align*}

We now present analytical expressions for the truncated moments
$\mathcal{M}_{pq}(u,\eta)$ defined in the main text.
For notational simplicity, we write
\begin{align*}
    d_u
    &:=
    d(t,u),
    \\
    d_{\eta}
    &:=
    d(t,\eta),
    \\
    \rho_x
    &:=
    \rho_{xx}(t,u,\eta),
    \\
    \bar{\rho}_x
    &:=
    \sqrt{
        1-\rho_x^2
    }.
\end{align*}
Let $\Phi_2(\cdot,\cdot;\rho_x)$ and
$\varphi_2(\cdot,\cdot;\rho_x)$ denote the cumulative distribution function and probability density function, respectively, of a standard bivariate normal distribution with correlation coefficient $\rho_x$.
We also define
\begin{align*}
    P
    &:=
    \Phi_2
    \left(
        d_u,
        d_{\eta};
        \rho_x
    \right),
    \\
    K_u
    &:=
    \varphi(d_u)
    \Phi
    \left(
        \frac{
            d_{\eta}-\rho_x d_u
        }{
            \bar{\rho}_x
        }
    \right),
    \\
    K_{\eta}
    &:=
    \varphi(d_{\eta})
    \Phi
    \left(
        \frac{
            d_u-\rho_x d_{\eta}
        }{
            \bar{\rho}_x
        }
    \right),
    \\
    K_{u\eta}
    &:=
    \varphi_2
    \left(
        d_u,
        d_{\eta};
        \rho_x
    \right).
\end{align*}
By repeatedly applying integration by parts to the bivariate normal density, the truncated moments required for the calculations in the main text can be expressed as follows:
\begin{align*}
    \mathcal{M}_{00}
    =&
    P,
    \\
    \mathcal{M}_{10}
    =&
    -K_u
    -
    \rho_x K_{\eta},
    \\
    \mathcal{M}_{01}
    =&
    -K_{\eta}
    -
    \rho_x K_u,
    \\
    \mathcal{M}_{11}
    =&
    \rho_x P
    -
    \rho_x d_u K_u
    -
    \rho_x d_{\eta}K_{\eta}
    +
    \bar{\rho}_x^2 K_{u\eta},
    \\
    \mathcal{M}_{20}
    =&
    P
    -
    d_u K_u
    -
    \rho_x^2 d_{\eta}K_{\eta}
    +
    \rho_x\bar{\rho}_x^2 K_{u\eta},
    \\
    \mathcal{M}_{02}
    =&
    P
    -
    d_{\eta}K_{\eta}
    -
    \rho_x^2 d_u K_u
    +
    \rho_x\bar{\rho}_x^2 K_{u\eta},
    \\
    \mathcal{M}_{21}
    =&
    \mathcal{M}_{01}
    +
    \rho_x\mathcal{M}_{10}
    -
    \rho_x d_u^2K_u
    -
    \rho_x^2d_{\eta}^2K_{\eta}
    \\
    &
    +
    \bar{\rho}_x^2
    \left(
        d_u+\rho_x d_{\eta}
    \right)
    K_{u\eta},
    \\
    \mathcal{M}_{12}
    =&
    \mathcal{M}_{10}
    +
    \rho_x\mathcal{M}_{01}
    -
    \rho_x^2d_u^2K_u
    -
    \rho_x d_{\eta}^2K_{\eta}
    \\
    &
    +
    \bar{\rho}_x^2
    \left(
        \rho_x d_u+d_{\eta}
    \right)
    K_{u\eta},
    \\
    \mathcal{M}_{30}
    =&
    \rho_x\mathcal{M}_{21}
    +
    2\bar{\rho}_x^2\mathcal{M}_{10}
    -
    \bar{\rho}_x^2d_u^2K_u,
    \\
    \mathcal{M}_{22}
    =&
    2\rho_x\mathcal{M}_{11}
    +
    \mathcal{M}_{20}
    -
    \rho_x^2d_u^3K_u
    \\
    &
    -
    d_{\eta}
    \left(
        \rho_x^2d_{\eta}^2
        +
        \bar{\rho}_x^2
    \right)
    K_{\eta}
    \\
    &
    +
    \bar{\rho}_x^2
    \left\{
        \rho_x d_u^2
        +
        d_{\eta}
        \left(
            d_u+\rho_x d_{\eta}
        \right)
    \right\}
    K_{u\eta},
    \\
    \mathcal{M}_{31}
    =&
    \rho_x\mathcal{M}_{22}
    +
    2\bar{\rho}_x^2\mathcal{M}_{11}
    -
    \bar{\rho}_x^2\rho_x d_u^3K_u
    +
    \bar{\rho}_x^4d_u^2K_{u\eta}.
\end{align*}
Here, we have suppressed the arguments $(u,\eta)$ and written
$\mathcal{M}_{pq}=\mathcal{M}_{pq}(u,\eta)$
for notational simplicity.

Expanding the expectation using the truncated moments derived above, we obtain
\begin{align*}
    &
    \mathbb{E}^{\mathbb{Q}_s}
    \left[
        \left.
        y(u)y(\eta)
        \left(
            r_{\mathrm{flr}}(u)-r(u)
        \right)^+
        \left(
            r_{\mathrm{flr}}(\eta)-r(\eta)
        \right)^+
        \right|
        \mathcal{F}_t
    \right]
    \\
    =&
    C_{00}(u,\eta)
    \Big[
        \Delta_r^{\mathbb{Q}_s}(t,u)
        \Delta_r^{\mathbb{Q}_s}(t,\eta)
        \mathcal{M}_{00}(u,\eta)
    \\
    &\hspace{22mm}
        -
        \Delta_r^{\mathbb{Q}_s}(t,u)
        \sigma_r^{\mathbb{Q}_s}(t,\eta)
        \mathcal{M}_{01}(u,\eta)
    \\
    &\hspace{22mm}
        -
        \sigma_r^{\mathbb{Q}_s}(t,u)
        \Delta_r^{\mathbb{Q}_s}(t,\eta)
        \mathcal{M}_{10}(u,\eta)
    \\
    &\hspace{22mm}
        +
        \sigma_r^{\mathbb{Q}_s}(t,u)
        \sigma_r^{\mathbb{Q}_s}(t,\eta)
        \mathcal{M}_{11}(u,\eta)
    \Big]
    \\
    &+
    C_{10}(u,\eta)
    \Big[
        \Delta_r^{\mathbb{Q}_s}(t,u)
        \Delta_r^{\mathbb{Q}_s}(t,\eta)
        \mathcal{M}_{10}(u,\eta)
    \\
    &\hspace{22mm}
        -
        \Delta_r^{\mathbb{Q}_s}(t,u)
        \sigma_r^{\mathbb{Q}_s}(t,\eta)
        \mathcal{M}_{11}(u,\eta)
    \\
    &\hspace{22mm}
        -
        \sigma_r^{\mathbb{Q}_s}(t,u)
        \Delta_r^{\mathbb{Q}_s}(t,\eta)
        \mathcal{M}_{20}(u,\eta)
    \\
    &\hspace{22mm}
        +
        \sigma_r^{\mathbb{Q}_s}(t,u)
        \sigma_r^{\mathbb{Q}_s}(t,\eta)
        \mathcal{M}_{21}(u,\eta)
    \Big]
    \\
    &+
    C_{01}(u,\eta)
    \Big[
        \Delta_r^{\mathbb{Q}_s}(t,u)
        \Delta_r^{\mathbb{Q}_s}(t,\eta)
        \mathcal{M}_{01}(u,\eta)
    \\
    &\hspace{22mm}
        -
        \Delta_r^{\mathbb{Q}_s}(t,u)
        \sigma_r^{\mathbb{Q}_s}(t,\eta)
        \mathcal{M}_{02}(u,\eta)
    \\
    &\hspace{22mm}
        -
        \sigma_r^{\mathbb{Q}_s}(t,u)
        \Delta_r^{\mathbb{Q}_s}(t,\eta)
        \mathcal{M}_{11}(u,\eta)
    \\
    &\hspace{22mm}
        +
        \sigma_r^{\mathbb{Q}_s}(t,u)
        \sigma_r^{\mathbb{Q}_s}(t,\eta)
        \mathcal{M}_{12}(u,\eta)
    \Big]
    \\
    &+
    C_{20}(u,\eta)
    \Big[
        \Delta_r^{\mathbb{Q}_s}(t,u)
        \Delta_r^{\mathbb{Q}_s}(t,\eta)
        \mathcal{M}_{20}(u,\eta)
    \\
    &\hspace{22mm}
        -
        \Delta_r^{\mathbb{Q}_s}(t,u)
        \sigma_r^{\mathbb{Q}_s}(t,\eta)
        \mathcal{M}_{21}(u,\eta)
    \\
    &\hspace{22mm}
        -
        \sigma_r^{\mathbb{Q}_s}(t,u)
        \Delta_r^{\mathbb{Q}_s}(t,\eta)
        \mathcal{M}_{30}(u,\eta)
    \\
    &\hspace{22mm}
        +
        \sigma_r^{\mathbb{Q}_s}(t,u)
        \sigma_r^{\mathbb{Q}_s}(t,\eta)
        \mathcal{M}_{31}(u,\eta)
    \Big]
    \\
    &+
    C_{11}(u,\eta)
    \Big[
        \Delta_r^{\mathbb{Q}_s}(t,u)
        \Delta_r^{\mathbb{Q}_s}(t,\eta)
        \mathcal{M}_{11}(u,\eta)
    \\
    &\hspace{22mm}
        -
        \Delta_r^{\mathbb{Q}_s}(t,u)
        \sigma_r^{\mathbb{Q}_s}(t,\eta)
        \mathcal{M}_{12}(u,\eta)
    \\
    &\hspace{22mm}
        -
        \sigma_r^{\mathbb{Q}_s}(t,u)
        \Delta_r^{\mathbb{Q}_s}(t,\eta)
        \mathcal{M}_{21}(u,\eta)
    \\
    &\hspace{22mm}
        +
        \sigma_r^{\mathbb{Q}_s}(t,u)
        \sigma_r^{\mathbb{Q}_s}(t,\eta)
        \mathcal{M}_{22}(u,\eta)
    \Big].
\end{align*}
Finally, using Equation~\eqref{eq:second_lambda_br_exp}, we obtain Equation~\eqref{eq:second_order_lambda_final_expression}.

\subsection{Numerical Validation of the First and Second Cumulants}
\label{app:lambda_expectations_numerical_validation}

In this subsection, we numerically validate the analytical expressions used to derive the first and second cumulants.
First, we compare the analytical and Monte Carlo evaluations of the two expectations appearing on the right-hand side of Equation~\eqref{eq:lmd_br_Qs_analytic}.
Next, we perform the same comparison for the first cumulant $\kappa_1(t,s)$ and the second cumulant $\kappa_2(t,s)$.

\begin{figure}[bhtp]
    \begin{center}
        \includegraphics[width=17cm]{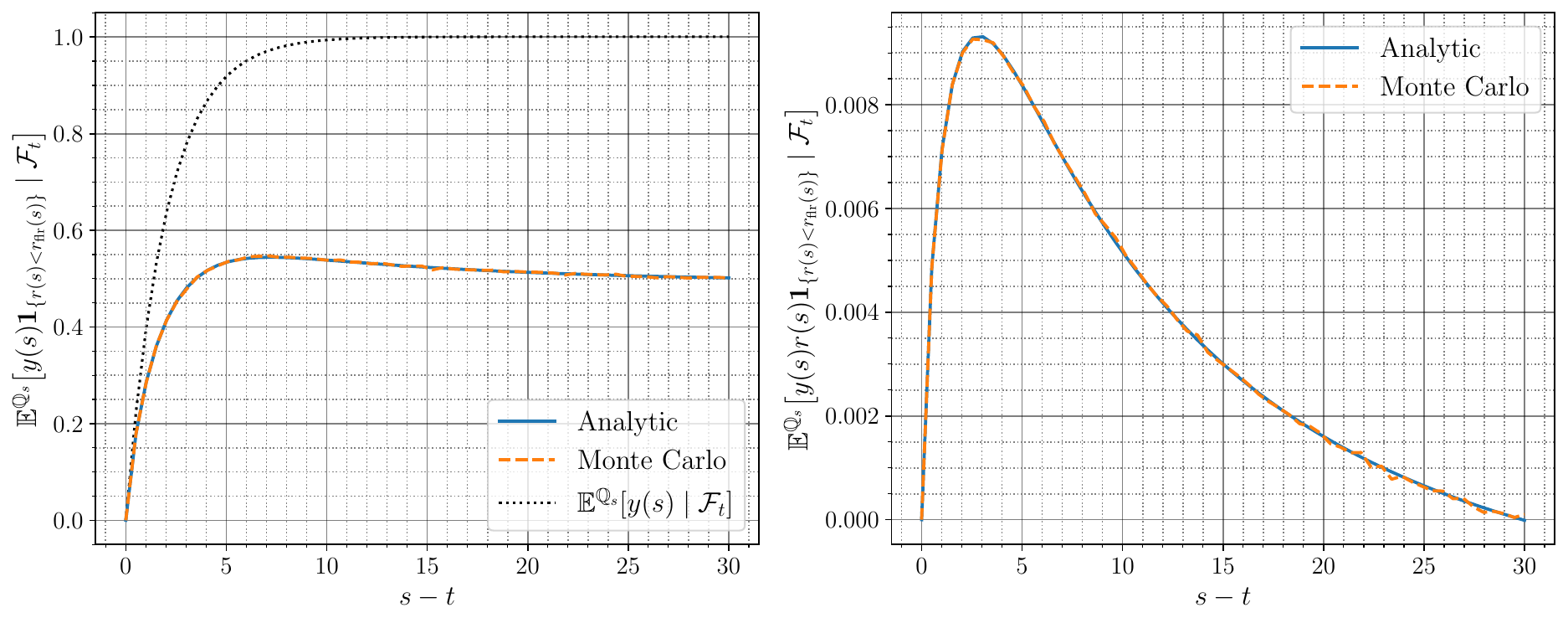}
        \caption{
            Comparison between the analytical and Monte Carlo evaluations of the two expectations appearing on the right-hand side of Equation~\eqref{eq:lmd_br_Qs_analytic}.
            The left panel shows $\mathbb{E}^{\mathbb{Q}_s}\left[\left.y(s)\mathbf{1}_{\{r(s)\leq r_{\mathrm{flr}}(s)\}}\right|\mathcal{F}_t\right]$,
            and the right panel shows $\mathbb{E}^{\mathbb{Q}_s}\left[\left.y(s)r(s)\mathbf{1}_{\{r(s)\leq r_{\mathrm{flr}}(s)\}}\right|\mathcal{F}_t\right]$.
            Analytic (solid blue lines) represents the results obtained by analytically evaluating each expectation using the properties of the joint normal distribution of $r(s)$ and $y(s)$ under $\mathbb{Q}_s$.
            Monte Carlo (dashed orange lines) represents the results obtained by Monte Carlo simulation, where $(r(s),y(s))$ is sampled directly from the joint normal distribution determined by the conditional means and standard deviations given in Equations~\eqref{eq:analytic_ry_01}--\eqref{eq:analytic_ry_04} and the conditional covariance.
            This method does not involve time discretization of the stochastic differential equations.
            The black dotted line in the left panel represents $\mathbb{E}^{\mathbb{Q}_s}\left[\left.y(s)\right|\mathcal{F}_t\right]$, evaluated analytically using Equation~\eqref{eq:analytic_ry_03}.
            The model parameters and the number of Monte Carlo simulation paths are given in Table~\ref{tab:base_parameters}.
        }
        \label{fig:two_type_expectation}
    \end{center}
\end{figure}

Figure~\ref{fig:two_type_expectation} presents the results of evaluating these expectations as functions of time to maturity $s-t$.
As shown in Figure~\ref{fig:two_type_expectation}, the analytical and Monte Carlo evaluations are in close agreement for both expectations, supporting the validity of the analytical expressions.
Furthermore, comparison with the black dotted line in the left panel shows that the indicator function restricts the expectation to the region where $r(s)\leq r_{\mathrm{flr}}(s)$, thereby reducing its value.

\begin{figure}[bhtp]
    \begin{center}
        \includegraphics[width=17cm]{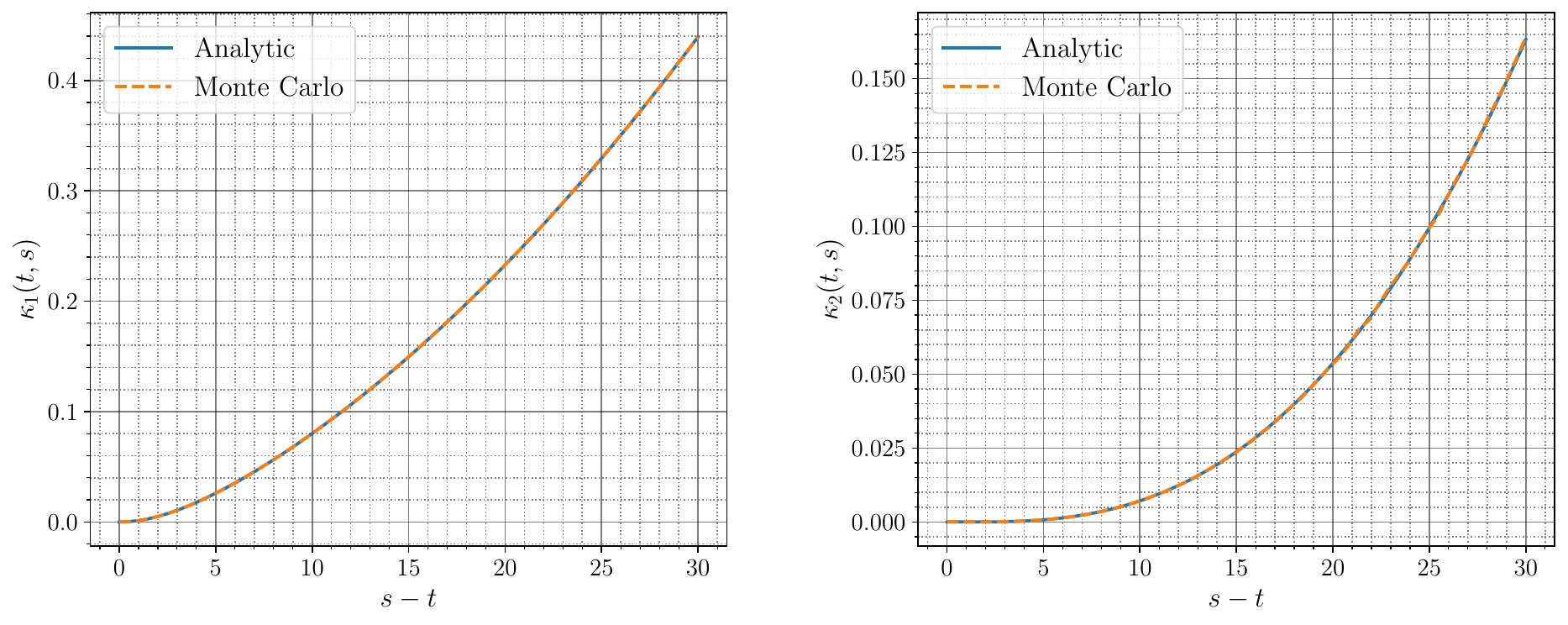}
        \caption{
            Comparison between the analytical and Monte Carlo evaluations of the first cumulant $\kappa_1(t,s)$ and the second cumulant $\kappa_2(t,s)$.
            The left panel shows $\kappa_1(t,s)$, and the right panel shows $\kappa_2(t,s)$.
            Analytic (solid blue lines) represents the results obtained by numerically integrating the analytical expressions derived using the properties of the joint normal distribution of the interest rate and prepayment sensitivity processes under $\mathbb{Q}_s$.
            Monte Carlo (dashed orange lines) represents the results obtained by evaluating each cumulant through Monte Carlo simulation, where random variables are sampled directly from the conditional normal distributions under $\mathbb{Q}_s$.
            This method does not involve time discretization of the stochastic differential equations.
            For both evaluations, the trapezoidal rule is used for numerical integration over time, with the number of integration points chosen so that the step size does not exceed $1/12$ year.
            The model parameters and the number of Monte Carlo simulation paths are given in Table~\ref{tab:base_parameters}.
        }
        \label{fig:kappa_1_kappa_2_Qs_exact_mc}
    \end{center}
\end{figure}

Next, we compare the analytical and Monte Carlo evaluations of the first cumulant $\kappa_1(t,s)$ and the second cumulant $\kappa_2(t,s)$ to numerically validate the analytical expressions.
Figure~\ref{fig:kappa_1_kappa_2_Qs_exact_mc} presents the results of evaluating $\kappa_1(t,s)$ and $\kappa_2(t,s)$ as functions of time to maturity $s-t$.
Figure~\ref{fig:kappa_1_kappa_2_Qs_exact_mc} compares the analytical and Monte Carlo evaluations of the first and second cumulants.

\section{Pathwise Monte Carlo Evaluation of MBS Sensitivities}
\label{app:pathwise_greeks}

In this appendix, we present a method for evaluating MBS interest rate sensitivities using Monte Carlo simulation.
In particular, we consider the Pathwise method, which evaluates sensitivities to parallel shifts in the initial zero-yield curve along individual simulation paths, without relying on finite differences of prices at different interest rate levels.
In Appendix~\ref{app:pathwise_duration}, we examine how parallel shifts in the initial zero-yield curve affect the interest rate process and the prepayment rate, and derive a method for evaluating the first-order interest rate sensitivity of MBS prices using the Pathwise method.
In Appendix~\ref{app:pathwise_convexity}, we extend the Pathwise method to second-order interest rate sensitivities and show how to evaluate MBS convexity without finite differences by accounting for contributions from the nondifferentiable boundary arising from the nonlinearity of the prepayment rate using conditional Monte Carlo simulation.

For methods of evaluating price sensitivities using simulation, see \textcite{Broadie_Glasserman:1996}.
As in the numerical analysis in the main text, we set the valuation time to $t=0$ and assume an initial discount curve given by $D(0,u)=e^{-Ru}$.
We also hold all model parameters other than $R$, as well as the initial states $y(0)$ and $\lambda_{\mathrm{c}}(0)$, fixed.

\subsection{Pathwise Evaluation of Duration}
\label{app:pathwise_duration}

First, we examine how a parallel shift in $R$ affects future interest rate paths.
When the initial zero-yield curve is flat, Equation~\eqref{eq:vartheta} implies that the drift term of the Hull--White model is given by
\begin{align*}
    \vartheta(u)
    =
    a_r R
    +
    \frac{\sigma_r^2}{2a_r}
    \left(1-e^{-2a_ru}\right)
\end{align*}
To make the dependence on $R$ explicit, we denote the corresponding short rate by $r_R(u)$ and define
\begin{align*}
    x(u):=r_R(u)-R.
\end{align*}
From $r_R(0)=R$ and Equation~\eqref{eq:HW}, we obtain
\begin{align*}
    \mathrm{d}x(u)
    =&
    \left\{
        \frac{\sigma_r^2}{2a_r}
        \left(1-e^{-2a_ru}\right)
        -a_rx(u)
    \right\}\mathrm{d}u
    +\sigma_r\mathrm{d}W_r^{\mathbb{Q}}(u),
    \qquad x(0)=0.
\end{align*}
Since this stochastic differential equation does not depend on $R$, constructing the paths corresponding to different values of $R$ using the same Brownian motion yields
\begin{align*}
    \frac{\partial r_R(u)}{\partial R}=1
\end{align*}
for each simulation path.

Next, using Equations~\eqref{eq:D_sol_HW_01}--\eqref{eq:D_sol_HW_03}, the 10-year zero yield at time $u$ can be expressed as
\begin{align*}
    R_{10,R}(u)
    =&
    R
    +\frac{B_{r,10}}{10}x(u)
    +\frac{\sigma_r^2 B_{r,10}^2}{40a_r}
      \left(1-e^{-2a_ru}\right)
\end{align*}
\begin{align}
    \frac{\partial R_{10,R}(u)}{\partial R}
    =&1
    \label{eq:pathwise_R10_derivative}
\end{align}
where $B_{r,10}=B_r(10)$.
This result implies that when the initial zero-yield curve is shifted in parallel from $R$ to $R+\delta R$, the entire path of the 10-year zero yield generated using the same Brownian motion also shifts by $\delta R$.
Note that $r_{\mathrm{flr}}(u)$, defined in Equation~\eqref{eq:lmd_r_written_on_rt_02}, also changes with $R$.
Therefore, the strike is not held fixed when taking derivatives with respect to $R$.

We denote the interest-rate-dependent prepayment rate by
\begin{align*}
    \lambda_{\mathrm{r},R}(u)
    :=
    y(u)\left(m-a-bR_{10,R}(u)\right)^+
\end{align*}
From Equation~\eqref{eq:pathwise_R10_derivative}, for paths and time points satisfying $m-a-bR_{10,R}(u)\neq0$, we obtain
\begin{align}
    \frac{\partial\lambda_{\mathrm{r},R}(u)}{\partial R}
    =&
    -b\,y(u)
    \mathbf{1}_{\{m-a-bR_{10,R}(u)>0\}}
    \label{eq:pathwise_lambda_derivative}
\end{align}
For any fixed $u>0$, $R_{10,R}(u)$ follows a nondegenerate normal distribution.
Therefore, under $b\neq0$, the probability that $m-a-bR_{10,R}(u)=0$ is zero.
Consequently, $\lambda_{\mathrm{r},R}(u)$ is almost surely differentiable with respect to $R$, and its derivative is given by Equation~\eqref{eq:pathwise_lambda_derivative}.
In the following, we assume the integrability conditions necessary to interchange differentiation with expectation and time integration.

For $Z_{\mathrm{r}}(0,s)$, we can similarly differentiate the integrand in Equation~\eqref{eq:Z_closed_form_03} pathwise for fixed $s$.
Indeed, substituting $r_R(u)=R+x(u)$ into Equations~\eqref{eq:r_t_under_Q_s} and \eqref{eq:y_t_under_Q_s} shows that the dynamics of $x(u)$ and $y(u)$ under $\mathbb{Q}_s$ are also independent of $R$.
Thus, we obtain
\begin{align*}
    \frac{\partial Z_{\mathrm{r}}(0,s)}{\partial R}
    ={}&
    b\,\mathbb{E}^{\mathbb{Q}_s}
    \left[
        e^{-\int_0^s\lambda_{\mathrm{r},R}(u)\mathrm{d}u}
        \int_0^s
        y(u)\mathbf{1}_{\{m-a-bR_{10,R}(u)>0\}}
        \mathrm{d}u
    \right].
\end{align*}

Next, we derive the duration of the overall MBS price.
Using Equations~\eqref{eq:MBS_value_result_01} and \eqref{eq:Z_r_mc_05}, and noting the independence of $\lambda_{\mathrm{c}}$ from $r(t)$ and $y(t)$, the MBS price at $t=0$ can be expressed as
\begin{align*}
    V(0,T)
    ={}&
    \int_0^T D(0,s)Z_{\mathrm{c}}(0,s)
    \mathbb{E}^{\mathbb{Q}_s}
    \left[
        e^{-\int_0^s\lambda_{\mathrm{r}}(u)\mathrm{d}u}
        \left\{
            c+\tilde{P}(s)
            \bigl(\bar{\lambda}_{\mathrm{c}}(0,s)
            +\lambda_{\mathrm{r}}(s)\bigr)
        \right\}
    \right]\mathrm{d}s
\end{align*}
In the following, for notational simplicity, we omit the subscript $R$ from $r_R(u)$, $R_{10,R}(u)$, and $\lambda_{\mathrm{r},R}(u)$, and define
\begin{align*}
    A(s)
    :=&
    \int_0^s\lambda_{\mathrm{r}}(u)\mathrm{d}u,
    \\
    J(s)
    :=&
    \int_0^s
    y(u)\mathbf{1}_{\{m-a-bR_{10}(u)>0\}}
    \mathrm{d}u,
    \\
    G(s)
    :=&
    c+\tilde{P}(s)
    \bigl(\bar{\lambda}_{\mathrm{c}}(0,s)+\lambda_{\mathrm{r}}(s)\bigr)
\end{align*}
Since $D(0,s)=e^{-Rs}$ and $\bar{\lambda}_{\mathrm{c}}(0,s)$ does not depend on $R$, we have
\begin{align*}
    \frac{\partial}{\partial R}
    \left\{D(0,s)e^{-A(s)}\right\}
    =&
    D(0,s)e^{-A(s)}\left(-s+bJ(s)\right),
    \\
    \frac{\partial G(s)}{\partial R}
    =&
    -b\tilde{P}(s)y(s)
    \mathbf{1}_{\{m-a-bR_{10}(s)>0\}}
\end{align*}
Therefore, applying the product rule yields the following expression for the first derivative of the MBS price with respect to $R$:
\begin{align}
    \frac{\partial V}{\partial R}(0,T)
    ={}&
    \int_0^T D(0,s)Z_{\mathrm{c}}(0,s)
    \mathbb{E}^{\mathbb{Q}_s}
    \Bigg[
        e^{-A(s)}
        \Bigg\{
            \left(-s+bJ(s)\right)G(s)
    \nonumber\\
    &\hspace{32mm}
            -b\tilde{P}(s)y(s)
            \mathbf{1}_{\{m-a-bR_{10}(s)>0\}}
        \Bigg\}
    \Bigg]\mathrm{d}s.
    \label{eq:pathwise_value_derivative}
\end{align}
In this expression, each term inside the expectation on the right-hand side can be evaluated along individual simulation paths.
The term $-sG(s)$ corresponds to the effect of a parallel shift in discount rates, $bJ(s)G(s)$ captures the change in the outstanding principal fraction resulting from changes in prepayments, and the final term represents the change in prepaid principal payments at time $s$.

Thus, the duration can be evaluated as
\begin{align*}
    \mathcal{D}(0,T)
    =&
    -\frac{\frac{\partial V}{\partial R}(0,T)}{V(0,T)}
\end{align*}
The right-hand side of Equation~\eqref{eq:pathwise_value_derivative} can be evaluated directly from simulated paths, without using finite differences to approximate the first derivative of the MBS price with respect to $R$.

\subsection{Conditional Monte Carlo Evaluation of Convexity}
\label{app:pathwise_convexity}

In this subsection, we extend the Pathwise method developed in the preceding subsection to second-order interest rate sensitivities and present a method for evaluating MBS convexity without using finite differences.
However, the interest-rate-dependent prepayment rate $\lambda_{\mathrm{r}}(u)=y(u)(m-a-bR_{10}(u))^+$ is not differentiable at the point where the refinancing incentive is zero.
Consequently, simply computing ordinary second derivatives along individual simulation paths would omit the contribution arising from this boundary.
In the following, we express the boundary contribution in terms of a conditional expectation and evaluate it using Monte Carlo simulation, exploiting the Gaussian properties of the Hull--White model.
As in the preceding subsection, we set $t=0$ and $D(0,s)=e^{-Rs}$, and assume $b>0$ and $\sigma_r>0$.

First, for notational simplicity, in addition to $A(s)$, $J(s)$, and $G(s)$ defined in the preceding subsection, we define
\begin{align*}
    I(u)
    :=&
    \mathbf{1}_{\{m-a-bR_{10}(u)>0\}},
    \\
    \Gamma(s)
    :=&
    -s+bJ(s)
\end{align*}
Using Equation~\eqref{eq:pathwise_lambda_derivative}, we obtain
\begin{align*}
    \frac{\partial A(s)}{\partial R}
    =&-bJ(s),
    \\
    \frac{\partial G(s)}{\partial R}
    =&-b\tilde{P}(s)y(s)I(s)
\end{align*}
On the other hand, the second derivative of the positive-part function can be expressed in the distributional sense using the Dirac delta function $\delta_0$.
Specifically,
\begin{align*}
    \frac{\partial^2 A(s)}{\partial R^2}
    =&
    b^2\int_0^s
    y(u)\delta_0\!\left(m-a-bR_{10}(u)\right)
    \mathrm{d}u,
    \\
    \frac{\partial^2 G(s)}{\partial R^2}
    =&
    b^2\tilde{P}(s)y(s)
    \delta_0\!\left(m-a-bR_{10}(s)\right)
\end{align*}
Therefore, differentiating $D(0,s)e^{-A(s)}G(s)$ twice with respect to $R$ yields
\begin{align*}
    &\frac{\partial^2}{\partial R^2}
    \left\{D(0,s)e^{-A(s)}G(s)\right\}
    \nonumber\\
    &\quad=
    D(0,s)e^{-A(s)}
    \Bigg[
        \Gamma(s)^2G(s)
        -2b\Gamma(s)\tilde{P}(s)y(s)I(s)
    \nonumber\\
    &\qquad\qquad
        +b^2\tilde{P}(s)y(s)
        \delta_0\!\left(m-a-bR_{10}(s)\right)
    \nonumber\\
    &\qquad\qquad
        -b^2G(s)\int_0^s
        y(u)\delta_0\!\left(m-a-bR_{10}(u)\right)
        \mathrm{d}u
    \Bigg].
\end{align*}
This expression is formal because it involves Dirac delta functions.
In the following, we evaluate these terms using conditional probability densities after taking expectations.

For fixed $s>0$, let $0<u\leq s$.
Under $\mathbb{Q}_s$, we define $p_s(u)$ as the probability density of $m-a-bR_{10}(u)$ evaluated at zero.
From Equations~\eqref{eq:lmd_r_written_on_rt} and \eqref{eq:lmd_r_written_on_rt_02}, we have
\begin{align*}
    m-a-bR_{10}(u)
    =&
    \frac{bB_{r,10}}{10}
    \left(r_{\mathrm{flr}}(u)-r(u)\right)
\end{align*}
Since $r(u)$ follows a normal distribution under $\mathbb{Q}_s$, $p_s(u)$ admits the following analytical expression:
\begin{align}
    p_s(u)
    =&
    \frac{10}{bB_{r,10}\sigma_r^{\mathbb{Q}_s}(0,u)}
    \varphi\!\left(d_s(u)\right),
    \nonumber\\
    \text{where}\quad
    d_s(u)
    :=&
    \frac{
        r_{\mathrm{flr}}(u)-\mu_r^{\mathbb{Q}_s}(0,u)
    }{
        \sigma_r^{\mathbb{Q}_s}(0,u)
    }.
    \label{eq:pathwise_boundary_density}
\end{align}
Here, $\varphi$ denotes the standard normal probability density function, while $\mu_r^{\mathbb{Q}_s}$ and $\sigma_r^{\mathbb{Q}_s}$ are the conditional mean and standard deviation defined in the main text.
Equation~\eqref{eq:pathwise_boundary_density} provides an analytical expression for the probability density at the boundary used in the conditional Monte Carlo method.

Using the properties of the Dirac delta function, the second derivative of the MBS price can be expressed as
\begin{align}
    \frac{\partial^2V}{\partial R^2}(0,T)
    ={}&
    \int_0^T D(0,s)Z_{\mathrm{c}}(0,s)
    \Bigg\{
        \mathbb{E}^{\mathbb{Q}_s}
        \left[
            e^{-A(s)}
            \left\{
                \Gamma(s)^2G(s)
                -2b\Gamma(s)\tilde{P}(s)y(s)I(s)
            \right\}
        \right]
    \nonumber\\
    &\quad+
        b^2\tilde{P}(s)p_s(s)
        \mathbb{E}^{\mathbb{Q}_s}
        \left[
            e^{-A(s)}y(s)
            \mid r(s)=r_{\mathrm{flr}}(s)
        \right]
    \nonumber\\
    &\quad-
        b^2\int_0^s p_s(u)
        \mathbb{E}^{\mathbb{Q}_s}
        \left[
            e^{-A(s)}G(s)y(u)
            \mid r(u)=r_{\mathrm{flr}}(u)
        \right]
        \mathrm{d}u
    \Bigg\}\mathrm{d}s.
    \label{eq:pathwise_convexity_value_derivative}
\end{align}
The first term on the right-hand side corresponds to ordinary pathwise differentiation, while the second and third terms represent contributions arising from the nondifferentiable boundary of the prepayment rate.
In particular, the second term corresponds to changes in prepaid principal payments at time $s$, while the third term corresponds to changes in the outstanding principal fraction resulting from prepayments up to time $s$.

Next, to evaluate the conditional expectations in Equation~\eqref{eq:pathwise_convexity_value_derivative}, we describe a method for generating conditional Gaussian paths.
For fixed $s$, we generate paths of $r(w)$ and $y(w)$ under $\mathbb{Q}_s$ and define the following for $v\in(0,s]$:
\begin{align}
    r^{[v]}(w)
    :=&
    r(w)
    +\frac{
        \operatorname{Cov}^{\mathbb{Q}_s}[r(w),r(v)]
    }{
        \left(\sigma_r^{\mathbb{Q}_s}(0,v)\right)^2
    }
    \left(r_{\mathrm{flr}}(v)-r(v)\right),
    \nonumber\\
    y^{[v]}(w)
    :=&
    y(w)
    +\frac{
        \operatorname{Cov}^{\mathbb{Q}_s}[y(w),r(v)]
    }{
        \left(\sigma_r^{\mathbb{Q}_s}(0,v)\right)^2
    }
    \left(r_{\mathrm{flr}}(v)-r(v)\right).
    \label{eq:pathwise_conditional_bridge}
\end{align}
The processes $r^{[v]}(w)$ and $y^{[v]}(w)$ have the same finite-dimensional distributions as the processes $r(w)$ and $y(w)$ conditional on $r(v)=r_{\mathrm{flr}}(v)$.
Letting $h:=\min(w,v)$, the required covariances are given by
\begin{align*}
    \operatorname{Cov}^{\mathbb{Q}_s}[r(w),r(v)]
    =&
    \frac{\sigma_r^2}{2a_r}
    e^{-a_r(w+v-2h)}
    \left(1-e^{-2a_rh}\right),
    \\
    \operatorname{Cov}^{\mathbb{Q}_s}[y(w),r(v)]
    =&
    \frac{\rho\sigma_y\sigma_r}{\kappa_y+a_r}
    e^{-\kappa_y(w-h)-a_r(v-h)}
    \left(1-e^{-(\kappa_y+a_r)h}\right)
\end{align*}
These covariances do not depend on the deterministic drift adjustments arising from the change to the forward measure.
Therefore, Equation~\eqref{eq:pathwise_conditional_bridge} allows us to generate conditional paths simply by transforming ordinary Gaussian paths.
In the following, we use the superscript $[v]$ to denote quantities calculated from these conditional paths.
For example, we define
\begin{align*}
    A^{[v]}(s)
    :=&
    \int_0^s
    y^{[v]}(w)
    \left(m-a-bR_{10}^{[v]}(w)\right)^+
    \mathrm{d}w,
    \\
    G^{[v]}(s)
    :=&
    c+\tilde{P}(s)
    \left(
        \bar{\lambda}_{\mathrm{c}}(0,s)
        +y^{[v]}(s)
        \left(m-a-bR_{10}^{[v]}(s)\right)^+
    \right)
\end{align*}
where $R_{10}^{[v]}(w)$ denotes the 10-year zero yield calculated using the conditional short rate $r^{[v]}(w)$.

In this case, the finite-dimensional distributions of the conditional paths coincide with those of the original paths conditional on $r(v)=r_{\mathrm{flr}}(v)$.
Therefore, for example,
\begin{align*}
    &\mathbb{E}^{\mathbb{Q}_s}
    \left[
        e^{-A(s)}G(s)y(v)
        \mid r(v)=r_{\mathrm{flr}}(v)
    \right]
    \\
    &\quad=
    \mathbb{E}^{\mathbb{Q}_s}
    \left[
        e^{-A^{[v]}(s)}G^{[v]}(s)y^{[v]}(v)
    \right]
\end{align*}
This allows us to evaluate the boundary contributions in Equation~\eqref{eq:pathwise_convexity_value_derivative} as ordinary expectations of functionals of the conditional Gaussian paths.
Thus, the convexity can be evaluated as
\begin{align*}
    \mathcal{C}(0,T)
    =&
    \frac{\frac{\partial^2V}{\partial R^2}(0,T)}{V(0,T)}
\end{align*}
This method allows us to evaluate the second-order interest rate sensitivity without using finite differences of MBS prices at different interest rate levels, while accounting for the boundary contributions arising from the nondifferentiability of the prepayment rate.

\section{Closed-Form Approximations for MBS Duration and Convexity}
\label{app:mbs_duration_convexity_closed_form}

In this section, we derive analytical expressions for MBS duration and convexity
based on the closed-form approximation for MBS prices
developed in Section~\ref{sec:mbs_approximation}.
Specifically, by taking the first and second derivatives
of the MBS price $V^{(2)}(t,T)$,
obtained using the cumulant expansion truncated at the second cumulant,
with respect to the interest rate level $R$,
we analytically evaluate $\mathcal{D}^{(2)}(t,T)$
and $\mathcal{C}^{(2)}(t,T)$,
as defined in Equations~\eqref{eq:duration_dcmp}
and \eqref{eq:convexity_dcmp}, respectively.

As in the numerical analysis in the main text,
we assume that the initial zero-yield curve is flat
and that $D(0,s)=e^{-Rs}$.
When differentiating with respect to the interest rate level $R$,
we hold the mortgage rate $m$ and all other model parameters fixed
and apply a parallel shift to the entire initial zero-yield curve.
Under this setting, not only the discount factors
but also the future interest rate distributions implied by the Hull--White model
and the interest-rate-dependent prepayment rates depend on $R$.
Therefore, the following derivations account for
both the interest rate sensitivity of the discount factors
and that of future cash flows through the prepayment process.

In Appendix~\ref{app:mbs_duration_closed_form},
we derive an analytical expression for duration $\mathcal{D}^{(2)}(t,T)$
by taking the first derivatives of the individual price components
in Equation~\eqref{eq:mbs_closed_form_02} with respect to $R$.
Next, in Appendix~\ref{app:mbs_convexity_closed_form},
we derive an analytical expression for convexity $\mathcal{C}^{(2)}(t,T)$
by evaluating the second derivatives of these price components.
In both cases, we analytically evaluate the interest rate sensitivities
of the first and second cumulants associated with interest-rate-dependent prepayments,
allowing the resulting sensitivities to be expressed
in terms of deterministic time integrals.

\subsection{Derivation of a Closed-Form Approximation for MBS Duration}
\label{app:mbs_duration_closed_form}

In this subsection, we derive a closed-form approximation for MBS duration
by differentiating the second-order cumulant approximation
$V^{(2)}(t,T)$ in Equation~\eqref{eq:mbs_closed_form_02}
with respect to the interest rate level $R$.
As in the numerical analysis in the main text,
we assume a flat initial zero-yield curve,
with $r(t)=R$ and $D(t,s)=e^{-R(s-t)}$.
When differentiating, we hold the mortgage rate $m$
and all other model parameters fixed.
Under these assumptions,
$\partial_R D(t,s)=-(s-t)D(t,s)$,
while $Z_{\mathrm{c}}(t,s)$ and $\bar c_0(t,s)$
are independent of $R$.

First, for the conditional means under the forward measure
and the refinancing incentive, we have
\begin{align*}
    \partial_R\mu_r^{\mathbb{Q}_s}(t,u)=1,\quad
    \partial_R\mu_y^{\mathbb{Q}_s}(t,u)=0,\\
    \partial_R r_{\mathrm{flr}}(u)
    =1-\frac{10}{B_{r,10}},\quad
    \partial_R\Delta_r^{\mathbb{Q}_s}(t,u)
    =-\frac{10}{B_{r,10}}
\end{align*}
Since the conditional variances and covariances
are independent of $R$,
the derivative of the integrand of the first cumulant is given by
\begin{align}
    \partial_R\bar\lambda_{\mathrm{r},1}^{\mathbb{Q}_s}(u)
    =-b\left[
        \mu_y^{\mathbb{Q}_s}(t,u)\Phi(d(t,u))
        -\frac{\operatorname{Cov}^{\mathbb{Q}_s}
          [r(u),y(u)\mid\mathcal F_t]}
          {\sigma_r^{\mathbb{Q}_s}(t,u)}\varphi(d(t,u))
    \right].
    \label{eq:app_duration_lambda1_derivative}
\end{align}
Similarly, differentiating Equation~\eqref{eq:second_order_lambda_final_expression}
yields the following expression for the derivative
of the integrand of the second cumulant:
\begin{align}
    &\partial_R\bar\lambda_{\mathrm{r},2}^{\mathbb{Q}_s}(u,\eta)
    \nonumber\\
    =&-\frac{b^2B_{r,10}}{10}
    \sum_{(p,q)\in\mathcal I}C_{pq}(u,\eta)
    \Big[
        (\Delta_u+\Delta_\eta)\mathcal M_{pq}(u,\eta)
    \nonumber\\
    &\qquad
        -\sigma_u\mathcal M_{p+1,q}(u,\eta)
        -\sigma_\eta\mathcal M_{p,q+1}(u,\eta)
    \Big]
    \label{eq:app_duration_lambda2_derivative}
\end{align}
Here, $\mathcal I$, $C_{pq}$, $\Delta_v$, $\sigma_v$,
and the truncated moments $\mathcal M_{pq}$
are defined in the derivation of the second cumulant in the main text.
These coefficients and conditional covariances are independent of $R$,
and the differentiation accounts for changes in
$\Delta_u$, $\Delta_\eta$, and the truncation region.

The derivatives of the first and second cumulants
are therefore given by
\begin{align}
    \partial_R\kappa_1(t,s)
    =&\int_t^s\partial_R\bar\lambda_{\mathrm{r},1}^{\mathbb{Q}_s}(u)\,\mathrm{d}u,
    \nonumber\\
    \partial_R\kappa_2(t,s)
    =&2\int_t^s\int_u^s\Big[
        \partial_R\bar\lambda_{\mathrm{r},2}^{\mathbb{Q}_s}(u,\eta)
    \nonumber\\
    &\qquad-\partial_R\left\{
        \bar\lambda_{\mathrm{r},1}^{\mathbb{Q}_s}(u)
        \bar\lambda_{\mathrm{r},1}^{\mathbb{Q}_s}(\eta)
        \right\}\Big] \,\mathrm{d}\eta\mathrm{d}u.
    \nonumber
\end{align}
We define
\begin{align}
    H^{(2)}(t,s)
    :=&\partial_R\kappa_1(t,s)
        -\frac12\partial_R\kappa_2(t,s),
    \label{eq:app_duration_H_def}\\
    \partial_R Z_{\mathrm{r}}^{(2)}(t,s)
    =&-H^{(2)}(t,s)Z_{\mathrm{r}}^{(2)}(t,s)
    \nonumber
\end{align}

Next, we derive the interest rate sensitivity
of the cash flows arising from interest-rate-dependent prepayments.
The effective prepayment rate under the second-order cumulant approximation
is given by
\begin{align}
    \bar\lambda_{\mathrm{r}}^{(2)}(t,s)
    :=&\bar\lambda_{\mathrm{r},1}^{\mathbb{Q}_s}(s)
    -\int_t^s\Big[
        \bar\lambda_{\mathrm{r},2}^{\mathbb{Q}_s}(u,s)
    \nonumber\\
    &\qquad-\bar\lambda_{\mathrm{r},1}^{\mathbb{Q}_s}(u)
        \bar\lambda_{\mathrm{r},1}^{\mathbb{Q}_s}(s)
    \Big] \,\mathrm{d}u
    \label{eq:app_duration_lambda_eff}
\end{align}
where this expression is obtained
by holding the $s$-forward measure fixed.
Consequently,
\begin{align}
    \partial_R\bar\lambda_{\mathrm{r}}^{(2)}(t,s)
    =&\partial_R\bar\lambda_{\mathrm{r},1}^{\mathbb{Q}_s}(s)
    -\int_t^s\Big[
        \partial_R\bar\lambda_{\mathrm{r},2}^{\mathbb{Q}_s}(u,s)
    \nonumber\\
    &\qquad-\partial_R\left\{
        \bar\lambda_{\mathrm{r},1}^{\mathbb{Q}_s}(u)
        \bar\lambda_{\mathrm{r},1}^{\mathbb{Q}_s}(s)
        \right\}\Big] \,\mathrm{d}u,
    \label{eq:app_duration_lambda_eff_derivative}\\
    \partial_R\bar c_{\mathrm{r}}^{(2)}(t,s)
    =&e^{-\int_0^t\lambda(u)\mathrm{d}u}\tilde P(s)
      \partial_R\bar\lambda_{\mathrm{r}}^{(2)}(t,s).
    \label{eq:app_duration_cr_derivative}
\end{align}

Using these results, the first-order interest rate sensitivities
of the individual price components are given by
\begin{align}
    \partial_R V_0(t,T)
    =&-\int_t^T (s-t)\bar c_0 DZ_{\mathrm c}\,\mathrm{d}s,
    \nonumber\\
    \partial_R V_{\mathrm r}^{(2)}(t,T)
    =&\int_t^T DZ_{\mathrm c}Z_{\mathrm r}^{(2)}
      \left[\partial_R\bar c_{\mathrm r}^{(2)}
      -(s-t+H^{(2)})\bar c_{\mathrm r}^{(2)}\right]\mathrm{d}s,
    \nonumber\\
    \partial_R V_{\mathrm{op}}^{(2)}(t,T)
    =&\int_t^T\bar c_0DZ_{\mathrm c}
       \left[-(s-t)(1-Z_{\mathrm r}^{(2)})
       +H^{(2)}Z_{\mathrm r}^{(2)}\right]\mathrm{d}s.
    \nonumber
\end{align}
Here, the arguments $(t,s)$ of the functions
within the integrals have been omitted for notational simplicity.
Substituting these expressions into Equation~\eqref{eq:duration_dcmp}
yields the duration contributions of the individual price components.
Alternatively, directly differentiating
the product representation of the MBS price gives
\begin{align}
    &\mathcal D^{(2)}(t,T)
    \nonumber\\
    =&\frac{1}{V^{(2)}(t,T)}
    \int_t^T DZ_{\mathrm c}Z_{\mathrm r}^{(2)}
    \Big[ (s-t+H^{(2)})\bar c^{(2)}
      -\partial_R\bar c_{\mathrm r}^{(2)}\Big]\,\mathrm{d}s.
    \label{eq:app_duration_closed_form_final}
\end{align}
Thus, $\mathcal D^{(2)}(t,T)$ can be evaluated
using analytical moments of truncated normal distributions
and deterministic time integrals,
without relying on finite differences or Monte Carlo simulation.

\subsection{Derivation of a Closed-Form Approximation for MBS Convexity}
\label{app:mbs_convexity_closed_form}

In this subsection, we derive an analytical expression
for the convexity $\mathcal{C}^{(2)}(t,T)$ of the MBS price
under the second-order cumulant approximation
by further differentiating the first-order interest rate sensitivities
derived in the preceding subsection.

First, we consider the second-order interest rate sensitivities
of the first and second cumulants.
For notational simplicity, we fix the $s$-forward probability measure
$\mathbb{Q}_s$ and define
\begin{align*}
    L_1(u)
    :=&
    \bar{\lambda}_{\mathrm{r},1}^{\mathbb{Q}_s}(u),
    \\
    L_2(u,\eta)
    :=&
    \bar{\lambda}_{\mathrm{r},2}^{\mathbb{Q}_s}(u,\eta)
\end{align*}
Throughout this subsection, primes denote derivatives
with respect to the interest rate level $R$.

For the integrand of the first cumulant,
further differentiating Equation~\eqref{eq:app_duration_lambda1_derivative}
yields, for $u>t$,
\begin{align}
    L_1''(u)
    =&
    \frac{10b}{B_{r,10}}
    \frac{\varphi(d(t,u))}
    {\sigma_r^{\mathbb{Q}_s}(t,u)}
    \Bigg[
        \mu_y^{\mathbb{Q}_s}(t,u)
        \nonumber\\
    &\quad+
        \frac{
            \operatorname{Cov}^{\mathbb{Q}_s}
            [r(u),y(u)\mid\mathcal{F}_t]
        }{
            \sigma_r^{\mathbb{Q}_s}(t,u)
        }
        d(t,u)
    \Bigg]
    \label{eq:app_convexity_lambda1_second_derivative}
\end{align}

Similarly, $L_2''(u,\eta)$ can be obtained
by further differentiating Equation~\eqref{eq:app_duration_lambda2_derivative}.
In this differentiation, the second derivative of the positive-part function
gives rise to contributions from the boundaries
at which the refinancing incentive is zero.
Specifically, defining $g(u):=m-a-bR_{10}(u)$,
we can express $L_2''(u,\eta)$ as
\begin{align}
    L_2''(u,\eta)
    =&
    b^2\mathbb{E}^{\mathbb{Q}_s}
    \Big[
        2y(u)y(\eta)
        \mathbf{1}_{\{g(u)>0\}}
        \mathbf{1}_{\{g(\eta)>0\}}
        \nonumber\\
    &\quad+
        y(u)\delta_0(g(u))
        \lambda_{\mathrm{r}}(\eta)
        \nonumber\\
    &\quad+
        \lambda_{\mathrm{r}}(u)
        y(\eta)\delta_0(g(\eta))
        \mid\mathcal{F}_t
    \Big]
    \label{eq:app_convexity_lambda2_second_derivative}
\end{align}
Here, $\delta_0$ denotes the Dirac delta function,
and the above expression is understood in the distributional sense
when taking expectations.
The second and third terms can be expressed
as products of conditional probability densities at the boundaries
and conditional expectations.
Under the Hull--White model,
$(r(u),r(\eta),y(u),y(\eta))$ follows a joint normal distribution.
Therefore, these expectations can be evaluated analytically
using moments of truncated normal distributions.

Consequently, the second derivatives
of the first and second cumulants are given by
\begin{align}
    \partial_R^2\kappa_1(t,s)
    =&
    \int_t^s L_1''(u)\mathrm{d}u,
    \label{eq:app_convexity_kappa1_second_derivative}
    \\
    \partial_R^2\kappa_2(t,s)
    =&
    2\int_t^s\int_u^s
    \Big[
        L_2''(u,\eta)
        -
        L_1''(u)L_1(\eta)
        \nonumber\\
    &\quad-
        2L_1'(u)L_1'(\eta)
        -
        L_1(u)L_1''(\eta)
    \Big]
    \mathrm{d}\eta\mathrm{d}u
    \label{eq:app_convexity_kappa2_second_derivative}
\end{align}

In addition to $H^{(2)}(t,s)$ defined in the preceding subsection,
we introduce
\begin{align}
    K^{(2)}(t,s)
    :=&
    \partial_R^2\kappa_1(t,s)
    -
    \frac{1}{2}\partial_R^2\kappa_2(t,s)
    \label{eq:app_convexity_K_def}
\end{align}
From Equation~\eqref{eq:Z_r_cumulant_2}, we obtain
\begin{align}
    \partial_R^2 Z_{\mathrm{r}}^{(2)}(t,s)
    =&
    Z_{\mathrm{r}}^{(2)}(t,s)
    \left[
        \left(H^{(2)}(t,s)\right)^2
        -
        K^{(2)}(t,s)
    \right]
    \label{eq:app_convexity_Zr_second_derivative}
\end{align}

Furthermore, the second derivative of the effective prepayment rate
defined in Equation~\eqref{eq:app_duration_lambda_eff}
is given by
\begin{align}
    &\partial_R^2
    \bar{\lambda}_{\mathrm{r}}^{(2)}(t,s)
    \nonumber\\
    =&
    L_1''(s)
    -
    \int_t^s
    \Big[
        L_2''(u,s)
        -
        L_1''(u)L_1(s)
        \nonumber\\
    &\qquad-
        2L_1'(u)L_1'(s)
        -
        L_1(u)L_1''(s)
    \Big]\mathrm{d}u
    \label{eq:app_convexity_lambda_eff_second_derivative}
\end{align}
Therefore,
\begin{align}
    \partial_R^2
    \bar{c}_{\mathrm{r}}^{(2)}(t,s)
    =&
    e^{-\int_0^t\lambda(u)\mathrm{d}u}
    \tilde{P}(s)
    \partial_R^2
    \bar{\lambda}_{\mathrm{r}}^{(2)}(t,s)
    \label{eq:app_convexity_cr_second_derivative}
\end{align}

Using these results, we derive the second-order interest rate sensitivities
of the individual price components.
In the following, we set $\tau=s-t$
and omit the arguments $(t,s)$ of the functions
within the integrals where appropriate.

First, for $V_0(t,T)$, we obtain
\begin{align}
    \partial_R^2 V_0(t,T)
    =&
    \int_t^T
    \tau^2\bar{c}_0DZ_{\mathrm{c}}
    \mathrm{d}s
    \label{eq:app_convexity_V0_second_derivative}
\end{align}

Next, applying the product rule to $V_{\mathrm{r}}^{(2)}(t,T)$ gives
\begin{align}
    &\partial_R^2 V_{\mathrm{r}}^{(2)}(t,T)
    \nonumber\\
    =&
    \int_t^T DZ_{\mathrm{c}}Z_{\mathrm{r}}^{(2)}
    \Big[
        \partial_R^2\bar{c}_{\mathrm{r}}^{(2)}
        -
        2(\tau+H^{(2)})
        \partial_R\bar{c}_{\mathrm{r}}^{(2)}
        \nonumber\\
    &\qquad+
        \left\{
            (\tau+H^{(2)})^2-K^{(2)}
        \right\}
        \bar{c}_{\mathrm{r}}^{(2)}
    \Big]\mathrm{d}s
    \label{eq:app_convexity_Vr_second_derivative}
\end{align}

Similarly, for $V_{\mathrm{op}}^{(2)}(t,T)$, we obtain
\begin{align}
    &\partial_R^2 V_{\mathrm{op}}^{(2)}(t,T)
    \nonumber\\
    =&
    \int_t^T
    \bar{c}_0DZ_{\mathrm{c}}
    \Big[
        \tau^2(1-Z_{\mathrm{r}}^{(2)})
        \nonumber\\
    &\qquad-
        Z_{\mathrm{r}}^{(2)}
        \left\{
            2\tau H^{(2)}
            +(H^{(2)})^2-K^{(2)}
        \right\}
    \Big]\mathrm{d}s
    \label{eq:app_convexity_Vop_second_derivative}
\end{align}

Substituting these expressions into Equation~\eqref{eq:convexity_dcmp}
allows us to decompose the overall MBS convexity
into the contributions of the individual price components.
Alternatively, directly differentiating the product representation
of the MBS price yields the following expression for
$\mathcal{C}^{(2)}(t,T)$:
\begin{align}
    &\mathcal{C}^{(2)}(t,T)
    \nonumber\\
    =&
    \frac{1}{V^{(2)}(t,T)}
    \int_t^T DZ_{\mathrm{c}}Z_{\mathrm{r}}^{(2)}
    \Big[
        \partial_R^2\bar{c}_{\mathrm{r}}^{(2)}
        -
        2(\tau+H^{(2)})
        \partial_R\bar{c}_{\mathrm{r}}^{(2)}
        \nonumber\\
    &\qquad+
        \left\{
            (\tau+H^{(2)})^2-K^{(2)}
        \right\}
        \bar{c}^{(2)}
    \Big]\mathrm{d}s.
    \label{eq:app_convexity_closed_form_final}
\end{align}

Equation~\eqref{eq:app_convexity_closed_form_final}
provides a closed-form approximation for MBS convexity
obtained by taking the second derivative
of the second-order cumulant approximation for the MBS price
with respect to the interest rate level $R$.
Since the required expectations can be evaluated analytically
using moments of truncated normal distributions,
convexity can be computed through deterministic time integrals,
without relying on Monte Carlo simulation
or finite differences with respect to the interest rate level.

\end{document}